\documentclass[twocolumn,
               aps,
               prd,   
               preprintnumbers,
               numbers,
               sort&compress,
               nofootinbib,
               showpacs,
               colorlinks,
               linkcolor=blue,   
               citecolor=blue]{revtex4-1}

\usepackage{graphicx,amsmath,amssymb,bm}
\usepackage{psfrag}
\usepackage{feynmp}
\usepackage{hyperref}
\usepackage{graphicx}
\usepackage{subcaption}
\usepackage{placeins}
\usepackage{xcolor} 

\def\<{\langle}
\def\>{\rangle}

\def\+{\dagger}

\def\U1A{U(1)$_{\rm A}$}

\def\ra{\rangle}
\def\la{\langle}
\def\rmd{\mathrm{d}}
 
\newcommand{\be}{\begin{eqnarray}}
\newcommand{\ee}{\end{eqnarray}}
\newcommand{\beq}{\begin{equation}}
\newcommand{\eeq}{\end{equation}}
\newcommand{\exclude}[1]{}
\newcommand{\jcap}{JCAP}

\newcommand{\physrep}{Physics Reports}

\newcommand{\Mead}[1]{{\color{red}(Mead: {#1})}}
\newcommand{\Liang}[1]{{\color{teal}(Liang: {#1})}}

\newcommand{\kmps}{\,\mathrm{km}\,\mathrm{s}^{-1}}
\newcommand{\Gyr}{\,\mathrm{Gyr}}

\begin{document}

\title{ Gravitationally trapped axions on Earth}

\author{Kyle Lawson$^{1}$}
\email{klawson@phas.ubc.ca}
\author{Xunyu Liang$^{1}$}
\email{xunyul@phas.ubc.ca}
\author{Alexander Mead$^{1,2}$}
\email{alexander.j.mead@googlemail.com}
\author{Md Shahriar Rahim Siddiqui$^{1}$}
\email{shahriar.naf07@gmail.com}
\author{Ludovic Van Waerbeke$^{1}$}
\email{waerbeke@phas.ubc.ca}
\author{Ariel Zhitnitsky$^{1}$}
\email{arz@phas.ubc.ca}
\affiliation{\\
1 Department of Physics and Astronomy, University of British Columbia, Vancouver, Canada\\
2 Institut de Ci\`encies del Cosmos, Universitat de Barcelona, Barcelona, Spain
}

\label{firstpage}

\begin{abstract}
 We advocate for the idea that there is a fundamentally new mechanism for axion production on Earth, as recently suggested in \cite{Fischer:2018niu,Liang:2018ecs}. We specifically focus on production of axions within Earth, with low velocities such that they will be trapped in the gravitational field. Our computations are based on the so-called Axion Quark Nugget (AQN) dark matter model, which was originally invented to explain the similarity of the dark and visible cosmological matter densities. This occurs in the model irrespective of the axion mass $m_\mathrm{a}$ or initial misalignment angle $\theta_0$. Annihilation of antimatter AQNs with visible matter inevitably produce axions when AQNs hit Earth. The emission rate of axions with velocities below escape velocity is very tiny compared to the overall emission, however these axions will be accumulated over the 4.5 billion year life time of the Earth, which greatly enhances the discovery potential. We perform numerical simulations with a realistically modeled incoming AQN velocity and mass distribution, and explore how AQNs interact as they travel through the interior of the Earth. We use this to estimate the axion flux on the surface of the Earth, the velocity-spectral features of trapped axions, the typical annihilation pattern of AQN, and the density profile of the axion halo around the Earth. Knowledge of these properties is necessary to make predictions for the observability of trapped axions using CAST, ADMX, MADMAX,  CULTASK, ORPHEUS, ARIADNE, CASPEr, ABRACADABRA, QUAX, ORGAN, TOORAD, DM Radio. 
   
\end{abstract}
\vspace{0.1in}


\maketitle
\section{Introduction}
\label{sec:1. introuction}


The Peccei-Quinn mechanism, accompanied by axions, remains the most compelling resolution of the strong charge-parity (CP) problem 
\cite[orignal papers:][]{1977PhRvD..16.1791P,1978PhRvL..40..223W,1978PhRvL..40..279W,KSVZ1,KSVZ2,DFSZ1,DFSZ2}\cite[recent reviews:][]{vanBibber:2006rb, Asztalos:2006kz,Sikivie:2008,Raffelt:2006cw,Sikivie:2009fv,Rosenberg:2015kxa,Marsh:2015xka,Graham:2015ouw,Ringwald:2016yge,Battesti:2018bgc}. In this model, conventional dark-matter (DM) axions are produced either by the misalignment mechanism \cite{1983PhLB..120..127P,1983PhLB..120..133A,1983PhLB..120..137D}, when the cosmological field $\theta(t)$ oscillates and emits cold axions before it settles at a minimum, or via the decay of topological objects \cite{Chang:1998tb,2012PhRvD..85j5020H,2012PhRvD..86h9902H,Kawasaki:2014sqa,Fleury:2015aca,Gorghetto:2018myk,Klaer:2017ond}. There are uncertainties in estimates of the axion abundance from these two channels,\footnote{\label{DM} According to recent computations \cite{Klaer:2017ond} the axion contribution to $\Omega_{\rm DM}$ as a result of decay of topological objects can saturate the observed DM density today if the axion mass is in the range $m_\mathrm{a}=(2.62\pm0.34)10^{-5} {\rm eV}$, while earlier estimates suggest that saturation occurs at a larger axion mass. There are additional uncertainties in this result, and we refer to the original studies \cite{Gorghetto:2018myk} on this matter. One should also emphasize that the computations  \cite{Chang:1998tb,2012PhRvD..85j5020H,2012PhRvD..86h9902H,Kawasaki:2014sqa,Fleury:2015aca,Gorghetto:2018myk,Klaer:2017ond} have been performed with assumption that Peccei-Quinn symmetry was broken after inflation.}, and we refer the reader to \cite{Chang:1998tb,2012PhRvD..85j5020H,2012PhRvD..86h9902H,Kawasaki:2014sqa,Fleury:2015aca,Gorghetto:2018myk,Klaer:2017ond} for a full discussion. In both of the traditional production mechanisms axions are produced as non-relativistic particles with typical $v_{\rm axion}/c\sim 10^{-3}$, and their contribution to the cosmological DM density scales as $\Omega_{\rm axion}\sim m_\mathrm{a}^{-7/6}$. This scaling implies that the axion mass must be tuned to $m_\mathrm{a}\simeq 10^{-5}$ eV in order to produce the observed cosmological DM density today. Higher axion masses will contribute very little to $\Omega_{\rm DM}$ and lower axion masses will over-saturate the density, resulting in a closed Universe which would be in strong conflict with cosmological data \cite{Planck2018}. Cavity-type experiments have the potential to discover these non-relativistic axions. 

Axions may also be produced via the Primakoff effect in stellar plasmas at high temperature \cite{Sikivie:1983ip}. These axions are ultra-relativistic; with a typical average energy of axions emitted by the Sun of $\la E\ra =4.2$ keV \cite{Andriamonje:2007ew}. Searches for Solar axions are based on helioscope instruments like CAST (CERN Axion Search Telescope) \cite{Andriamonje:2007ew}. 
 
Recent works \cite{Fischer:2018niu,Liang:2018ecs} have suggested a fundamentally novel mechanism for axion production in planets and stars, with a mechanism rooted in the so-called axion quark nugget (AQN) DM model \cite{Zhitnitsky:2002qa}, similar to the original quark nugget model by Witten \cite{Witten:1984rs} (see \citep{Madsen:1998uh} for a review). AQN are ``cosmologically dark", not because of the weakness of their interactions, but due to their small cross-section-to-mass ratio, which scales down many observable consequences of an otherwise strongly-interacting DM candidate. There are two additional elements in our AQN model compared to \cite{Witten:1984rs,Madsen:1998uh}. First, there is an additional stabilization factor for the nuggets provided by axion domain walls that are produced in great quantity during the quantum chromodynamic (QCD) transition and that help to alleviate a number of problems with the original nugget model.\footnote{\label{first-order}In particular, a first-order phase transition is not required as the axion domain wall plays the role of the squeezer. Another problem with \cite{Witten:1984rs,Madsen:1998uh} is that nuggets will likely evaporate on a Hubble time-scale even. For the AQN model this argument is not applicable because the vacuum-ground-state energies inside (color-superconducting phase) and outside (hadronic phase) the nugget are drastically different. Therefore, these two systems can coexist only in the presence of an external pressure, provided by the axion domain wall. This contrasts with the original model \cite{Witten:1984rs,Madsen:1998uh}, which must be stable at zero external pressure.} Another feature of AQN is that nuggets can be made of {\it matter} as well as {\it antimatter} during the QCD transition. This element of the model completely changes the AQN framework \cite{Zhitnitsky:2002qa} because the DM density, $\Omega_{\rm DM}$, and the baryonic matter density, $\Omega_{\rm visible}$, automatically become similar to each other, i.e. $\Omega_{\rm DM}\sim \Omega_{\rm visible}$, because they have the same QCD origin in that the same generating mechanism ensures they are of the same order of magnitude without any fine tuning as they both proportional to one and the same fundamental $\Lambda_{\rm QCD}$ scale. The existence of both matter and antimatter AQNs explains the asymmetry between visible components made of matter and antimatter. In other words, 
in the AQN framework the conventional idea of  baryogenesis is replaced by the charge separation mechanism when the net baryon charge remains zero in the entire Universe. In short, the AQN model explains two fundamental questions at the same time: the nature of dark matter and the asymmetry between matter and antimatter. This is irrespective of specific details of the model, such as the axion mass $m_a$ or misalignment angle $\theta_0$. 

The presence of bound axions in the construction of AQNs plays a crucial role for the present work: Annihilation events will always happen when an AQN propagates in a medium, and this inevitably produces propagating axions along with the conventional products of baryon-antibaryon annihilation. Emitted axions will be released with a velocity spectrum, with a typical value $v_{\rm axion}^{\rm AQN}\simeq 0.5 c$. This can be contrasted with conventional galactic non-relativistic axions $v_{\rm axion}\sim 10^{-3}c$ and Solar ultra-relativistic axions with typical energies $\la E\ra =4.2$ keV. 

Some of the axions emitted when AQNs annihilate with Earth will have very small velocities, below the escape velocity of Earth ($v\leq v_{\rm esc}\simeq 11 \kmps$). These axions will be gravitationally trapped and represent the main subject of this work. The idea that axions can be trapped in a gravitational field dates back to 2003 paper by DiLella and Zioutas \cite{DiLella:2002ea}. The present work is a specific analysis of the gravitationally bound axions within the AQN framework, when the axions are produced in the interior of the Earth, as suggested in \cite{Fischer:2018niu,Liang:2018ecs}. The emission spectrum of axions from the AQN production mechanism is studied in Ref. \cite{Liang:2018ecs}. An important conclusion of this work is that a tiny fraction $\sim10^{-17}$ of the number of axions emitted within the Earth will be low-velocity ($v\leq v_{\rm esc}$),   and thus will be trapped by the Earth. Over 4.5 billion years of accumulation these bound axions may become dense enough for detection by instruments such as ADMX and ADMX-HF \cite{Stern:2016bbw}, upgraded CAST \cite{Anastassopoulos:2017ftl}, ORPHEUS \cite{Rybka:2014cya}, MADMAX \cite{TheMADMAXWorkingGroup:2016hpc}, ORGAN \cite{McAllister:2017lkb,2019PDU....25..306F},  CULTASK \cite{Chung:2018wms}, ARIADNE \cite{Geraci:2017bmq}, CASPEr \cite{JacksonKimball:2017elr}, ABRACADABRA \cite{Kahn:2016aff}, QUAX \cite{Barbieri:2016vwg}, X3 \cite{Brubaker:2016ktl}, DM Radio \cite{Chaudhuri:2018rqn}, TOORAD \cite{2018arXiv180708810M},
axion plasma haloscope\cite{Lawson:2019brd}.

In \cite{Liang:2018ecs} a number of order-of-magnitude estimates were made regarding properties of the trapped-axions. The goal of the present work is to upgrade these estimates by performing detailed numerical simulations of axion production within the Earth. Questions that we are interested in answering are:
\begin{enumerate}
    \itemsep0em
    \item What fraction of an AQN is annihilated when it passes through the interior of the Earth?\\
    \item What is the distribution of where annihilation events occur within the Earth?\\
    \item How is the trapped-axion density distributed around the Earth?\\
    \item What is the trapped-axion flux through the surface of the Earth, where they might be detected?\\
    \item What are the axion spectral features on the surface of the Earth?\\
    \item How do answers to the above questions depend on the AQN mass distribution and other model-dependent features?\\
\end{enumerate}

Assuming that AQNs saturate the cosmological DM density, and defining $\Delta B/B$ as the fraction of a typical AQN that is annihilated when passing through the interior of the Earth, one can estimate the total number of gravitationally trapped axions accumulated by the Earth over 4.5 billion years as follows  \cite{Liang:2018ecs}:
\begin{equation}
\label{eq:1. total trapped axion N}
\begin{aligned}
N_{\oplus, \rm 4.5Gyr}^{\rm(trap)}
&\sim 4\pi R_\oplus^2\frac{\rho_{\rm DM}v_{\rm AQN}}{E_\mathrm{a}}\xi_\oplus
\left(\frac{\Delta B}{B}\right)
\cdot{4.5\,\rm Gyr}  \\
&\sim 10^{40}\left(\frac{\xi_\oplus}{10^{-17}}\right)
\left(\frac{\Delta B}{B}\right)
\left(\frac{\rm 10^{-5}eV}{m_\mathrm{a}}\right)\ ,
\end{aligned}
\end{equation}
where $\xi_\oplus\sim 10^{-17} $ is the fraction of axions trapped in the gravitational field of the Earth compared to the total axion flux, estimated in \cite{Fischer:2018niu,Liang:2018ecs}.
The corresponding accumulated energy is
\begin{equation}
\label{eq:1. total trapped axion E}
\begin{aligned}
E_{\oplus, \rm 4.5Gyr}^{\rm(trap)}
&\simeq m_\mathrm{a}\cdot N_{\oplus, \rm 4.5Gyr}^{\rm(trap)} \\
&\sim 10^{35}\left(\frac{\xi_\oplus}{10^{-17}}\right)
\left(\frac{\Delta B}{B}\right)
\rm eV\ .
\end{aligned}
\end{equation}
The trapped energy \eqref{eq:1. total trapped axion E} can be expressed in terms of the trapped-axion mass $\Delta M_\oplus$ accumulated by the Earth
\begin{equation}
\label{eq:1. total trapped axion E in kg}
\Delta M_\oplus^{\rm (trap)}
\sim0.1\left(\frac{\xi_\oplus}{10^{-17}}\right)
\left(\frac{\Delta B}{B}\right)\mathrm{kg}\ ,
\end{equation}
which represents a tiny fraction of the mass of the Earth: $M_\oplus\simeq 5.9\times 10^{24}$ kg.

The key unknown element in estimates (\ref{eq:1. total trapped axion N}) and (\ref{eq:1. total trapped axion E}) is the parameter $\Delta B/B$: the fraction of the mass of a typical AQN that is annihilated within Earth. This parameter is difficult to estimate without numerical simulations because it depends on the trajectory of the AQN through Earth, which itself is affected by the the incoming AQN velocity distribution, the gravity of the Earth, the details of the density distribution within the Earth, as well as the axion-emission velocity spectrum.

If we make the assumption that axions are uniformally distributed within the Earth, we can estimate the average axion energy density within the Earth \cite{Liang:2018ecs}:
\begin{equation}
\label{eq:1. rho_a}
\rho_{a,\oplus}^{(\rm avg)}
\sim\frac{E_{\oplus, \rm 4.5Gyr}^{\rm(trap)}}{4\pi R_\oplus^3/3}
\sim 0.1
\left(\frac{\xi_\oplus}{10^{-17}}\right)
\left(\frac{\Delta B}{B}\right)
{\rm \frac{GeV}{cm^3}}\ .
\end{equation}
However, it is clear that the distribution of axions will not be uniform and instead, may be non-trivial, in a way determined by the position of the AQN when emission occurs and the direction of AQN velocity at the moment of emission. Evidently, the crude estimate of \eqref{eq:1. rho_a} is inadequate for planning experiments. To address this (and items 3, 4 and 5 from the list above) we use Monte-Carlo simulations to allow us to evaluate \eqref{eq:1. rho_a} precisely, as a function of a position from the Earth. This information is crucial for planning axion search experiments, which require a value for $\rho_{a,\oplus}(r=R_{\oplus})$ along with the spectral properties of the axions on the surface $r=R_{\oplus}$. Specifically, we simulate the entire process responsible for the generation of the axion halo around Earth. We start by simulating the propagation of AQNs through Earth, including the AQN annihilation processes. We then simulate the orbits of the low-velocity axions that are emitted by the AQN and calculate the accumulation over 4.5 Gyr. This axion-simulation process depends on the position of the AQN within the Earth where annihilation occurs and the axions are emitted, the velocity of the AQN at the moment of emission, and the velocities and directions of the emitted axions. This procedure allows us to compute $\rho_{a,\oplus}(r=R_{\oplus})$. 

The present work is a natural continuation of \cite{Fischer:2018niu,Liang:2018ecs}. To avoid repetition we refer the readers to \cite{Liang:2018ecs} for the derivation details of axion emission spectrum in the rest frame of the AQN and to \cite{Fischer:2018niu} for an overview of the AQN model, including motivation, consequences and predictions. We note that the AQN model is consistent with all available cosmological, astrophysical, satellite and ground-based constraints. While the model was invented to explain the observed relation $\Omega_{\rm DM}\sim \Omega_{\rm visible}$, it may also explain a number of other observed phenomena, such as the excess of diffuse galactic emission in different frequency bands (including the 511 keV line) as reviewed in \cite{Lawson:2013bya}. AQNs may also offer a resolution to the so-called ``Primordial Lithium Puzzle" \cite{Flambaum:2018ohm} and to the ``The Solar Corona Mystery"  \cite{Zhitnitsky:2017rop,Raza:2018gpb} and may also explain the recent EDGES observation \cite{Lawson:2018qkc}, which is in some tension with the standard cosmological model. It is the same set of physical AQN parameters that address the aforementioned phenomena that we will adopt for the present work. What differentiates the present work is that we are advocating a direct search for the gravitationally trapped axions: a \emph{direct} manifestation of the AQN model. An observation of axions with very distinct spectral properties compared to those predicted from conventional galactic axions would be a smoking gun for the AQN framework and this may then answer a fundamental question on the nature of DM.

This paper is organized as follows: In Sec. \ref{sec:2. overview}, we present a brief overview of AQN model: its accomplishments, generic consequences and observational constraints. We also overview the axion emission spectrum, which plays an important role for the present study. In Sec. \ref{sec:3. setup}, we develop analytical equations to describe the annihilation of AQNs impacting the Earth and the emission spectrum of axions in laboratory frame. The corresponding numerical simulations are described in Sec. \ref{sec:4. algorithm}, with results presented in Sec. \ref{sec:5. axion profiles}. Finally, we conclude with some thoughts on possible future developments in Sec. \ref{sec:6.conclusion}.

\section{Overview of the AQN model}
\label{sec:2. overview}

\subsection{Basic and generic features of the AQN framework}

The original AQN model was developed by \cite{Zhitnitsky:2002qa} as a natural explanation to two of the most fundamental problems in cosmology: namely, the nature of DM and the asymmetry between matter and antimatter. To be more specific, ``baryogenesis" in this model is replaced by ``baryon charge separation" such that the DM density $\Omega_{\rm DM}$ and the baryonic matter density $ \Omega_{\rm visible}$ become similar to each other. Indeed, the conservation of the baryon charge implies
 \be
  \label{eq:321}
    B_{\text{universe}} = 0 &=& B_{\text{nugget}}
    + B_{\text{visible}}- |{B}|_{\text{antinugget}}\nonumber \\
    |B|_{\text{DM}} &=& B_{\text{nugget}} +
   |{B}|_{\text{antinugget}}  
\ee
where $B_{\text{universe}}=0$ is the total number of baryons in the universe, $|B|_{\text{DM}}$ counts total number of baryons and antibaryons hidden in both the nuggets and antinuggets that make up the DM. $B_{\text{visible}}$ is the total number of residual ``visible'' baryons (regular matter). The energy per baryon charge is approximately the same for AQNs and visible matter, because both types of matter are formed during the same QCD transition, and both are proportional to the same dimensional parameter of proton mass, $m_\mathrm{p}$, which implies that 
\be
\label{Omega}
 \Omega_{\rm DM}\sim \Omega_{\rm visible}\ .
\ee
In other words, the nature of DM and the problem of the Universal asymmetry between matter and antimatter become two sides of the same coin in this framework. As has been argued in refs.   \cite{Liang:2016tqc,Ge:2017ttc,Ge:2017idw} the relation (\ref{Omega})  is a very generic outcome of the AQN framework, and it is not sensitive to specific details of the model.


To be more specific, from the observed ratio $ \Omega_{\rm DM}\simeq 5\cdot \Omega_{\rm visible}$, one can infer that number of antinuggets is larger than the number of nuggets by a factor of $\sim3/2$ at the conclusion of AQN formation. This results in a Universal matter budget containing baryons, quark nuggets, and antiquark nuggets in an approximate ratio,
\begin{equation}
\label{eq:2 AQN ratio}
|B_{\rm visible}|:|B_{\rm nuggets}|:|B_{\rm antinuggets}|
\simeq1:2:3
\end{equation}
with a net baryonic charge of zero. 

The axion field, which is a crucial element in the present work, plays an essential role during AQN formation in two ways: First, the existence of a coherent ${\cal CP}$-odd axion field $\theta$ is the fundamental mechanism to break the global symmetry of baryon charge between nuggets and antinuggets. This difference is always an order unity effect, irrespective of the parameters of the theory, as argued in \cite{Liang:2016tqc,Ge:2017ttc}. This is precisely the reason why the resulting visible and DM densities must be the same order of magnitude (\ref{Omega}) in this framework, as they are both proportional to the same fundamental $\Lambda_{\rm QCD}$ scale, and they both originated at the same QCD epoch. If these processes are not fundamentally related, the two components $\Omega_{\rm DM}$ and $\Omega_{\rm visible}$ could easily exist at vastly different scales. Second, the so-called ${ N}_{\rm DW}=1$ axion domain walls provide sufficient pressure to accumulate baryon charge inside the walls, and ultimately lead to formation of stable AQNs in color-superconducting phase of size $R\sim m_\mathrm{a}^{-1}$, discussed in detail in \cite{Liang:2016tqc}. 

It is precisely this energy, hidden in form of the bound axion domain wall since the moment of formation, that can be released in form of free-propagating axions when the baryon charge from antimatter AQNs are annihilated with matter in the interior of the Earth. The present work is devoted to an analysis of the fate of these, previously hidden, axions. In the following subsection we briefly review the basic features of the AQN model relevant for the present work.

\subsection{Observational constraints on $\la B\ra$}
\label{subsec:2.1 AQN structure}

An AQN is analogous to a gigantic atom, with a central ``nucleus'' made out of quark condensate in the color-superconducting phase enclosed by a thick domain wall made out of axion and $\eta'$ meson. Because the quark core is in general not neutral, it is surrounded by a cloud of leptons in what is known as the ``electrosphere''. Once an AQN made of antimatter impacts upon matter annihilation will occur. As a consequence, the AQN will loose mass and emits axions.

The flux of AQNs on Earth is determined by the AQN mass $M_N$, or equivalently by the baryon charge $B$, related via $M_N\simeq m_\mathrm{p} B$. Assuming that AQNs saturate the DM density the AQN flux can be approximated as 
\begin{equation}
\label{eq:flux1}
\Phi = n_{\rm N} v_{\rm N} \approx \frac{\rho_{\rm DM} v_{\rm N}}{M_{\rm N}} \approx 1~{\rm{km}}^{-2}{\rm{yr}}^{-1}
\left( \frac{\langle B \rangle}{10^{24}} \right)^{-1},
\end{equation} 
so that direct detection experiments impose lower limits on the value of $\langle B \rangle$ for the baryon-charge distribution of the AQN.

The strongest direct detection limit on AQN is likely set by the non-detection of an AQN flux \cite{Aartsen:2014awd} by the IceCube Observatory, which limits the AQN flux to $\Phi \lesssim 1$ km$^{-2}$yr$^{-1}$. If we take the local DM mass density to be $\rho \approx 0.3$ GeV/cm$^{3}$ and assume that AQN make up all DM we may translate the flux constraint obtained from IceCube into a lower limit with $3.5\sigma$ confidence on the mean baryon number of the AQN distribution of 
\be
\label{direct}
\la B \ra > 3\times 10^{24} ~~~[{\text{non-detection constraint}]},
\ee
see Appendix \ref{sec:Constraint from IceCube} for details. A similar constraint also comes from the Antarctic Impulsive Transient Antenna  (ANITA) \cite{Gorham:2012hy}. 

Limits on $\la B\ra$ may also be obtained from indirect astrophysical and cosmological observations. Photon production during matter-AQN interactions along a given line of sight implies an observable flux:
\begin{equation}
\label{eq:observable}
\Phi \sim R^2\int \rmd\Omega \rmd l [n_{\rm visible}(l)\cdot n_{\rm DM}(l)] \sim \frac{1}{\langle B \rangle^{1/3}},
\end{equation}
where $R\sim B^{1/3}$ is the typical size of the AQN, which determines the effective cross section of interaction between DM and visible matter, while $n_{\rm DM}\sim B^{-1}$ is the number density of the AQNs. Thus, astrophysical constraints impose a lower bound on $\langle B \rangle$. The annihilation contribution to the galactic spectrum has been analyzed for different frequencies from a radio bands to X-ray and $\gamma$-ray photons. In particular, if $\langle B \rangle\sim 10^{25}$ the AQN model could offer an explanation for the observed 511 keV emission features (including the width, morphology, $3\gamma$ continuum spectrum, etc). In all other cases, where computations have been performed in different frequency bands, the predicted emission is consistent with observations. In fact, in some cases it could explain some, or even the dominant, portion of the observed excess of radiation. All these  emissions in different frequency bands are expressed in terms of one and the same integral (\ref{eq:observable}), and therefore, the relative intensities are completely determined by internal structure of the AQNs, which is described by conventional nuclear physics and QED. For further details see the original works 
\cite{Oaknin:2004mn, Zhitnitsky:2006tu,Forbes:2006ba, Zhitnitsky:2006vt, Lawson:2007kp,
Forbes:2008uf,Forbes:2009wg,Lawson:2012zu} with specific computations in different frequency bands in galactic radiation, and a short overview \cite{Lawson:2013bya}. 


\subsection{Baryon charge distribution of AQN model}
\label{subsec:2.2 baryon charge distribution of AQNs}

Another AQN-connected observation might be related to the so-called ``Solar corona heating mystery". It has been a long-standing puzzle that the corona has a temperature $T\simeq 10^6$K that is 100 times hotter than the surface temperature of the Sun, and conventional astrophysical sources fail to explain the extreme UV (EUV) and soft X-ray radiation from the corona, 2000 km above the photosphere. This puzzle may find a resolution within the AQN framework, as recently argued in \cite{Zhitnitsky:2017rop,Zhitnitsky:2018mav,Raza:2018gpb}. The basic idea is that nanoflares, postulated by Parker to explain the Solar corona heating, can be identified with the AQN annihilation events. In \cite{Raza:2018gpb} it was shown that most of the Solar-incident AQN are are annihilated in the transition region between the photosphere and corona, specifically 2000 km above the photosphere, where drastic changes in temperature and pressure are known to occur. Solar flares are sensitive to the distribution $\rmd N/\rmd B$ because the AQN framework allows us to connect the nanoflare-size distribution, $\rmd N/\rmd W$, to the baryon-charge distribution. This is possible due to $\rmd N/\rmd W$ being modelled via magnetic-hydro-dynamics (MHD) simulations \cite{Pauluhn:2006ut,Bingert:2013} in such a way that the Solar observations match simulations. The connection with AQN is to ensure that the energy distribution of nanoflare events in the Solar corona equates with the baryon charge distribution within AQN framework \cite{Zhitnitsky:2017rop}, i.e.
\begin{equation}
\label{eq:2.2 W and B}
\rmd N\sim B^{-\alpha}\rmd B\sim W^{-\alpha}\rmd W
\end{equation}
where $\rmd N$ is the number of nanoflare events per unit time, with energy between $W$ and $W+\rmd W$, which occur as a result of complete annihilation of the antimatter AQN carrying the baryon charges between $B$ and $B+\rmd B$. The parameter $\alpha$ is fixed to match observations \cite{Pauluhn:2006ut,Bingert:2013} of the nanoflare distribution. \exclude{\Mead{Is this correct?}\Liang{I think it is correct. But should be confirmed by Ludo or Eric. They know this better. }\textbf{KYLE: Technically it is fixed to match MHD simulations of flare heating as there are no direct observations at the relevant energies.}}


In what follows we use the same baryon charge distribution $\rmd N/\rmd B$ with the purpose of analyzing the axion emission from AQN annihilated in the interior of the Earth. One should also note that it has been recently argued \cite{Ge:2019voa} that the algebraic scaling (\ref{eq:2.2 W and B}) is a generic feature of the AQN formation mechanism based on percolation theory. The phenomenological parameter $\alpha$ is determined by the properties of the domain wall formation during the QCD transition in the early Universe, but it cannot be theoretically computed in strongly coupled QCD. Instead, it will be constrained based on the observations as discussed below. 


If we assume that the baryon charge distribution follows a power-law as suggested in (\ref{eq:2.2 W and B}) the average baryon number of the distribution is
\begin{equation}
\label{eq:2.2 f(B)}
\langle B\rangle 
=\int_{B_{\rm min}}^{B_{\rm max}}\rmd B~B f(B),
\qquad f(B)\propto B^{-\alpha}
\end{equation}
where $f(B)$ is normalized and the power-law is taken to hold in the range from $B_{\rm min}$ to $B_{\rm max}$. This range is determined by the complicated formation and evolution of the AQN as discussed in \cite{Ge:2019voa} but is not strongly constrained from the theoretical side. Model independent constraints on heavy dark matter candidates requires an average mass above $\sim 55$ g (corresponding to $\la B \ra \sim 10^{25}$) while lunar seismology disfavours a significant AQN population in the 10 kg to one ton range ($B \sim 10^{28}-10^{30}$) as reviewed in \cite{Lawson:2013bya,Jacobs:2014yca}. However these constraints do not constrain the distribution of AQN within this range.
\exclude{\Mead{Is it possible to explain in a single extra sentence where the limits $10^{23}$ and $10^{28}$ actually come from? e.g., the lower limit arises from $x$ and the upper limit arises because too many high-mass AQNs would produce too many extreme Solar flares.}\Liang{I am feeling such statement might be too strong. Current observations constrain the value of $\langle B\rangle$, i.e. the value on the left hand side of Eq. (12), but tells not much about  $B_{\rm min}$ nor $B_{\rm max}$. Solar flares may be special and tell something about the distribution of $B$ especially the upper end  $B_{\rm max}$. But this constraint only applies based on assumption that AQN resolves the solar flare problem. If AQN has nothing to do with this problem, there is no such a constraint.} \Mead{I see, but if the reason we picked $10^{28}$ as an upper limit is due to the Solar-flare thing then we should probably mention this? But we could also mention that it could be higher?}\textbf{KYLE: The constraints on the mean value $\la B \ra$ come from a variety of DM non-detections. This is different from the actual distribution models used in fitting the nanoflare distributions. I'll try to rewrite this section to make that more clear. } \Mead{Great, I'll leave it to you guys to rewrite this. Please feel free to delete my comments when they are no longer relevant.}}
Rather than introducing a completely new set of distributions we will assume that the baryon number distribution follows the nanoflare distribution as implied by expression (\ref{eq:2.2 W and B}). Following the flare distribution models considered in \cite{Pauluhn:2006ut,Bingert:2013}, we investigate three different choices for the power-law index $\alpha$:
\begin{equation}
\label{eq:2.2 f(B) ass_alpha}
\alpha=2.5,~2.0,~{\rm or}\left\{
\begin{aligned}
&1.2  &B\lesssim 3\times 10^{26}  \\
&2.5  &B\gtrsim  3\times 10^{26}.
\end{aligned}\right.
\end{equation}
We also investigate two different choices of $B_{\rm min}$: $10^{23}$ and $3\times10^{24}$ while we fix $B_{\rm max}=10^{28}$. Therefore, we have a total of 6 different models for $f(B)$ corresponding to different models from \cite{Pauluhn:2006ut,Bingert:2013}. In Table \ref{table:2.2 mean B} we show the mean baryon charge $\langle B \rangle$ for each of the 6 models.
\begin{table} [h] 
	\caption{Values of the mean baryon charge $\langle B\rangle$ for different parameters of the AQN mass-distribution function.} 
	\centering 
	\begin{tabular}{c|ccc} 
		\hline\hline
		$(B_{\rm min},\alpha)$ & 2.5                 & 2.0                 & (1.2, 2.5)          \\ \hline
		$10^{23}$                   & $2.99\times10^{23}$ & $1.15\times10^{24}$ & ${4.25\times10^{25}}$ \\ 
		$3\times10^{24}$                   & $8.84\times10^{24}$ & $2.43\times10^{25}$ & ${1.05\times10^{26}}$ \\ \hline
	\end{tabular}
	\label{table:2.2 mean B} 
\end{table}


For the simulations in this work, we will only investigate parameters that give $\langle B\rangle\gtrsim10^{25}$ because present constraints require $\langle B\rangle > 3\cdot 10^{24}$ according to (\ref{direct}). This means excluding two models: that with $B_{\rm min} \sim 10^{23}$ and that with $\alpha=2, 5$ and $\alpha=2$.

\subsection{Axion emission mechanism}
\label{subsec:2.3 Spectral properties}

The goal of this subsection is to explain the process of axion emission when an AQN enters impacts upon the Earth and annihilates. We refer the readers to Refs. \cite{Fischer:2018niu,Liang:2018ecs} for the technical details on the computation of the axion spectrum. 

First, one should recall that the axion plays a key role in the AQN construction: due to the domain wall pressure the AQN remain stable on cosmological time scales. The same domain walls that provide the stability of AQN also contribute to their energy. The analysis of \cite{Ge:2017idw} suggests that the axion domain wall contributes $\sim1/3$ to its total energy. It is this energy in the form of free, propagating axions that can be released when the antibaryon charge from an AQN is annihilated by surrounding material. 

Second, note that an axion domain wall in the equilibrium does not emit any axions as a result of a purely kinematical constraint: the domain wall axions are off-shell axions in equilibrium. Now consider an AQN losing mass due to annihilation processes: the axion portion to the energy is initially unchanged, as it is not being directly annihilated. However, time-dependent perturbations due to annihilation processes obviously change this equilibrium configuration, and the configuration becomes unstable with respect to the axion emission because the total energy of the system is no longer at its minimum when some portion of the baryon charge in the core is annihilated. To retrieve the ground state, an AQN will therefore lower its domain wall contribution to the total energy by radiating axions. To summarize: the emission of axions is an inevitable consequence during the annihilation of antimatter AQNs and is a consequence of AQN minimizing their binding energy.

A mathematically consistent procedure that allows one to compute the emitted axion-velocity spectrum was developed in \cite{Liang:2018ecs} and can be explained as follows: Consider a general form of a domain wall solution as follows:
\begin{equation}
\label{eq:2.3 phi soln}
\phi(R_0)=\phi_w(R_0)+\chi
\end{equation}
where $R_0$ is the radius of the AQN, $\phi_w$ is the classical solution of the domain wall, while  $\chi$ describes excitations due to the time-dependent perturbation. Note that, $\phi_w$ is clearly off-shell time-independent classical solution, while $\chi$ describes on-shell propagating axions. Thus, whenever the domain wall is excited, corresponding to $\chi\neq0$, freely propagating axions will be produced and emitted by the excitation modes.

Suppose an AQN is travelling in vacuum where no annihilation events take place. We expect the solution stays in its ground state $\phi(R_0)=\phi_w(R_0)$, which corresponds to the minimum energy state. Since there is no excitation (i.e. $\chi=0$), no free axion can be produced. However, the scenario changes when baryon charge from the AQN is annihilated. As the AQN starts to loose a small amount of mass, its size consequentially shrinks from $R_0$ to a slightly smaller radius $R_{\rm new}=R_0-\Delta R$. The quantum state $\phi(R_0)=\phi_w(R_0)$ is then no longer the ground state, because a lower energy state $\phi_w(R_{\rm new})$ becomes available. Then, we may write the current state of the domain wall as $\phi(R_0)=\phi_w(R_{\rm new})+\phi_w'(R_{\rm new})\Delta R$, so the domain wall now has a nonzero exciting mode $\chi=\phi_w'(R_{\rm new})\Delta R$ and free axions can be produced during oscillations of the domain wall.

The annihilation processes inside the AQN force the surrounding domain wall to oscillate. These oscillations of the domain wall generate excitation modes and ultimately lead to radiation of the propagating axions. The corresponding oscillations have a typical frequency of order the axion mass $\sim m_\mathrm{a}$; the size of the AQN is also order of $m_\mathrm{a}^{-1}$, such that typical velocity of the oscillations of order $c$. These simple dimensional arguments suggest that the typical velocities of the emitted axions will be of order $c$, i.e. the most of the radiated axions will have relativistic velocities, in contrast with conventional galactic axions. However, the present work is devoted to the low energy portion  of this spectrum with $v\lesssim v_{\rm esc}$, which is suppressed as $v^3$ at $v/c\ll 1$ \cite{Liang:2018ecs}. Explicit computations presented in \cite{Liang:2018ecs} support all these claims of the basic features of the axion emission and the corresponding spectrum. 

\section{Annihilation modeling}
\label{sec:3. setup}

In this section we present all equations that will be used in our simulations and calculations of the axion-density distribution around the Earth and the speed spectrum of axions passing through the surface of the Earth. In what follows we adopt the following terminology for the sake of brevity: whenever ``per AQN'' is referred, we mean ``per $\langle B\rangle$ baryon charge''.

\subsection{Annihilation of AQNs impacting the Earth}
\label{subsec:3.1 annihilation of AQNs impacting the Earth}

The equations derived in this subsection are similar to equations discussed in \cite{Raza:2018gpb}, which was devoted to annihilation of AQNs in the Solar corona. However, there are two main differences between our work and that of \cite{Raza:2018gpb}: First, in our work, the motion of AQNs in the interior of the Earth will experience differing strengths of local gravity because we find that the average AQN will only be partially annihilated when passing through Earth. This is in contrast to Ref. \cite{Raza:2018gpb} where AQNs are almost completely annihilated in the transition region 2000 km above the Solar photosphere and there is an extremely small probability for an AQN to survive until the Solar surface. The second distinct feature of our modeling is related to the different structural properties of the Sun and Earth. The Solar corona is a highly-ioninized plasma, where protons, ions and electrons move freely and experience a long-range Coulomb interaction with AQNs due to the induced electric charge on the AQN at $T\neq 0$, see estimates in Appendix A in ref. \cite{Zhitnitsky:2017rop}. This should be contrasted with the structure of Earth, where most atoms are neutral or weakly ionized. Furthermore, most material on Earth are in the solid or liquid phase, with low mobility of the constituents. As a result, atoms from the interior of the Earth interact with AQNs on as a result of head-on collisions with the cross section $\sigma\sim \pi R^2$. This key difference ensures that AQN interactions with Earth have a much smaller effective cross section compared to interactions between AQN and the Solar corona.
 

The energy loss due to the collision of AQNs with the Earth can be expressed as follows \cite{DeRujula:1984axn}
\begin{equation}
\label{eq:3.1 energy loss_a}
\frac{\rmd E}{\rmd s}=\frac{1}{v}\frac{\rmd E}{\rmd t}=-\sigma\rho v^2
\end{equation}\ ,
where $E=mv^2/2$ is the kinetic energy, $s$ is the path length, $\sigma$ is the effective cross section of the AQN, $\rho$ is the density of the local environment and $v$ is the AQN velocity. Note that, following from the definition, we can express the time derivative of $E$ in the form
\begin{equation}
\label{eq:3.1 energy loss_b}
\frac{\rmd E}{\rmd t}
=mv\frac{\rmd v}{\rmd t}+\frac{1}{2}v^2\frac{\rmd m}{\rmd t}
=mv\frac{\rmd v}{\rmd t}-\frac{1}{2}\sigma\rho v^3\ ,
\end{equation}
where in the last step, we substitute the conventional rate of mass loss
\begin{equation}
\label{eq:3.1 mass loss}  
\dot{m}=-\sigma\rho v\ .
\end{equation}
Comparing Eqs. \eqref{eq:3.1 energy loss_a} and \eqref{eq:3.1 energy loss_b}, we conclude that the mass loss acts like a friction term in the equation of motion
\begin{equation}
\label{eq:3.1 friction term}
m\frac{\rmd v}{\rmd t}
=-\frac{1}{2}\sigma\rho v^2
\equiv-\frac{1}{2}\pi (\epsilon R)^2\rho v^2\ .
\end{equation}
In the last step, we use $\sigma\equiv\pi (\epsilon R)^2$, where $R$ is the radius of the AQN and $\epsilon$ is a dimensionless parameter that describes deviations from the geometrical cross section $\pi R^2$. In the studies of the Solar corona preformed in Ref. \cite{Raza:2018gpb}, $\epsilon_{\rm corona}\gg 1$ due to the high mobility of ions in plasma. In our case we expect $\epsilon\sim 1$, but we leave it as a free parameter to account for our ignorance regarding a number of physical effects that we have neglected but that we believe may be important. The main effects that $\epsilon \ne 1$ could describe are partial ionization of the AQN, the fact that AQN travel at supersonic speeds. Both effects can change the internal-core heating of the AQN and therefore change the cross-section. However, in our case, the rigidity of the surrounding material while the AQN traverses the Earth, implies that $\epsilon$ is likely to remain very close to one. We therefore set $\epsilon=1$ in most simulations by default, but we also investigate a much larger value of $\epsilon=3$ in order to assess how our results are modified when the AQN cross-section is significantly increased. 
\exclude{\textbf{KYLE: Is there a reason for the choice $\epsilon =3$? I'm not sure it should be considered a `high' value as it is still much smaller than in the solar corona.}\Liang{Yes we do not know the exact value of $\epsilon$ as it is a phenomeogical parameter. But we know the temperature of the Earth is a lot `colder' than the Sun. Also, typical speed of AQN is $\sim200\kmps$, it penetrates the Earth within seconds. There is little time for an AQN to heat up significantly, so we in general expect $\epsilon\simeq1$ and $\epsilon=3$ can be a `high' value from this point of view. }\textbf{KYLE: AQN in the corona have to burn up on a shorter timescale and after passing through a smaller column density of matter. I agree that turbulent flows in the plasma will make annihilation more efficient in the sun but the choice $\epsilon = 3$ still seems quite arbitrary. I assume that an almost order of magnitude boost in the cross-section means that the AQN almost totally annihilate which would put us close to the earth heating constraint. If we can give an observational upper bound on $\epsilon$ that might be relevant.}\Liang{I see your worry. I agree the choice of $\epsilon$ is somewhat arbitrary but still reasonable. First, $\epsilon$ is a phenomenological parameter and there is no way to obtain the exact value. Even the equation of $\epsilon$ in the corona papers are based on order of estimate, so we expect it is only qualitatively correct up of order of magnitude. Namely, $\epsilon=3$ or 5 or 8 cannot tell much fundamental difference. Boosting $\epsilon$ to order of magnitude is more likely a vague mathematical trick rather than telling any physical importance, as $\epsilon$ by it self is order of estimate.  Second, $\epsilon$ must be much smaller than the corona case. The timescale in the corona is 10 to 100 seconds, which is not shorter than the Earth, but at most comparable. Also, the temperature on corona ($\sim10^6$K) is 3 to 4 orders higher than the Earth ($\sim10^3$K). Additionally, note that the annihilation efficiency is a lot greater on corona due to abundance of Hydrogen element. On Earth, AQN collides with heavy elements, and therefore elastic collision can be more frequent rather than annihilation. In this sense, $\epsilon$ may be even overestimated in this work as we assume 100\% efficiency. Lastly, dependency of $\epsilon$ seems to drop quickly for mass loss. If you take a look at Table 2 in the second report, you can see $\epsilon=5$ produces no much difference than the case in $\epsilon=3$. So naively, at least we may expect larger $\epsilon$ produce no much difference than $\epsilon=3$.}\textbf{To be honest this is probably not worth putting too much more thought into. But I would say that the corona has a thickness of a couple thousand km and the AQNs there are moving faster than on earth so I can't see how the timescale could be appreciably longer, and the reason they are hotter there is that $\epsilon$ is larger, so that argument is kind of cyclic. I agree that there is not really any calculation that can be done here. I the corona case a very specific value $\epsilon\sim 10^4$ is required so that the AQN burn up precisely in the transition zone. That number is so much larger than 1 or three that the range of variation we considered here just seems very small. I've added the word ``slightly" in front of higher value but this discussion is probably more detailed than the paper requires. }\Liang{I guess it might be more proper to remove the word `slightly' in front of the higher value according the emails by Eric/Ludo? The `slightly' may weaken the simulation result too much.. }}


The complete dynamical equations of motion in vector form are:
\begin{subequations}
	\label{eq:3.1 AQN DEs}
	\begin{equation}
	\label{eq:3.1 AQN DEs_r}
	\frac{\rmd \mathbf{r}}{\rmd t}=\mathbf{v}\ ;
	\quad r=|\mathbf{r}|\ ;
	\quad v=|\mathbf{v}|\ ;
	\end{equation}
	\begin{equation}
	\label{eq:3.1 AQN DEs_v}
	\frac{\rmd \mathbf{v}}{\rmd t}
	=-\frac{1}{2}\pi(\epsilon R)^2\frac{\rho(r)}{m}v^2\mathbf{\hat{v}}
	-\frac{GM_{\rm eff}(r)}{r^2}\mathbf{\hat{r}}\ ,
	\end{equation}
\end{subequations} 
where $G$ is the gravitational constant, and we define
\begin{subequations}
	\label{eq:3.1 AQN DEs ass}
	\begin{equation}
	\label{eq:3.1 AQN DEs ass_R}
	R=\left(\frac{3m}{4\pi\rho_\mathrm{n}}\right)^{1/3}
	\simeq1.045\times10^{-13}B^{1/3}\,{\rm cm}\ ,
	\end{equation}
	\begin{equation}
	\label{eq:3.1 AQN DEs ass_f}
	\begin{aligned}
	&M_{\rm eff}(r)
	=\sum_{i=1}^{j}\frac{4\pi}{3}\rho_i(r_i^3-r_{i-1}^3)
	+\frac{4\pi}{3}\rho_{j+1}(r^3-r_j^3),  \\
	&(r_j\leq r< r_{j+1},~~{\rm with~} j=1, 2... 5~
	{\rm and~} r_0\equiv0)\ .
	\end{aligned}
	\end{equation}
\end{subequations}
In Eq. \eqref{eq:3.1 AQN DEs ass_R}, we follow the parameters adopted in \cite{Raza:2018gpb}, in which $\rho_\mathrm{n}=3.5\times10^{17}$\,kg\,m$^{-3}$ is the nuclear density and $m=m_\mathrm{p} B$ is the AQN mass. In Eq. \eqref{eq:3.1 AQN DEs ass_f}, we approximate the local environmental density $\rho(r)$ as discrete step functions due to the discontinuous geological structure of the Earth. The labels $i,j=1, ..., 5$ correspond to layers, summarized in Table \ref{table:3.1 Labels for each layers}. The parameter $r_i$ is the radius of the start of the layer as measured from the center of the Earth. $\rho_i$ is the average density of the corresponding shell respectively. We make the approximation that the density within each layer is uniform, and therefore take the average value from density at the top/bottom of the shell as the local density. The data in Table \ref{table:3.1 Labels for each layers} is taken from Ref.\footnote{Don L. Anderson, Theory of the Earth, Boston, Blackwell Publications (1989). Also see, http://pubs.usgs.gov/gip/interior.
}

\begin{table}[h]
	\caption{Our model for the density structure of the Earth. We consider the Earth to be made from 5 distinct layers, each of which we model as a constant-density shell, with density $\rho_i$. We show the outer radius for each shell, $r_i$, as well as the thickness of the shell.} 
	\centering 
	\begin{tabular}{ccccc}
		\hline\hline
		Label & Layer   & Thickness [km] & $r_i$ [km]	 & $\rho_i$ [$\rm g\,cm^{-3}$]     \\
		\hline
		1     & Inner core  &1221  	& 1221	 &  12.95			\\
		2     & Outer core 	& 2259 &3480 	 & 11.05			\\
		3     & Lower mantle  & 2171 &5651	 & ~5.00			\\
		4     & Upper mantle   & \, 720 &6371	 & ~3.90			\\
		5     & Crust 	& \,\,~~30 & 6401
			 & ~2.55 		    \\ 
		\hline
	\end{tabular}
	\label{table:3.1 Labels for each layers}
\end{table}

\subsection{Axion-emission spectrum in the observer frame}
\label{subsec:3.2 axion emission spectrum in laboratory frame}

We assume axion emission with non-relativistic velocities from AQN will be spherically symmetric in the AQN frame. However, AQNs move with velocities $v_{\rm AQN}\sim 220\kmps\sim 10^{-3}c$, which greatly exceeds the escape velocity of the axions we are interested in. Therefore, the angular distribution of the emission may be strongly affected by the motion of the AQN in the interior of Earth. This subsection is devoted to analysis of this modification in the angular distribution due to the frame change.

In general, an annihilating AQN will emit axions in a frame moving with respect to an observer on Earth. We introduce the notation $\tilde{K}$ and $K$ for the rest frame of the AQN and the frame of the observer respectively. The axion-emission velocity spectrum in the AQN rest frame is calculated in Ref. \cite{Liang:2018ecs} and we refer the reader to that paper for the calculation details. The velocity spectrum in the rest frame is, by definition, the derivative of the radiated flux $\Phi_{\rm rad}(\tilde{v}_\mathrm{a})$:
\begin{equation}
\label{eq:3.2 rho}
\begin{aligned}
\rho_{\rm rest}(\tilde{v}_\mathrm{a})
&\equiv
\frac{1}{\Phi_{\rm rad}^{\rm tot}}\frac{\rmd}{\rmd \tilde{v}_\mathrm{a}}\Phi_{\rm rad}(\tilde{v}_\mathrm{a})  \\
&\simeq\frac{\tilde{v}_\mathrm{a}^3}{N(\delta)}\left(\frac{\tilde{E}_\mathrm{a}}{m_\mathrm{a}}\right)^6|H_0(0,\delta)|^2
[1+{\cal O}(\tilde{v}_\mathrm{a}^2)],
\end{aligned}
\end{equation}
where $\tilde{v}_\mathrm{a}$ and $\tilde{E}_\mathrm{a}$ are the rest frame axion velocity and energy respectively. The function $H_l(\tilde{p},\delta)$ corresponds to partial wave expansion in the approximate solutions, as derived in \cite{Liang:2018ecs}. In the present work we are interested in non-relativistic AQNs and axions, therefore, only the $l=0$ mode and $\tilde{v}_a/c\ll 1$ are considered in Eq. \eqref{eq:3.2 rho}. The parameter $\delta\in (0,1)$ is a convenient factor introduced in Ref. \cite{Liang:2018ecs} as a result of approximations due to absence of simple analytic solutions of the general expressions. Tuning $\delta\in (0,1)$ leads to changes in the velocity spectrum \eqref{eq:3.2 rho} that do not exceed $\sim20\%$. The velocity spectrum is not known more accurately than this (see Appendix \ref{rest-frame}). The normalization factor  $N(\delta)$, depending on the parameter $\delta$, is also known and presented in Appendix \ref{rest-frame}.

In the frame of the observer, an AQN is moving with a velocity $v_{\rm AQN}\lesssim10^{-3}c$. Thus, we need only to consider a non-relativistic transformation of frames, with relations as follows: 
\begin{subequations}
	\label{eq:3.2 frame transformation}
	\begin{equation}
	\label{eq:3.2 frame transformation_vp}
	\mathbf{\tilde{v}}_\mathrm{a}=\mathbf{v}_\mathrm{a}-\mathbf{v}_{\rm AQN}\ ;
	\quad\mathbf{\tilde{p}}=\mathbf{p}-m_\mathrm{a}\mathbf{v}_{\rm AQN}\ ;
	\end{equation}
	\begin{equation}
	\label{eq:3.2 frame transformation_v}
	\tilde{v}_\mathrm{a}
	=\sqrt{v_\mathrm{a}^2+v_{\rm AQN}^2-2\mathbf{v}_\mathrm{a}\cdot\mathbf{v}_{\rm AQN}}\ ;
	\end{equation}
	\begin{equation}
	\label{eq:3.2 frame transformation_p}
	\tilde{p}
	=\sqrt{p^2+m_\mathrm{a}^2v_{\rm AQN}^2
	-2m_\mathrm{a}\mathbf{p}\cdot\mathbf{v}_{\rm AQN}}\ ;
	\end{equation}
	\begin{equation}
	\label{eq:3.2 frame transformation_E}
	\tilde{E}_\mathrm{a}
	=\sqrt{E_\mathrm{a}^2
		+m_\mathrm{a}^2v_{\rm AQN}^2
		-2m_\mathrm{a}\mathbf{p}\cdot\mathbf{v}_{\rm AQN}}\ .
	\end{equation}
\end{subequations}
Working in the manifold of $\mathbf{v}_\mathrm{a}$, we know $\rho(v_\mathrm{a})$ is normalized within a unit 3-ball $B^3$
\begin{equation}
\label{eq:3.2 rho transform_1}
1
=\int_{B^3} \rmd^3\mathbf{\tilde{v}}_\mathrm{a}
\frac{\rho_{\rm rest}(\tilde{v}_\mathrm{a})}{4\pi \tilde{v}_\mathrm{a}^2}
=\int_{B^3} \rmd^3\mathbf{v}_\mathrm{a}
\frac{\rho_{\rm rest}[\tilde{v}_\mathrm{a}(v_\mathrm{a})]}{4\pi (\tilde{v}_\mathrm{a}(v_\mathrm{a}))^2}\ ,
\end{equation}
where in the second step, we transform our coordinate from $\tilde{v}_\mathrm{a}$ to $v_\mathrm{a}$ using Eqs. \eqref{eq:3.2 frame transformation}. Note that such transformation produces a small error   related to a slight shift of spherical center by $\sim10^{-3}c$. However, the inconsistency is negligible comparing to the uncertainty of approximation in terms of $\delta$. Noting that the last equality in Eq. \eqref{eq:3.2 rho transform_1} is completely expressed in terms of $v_\mathrm{a}$, the velocity in the frame of the observer. Therefore, we read off the emission spectrum in the frame of the observer



\begin{equation}
\label{eq:3.2 rho transform_2}
\rho_{\rm obs}(v_\mathrm{a})
=\int \rmd\Omega~v_\mathrm{a}^2
\frac{\rho_{\rm rest}(\tilde{v}_\mathrm{a})}{4\pi \tilde{v}_\mathrm{a}^2}\ ,
\end{equation} 
where $\Omega$ is the solid angle made by $\langle\hat{\mathbf{v}}_\mathrm{a},\hat{\mathbf{v}}_{\rm AQN}\rangle$. For a moving AQN with axion emission, it is reasonable to assume azimuthal symmetry, which gives
\begin{equation}
\label{eq:3.2 rho transform_3}
\rho_{\rm obs}(v_\mathrm{a})
=\frac{1}{2}\int_{-1}^{1} \rmd u~\frac{v_\mathrm{a}^2\tilde{v}_\mathrm{a}}{N(\delta)}
\left(\frac{\tilde{E}_\mathrm{a}}{m_\mathrm{a}}\right)^6
|H_0(\tilde{p},\delta)|^2\ , 
\end{equation} 
where $\tilde{v}_\mathrm{a}$, $\tilde{p}$, and $\tilde{E}_\mathrm{a}$ are expressed as functions of $v_\mathrm{a}$ defined in Eqs. \eqref{eq:3.2 frame transformation}. In this work, we are specifically interested in the hierarchy when $v_\mathrm{a}\ll v_{\rm AQN}\ll c$, such that one can expand Eq. \eqref{eq:3.2 rho transform_3} to arrive at
\begin{equation}
\label{eq:3.2 number density}
\rho_{\rm obs}(v_\mathrm{a})
\simeq\frac{v_{\rm AQN}}{v_\mathrm{a}}\rho_{\rm rest}(v_\mathrm{a})
\left[1+\mathcal{O}\left(\frac{v_\mathrm{a}}{v_{\rm AQN}}\right)^2\right]\ , 
\end{equation}
and using the fact that $H_0(p,\delta)$ is locally quadratic near $p=0$, (Appendix \ref{rest-frame}). These corrections originate from the fact that axions that are eventually trapped are emitted with velocity in the rest frame that greatly exceeds the escape velocity, but they are emitted in the opposite direction to $\mathbf{v}_{\rm AQN}$, so cancellation occurs and these are the axions that end up with a low velocity in the observer frame and therefore end up trapped and contributing to $\rho_{\rm obs}(v_\mathrm{a})$. This transformation has a small dipole term, but we have numerically checked that the dipole contribution to the emission is small ($\pm6\%$ correction for the dipole, while the quadrupole gives $\pm0.2\%$), and so we neglect it and take axion emission to be isotropic in the observer frame.



To summarize: the emission spectrum the frame of the observer (\ref{eq:3.2 number density}) is directly expressed in terms of the spherically symmetric spectrum in the rest frame via equation (\ref{eq:3.2 rho}) and computed in \cite{Liang:2018ecs}. This is expressed in terms $H_0(0,\delta)$ given in Appendix \ref{rest-frame}. We emphasize that the axion angular distribution is not spherically symmetric; instead, for each incoming AQN, it is azimuthally symmetric with respect to the direction defined by the velocity of the AQN $\mathbf{v}_{\rm AQN}$. It is the integration over incoming directions in Eq. (\ref{eq:3.2 rho transform_2}) that leads to the simple relation in equation (\ref{eq:3.2 number density}) between the two distributions.

\subsection{The axion density distribution}
\label{subsec:3.3 axion emission}

The slowest of the axions that are emitted due to AQN annihilation while travelling through the Earth will be trapped by the gravitational field. As the axions are non-interacting particles, their dynamical equations of motion are simpler than that of the AQNs:
\begin{equation}
\label{eq:3.3 axion DEs}
\frac{\rmd ^2\mathbf{r}}{\rmd t}
=-g(r)\mathbf{\hat{r}}\ ; 
~~~~~~ 
g(r)\equiv \frac{GM_{\rm eff}(r)}{r^2}\ .
\end{equation}
In other words, the trajectories of axions are entirely determined by the gravitational force from the Earth. Axions are mutually non-interacting, so if they are generated trapped then they remain trapped for all time. The only difference in comparison with standard Kepler problem being that the effective acceleration, $g(r)$, does not scale like $r^{-2}$ in the interior of the Earth. We will use this set of equations and sample large numbers of axion trajectories with initial conditions set by the AQN annihilation events within Earth. This allows us to obtain a distribution function $p_\mathrm{a}(r)$ that gives us the probability to find an axion at a given distance $r$ from the center of the Earth. This is then directly related to the axion density profile, which builds up around the Earth over 4.5 Gyrs and is the crucial quantity for the axion search experiments reviewed in the introduction. A full description of the numerical simulations that allow us to compute $p_\mathrm{a}(r)$ will be described in Sec. \ref{sec:4. algorithm}, while here we relate $p_\mathrm{a}(r)$ to other observables that enter the numerical simulations. 

First, we want to relate $p_\mathrm{a}(r)$ to the energy density of gravitationally-trapped axions, $\rho_\mathrm{a}(r)$:
\begin{equation}
\label{eq:3.3 rho_a(r)}
\rho_\mathrm{a}(r)
=m_\mathrm{a}\cdot\langle N_{\rm AQN}^{\rm 4.5Gyr}\rangle\langle N_\mathrm{a}^{\rm trap}\rangle
\frac{p_\mathrm{a}(r)}{4\pi r^2}\ ,
\end{equation}
where $\langle N_{\rm AQN}^{\rm 4.5Gyr}\rangle$ is the expected total number of antimatter AQNs that impacted the Earth within 4.5 Gyr, and $\langle N_\mathrm{a}^{\rm trap}\rangle$ is the mean number of trapped axions emitted per AQN. The first term entering (\ref{eq:3.3 rho_a(r)}) is easy to estimate  
\begin{equation}
\label{eq:3.3 N_AQN_4.5Gyr}
\begin{aligned}
&\langle N_{\rm AQN}^{\rm 4.5Gyr}\rangle
=\langle\dot{N}\rangle\cdot4.5{\rm Gyr}  \\
&\simeq9.52\times 10^{16}
\left(\frac{\rho_{\rm DM}}{0.3\,{\rm GeV\,cm^{-3}}}\right)
\left(\frac{v_{\rm AQN}}{220\kmps}\right)
\left(\frac{\langle B\rangle}{10^{25}}\right)^{-1}\ ,
\end{aligned}
\end{equation}
where $\langle\dot{N}\rangle$ is the average flux of AQNs hitting the surface of the Earth, estimated in Appendix \ref{sec:A. estimation of AQN flux}.

The estimation of $\langle N_\mathrm{a}^{\rm trap}\rangle$ in (\ref{eq:3.3 rho_a(r)}) is more complicated. We choose to express this in terms of the ``heat'' distribution function $q(r,v_{\rm AQN})$ defined as the fraction of the mass loss of an AQN moving with local velocity $v_{\rm AQN}$ at a location $r$. We name this distribution ``heat" to emphasize that this function describes the annihilation events where locally one should expect about $2m_\mathrm{p} c^2$ energy release per single event of annihilation. We normalize the heat distribution as follows:
\begin{equation}
\label{eq:3.3 q(r,v)}
\langle \Delta m_{\rm AQN}\rangle
=\int_0^{R_\oplus} \rmd r\int_0^\infty \rmd v_{\rm AQN}~q(r,v_{\rm AQN})\ ,
\end{equation}
where $\langle \Delta m_{\rm AQN}\rangle$ is mean total mass loss per AQN as it passes through the Earth. The total number of axions emitted for a given mass loss $\Delta m_{\rm AQN}$ can be estimated as 
\begin{equation}
\label{eq:3.3 N_a_tot}
\la N_\mathrm{a}^{\rm tot}\ra
\simeq\frac{1}{3}\frac{\Delta m_{\rm AQN}}{\langle E_\mathrm{a}\rangle}
\simeq\frac{\Delta m_{\rm AQN}}{4m_\mathrm{a}}\ ,
\end{equation} 
where $\langle E_\mathrm{a}\rangle\simeq1.3m_\mathrm{a}$ is the average energy of axions computed from the spectrum in equation~\eqref{eq:3.2 rho}. The coefficient  $1/3$ in (\ref{eq:3.3 N_a_tot}) comes from the fact that approximately $1/3$ of the total AQN energy is stored in the form of axion energy at formation time. This energy is inevitably released as axions during the annihilation. 

We are interested in trapped axions with velocities below the local  escape velocity $v_{\rm esc}(r)$ in the interior of the Earth, with $v_{\rm esc}(r)$ given by:
\begin{equation}
\label{eq:3.3 v_esc}
v_{\rm esc}(r)
=\sqrt{\frac{-2U(r)}{m_\mathrm{a}}}\ ;\quad
\frac{U(r)}{m_\mathrm{a}}
=\int_{\infty}^{r}\rmd r'
\frac{GM_{\rm eff}(r')}{(r')^2}\ .
\end{equation}
Now we can express $\langle N_\mathrm{a}^{\rm trap}\rangle$ entering (\ref{eq:3.3 rho_a(r)}) in terms of the heat distribution $q(r,v_{\rm AQN})$ as follows    
\begin{equation}
\label{eq:3.3 N_a_trap}
\begin{aligned}
\langle N_\mathrm{a}^{\rm trap}\rangle
&\simeq \int_0^{R_\oplus} \rmd r\int_0^\infty \rmd v_{\rm AQN}~
\frac{q(r,v_{\rm AQN})}{4m_\mathrm{a}} \\
&\quad\times\int_0^{v_{\rm esc}(r)}\rmd v_\mathrm{a}~\rho_{\rm obs}(v_\mathrm{a})\ ,
\end{aligned}
\end{equation}
where $\rho_{\rm obs}(v_\mathrm{a})$ is the axion-velocity spectrum in the frame of the observer, computed in equation~(\ref{eq:3.2 number density}). The computation of $q(r,v_{\rm AQN})$ requires Monte Carlo simulations, which we now discuss.

\section{Algorithm and simulations}
\label{sec:4. algorithm}

\subsection{Simulating initial conditions of AQNs for the heat emission profile}
\label{subsec:4.1 AQNs initial}

\begin{figure}[h]
	\centering
	\captionsetup{justification=raggedright}
	\includegraphics[width=0.7\linewidth]{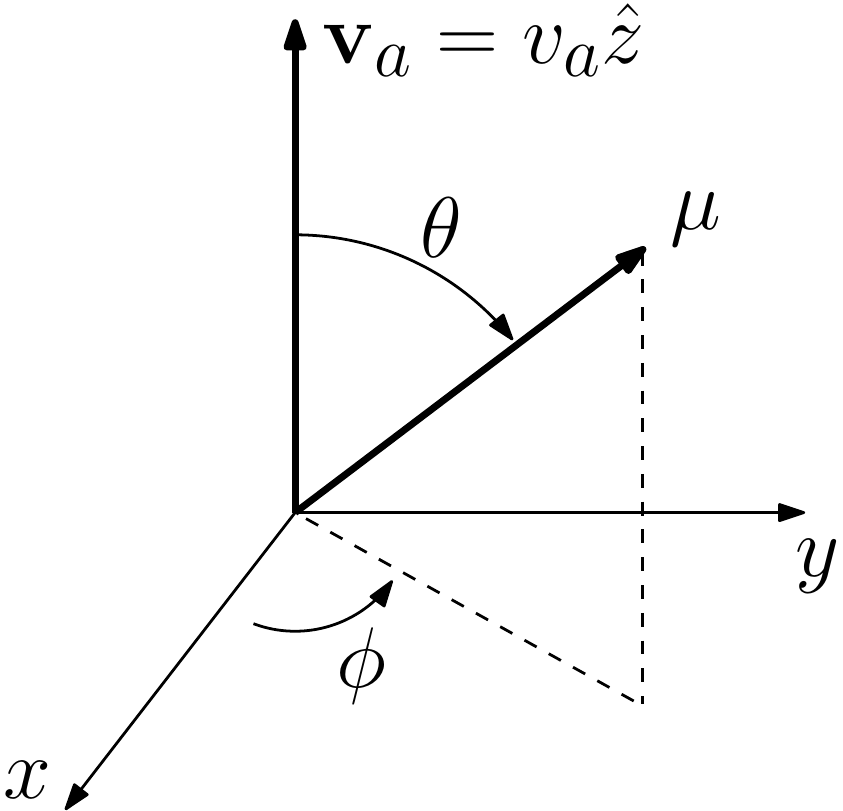}
	\caption{The coordinate system used in the calculation of the AQN flux when we isotropize the incoming wind direction. The solid-angle element $\rmd\Omega=\sin\theta\, \rmd\theta\rmd\phi$ is integrated out over $4\pi$ to obtain an isotropic average.}
	\label{fig:4.1 isotropic wind}
\end{figure}

To simulate the trajectories of AQNs through Earth we start from the velocity distribution function of AQNs in the vicinity of Earth. This distribution is Gaussian in each Cartesian direction with dispersion $\sigma\simeq 110\kmps$ and an additional `wind' component in one direction $\mu\simeq220\kmps$ due to the motion of the Solar system with respect to the DM halo\exclude{{\bf LUDO: we might consider having a 3D diagram showing the spherical coordinates used in this calculation.}{\bf XUNYU: the diagram is now added as Fig. 1}}:
\begin{equation}
\label{eq:4.1 v maxwell instant}
f_{\mathbf{v}}(\mathbf{v})\rmd^{3}\mathbf{v}
=\frac{\rmd^{3}\mathbf{v}}{(2\pi \sigma^2)^{3/2}}\exp
\left[-\frac{v_x^2+v_y^2+(v_z-\mu)^2}{2\sigma^2}\right].
\end{equation}
\exclude{Note that in the present work, an extremely long-time (4.5 Gyr) accumulation of trapped axions is studied. Therefore, it is advisable to use an isotropic distribution as a long-time average}
We consider the accumulation of axions trapped in orbit around Earth over the lifetime of the Solar System ($\sim 4.5\Gyr$). Over this time, due to the spin and orbit of the Earth, the orbit of the Solar system around the galactic centre, and perturbations to both of these due to secular effects of gravity, we consider the wind direction to have averaged over all $4\pi$ solid angle, see Fig. \ref{fig:4.1 isotropic wind}. Taking the average of Eq. (\ref{eq:4.1 v maxwell instant}) gives us an isotropic distribution for incoming AQN speed that includes the effect of the wind:
\begin{equation}
\label{eq:4.1 v maxwell}
\langle f_{\mathbf{v}}(\mathbf{v})\rangle_{\Omega}
=\frac{1}{(2\pi\sigma^2)^{3/2}}
\frac{\sinh(\mu v/\sigma^2)}{\mu v/\sigma^2}
\exp\left[-\frac{v^2+\mu^2}{2\sigma^2}\right]\ .
\end{equation}
To start our simulations we must use the axion \emph{flux} distribution through the surface of the Earth. The flux distribution is given by
\begin{equation}
\label{eq:4.1 flux distribution}
\begin{aligned}
\frac{\rmd}{\rmd v}\langle\dot{N}\rangle_\Omega
&=\frac{1}{4}n\cdot 4\pi R_\oplus^2\cdot 4\pi v^3
\langle f_{\rm v}({\rm v})\rangle_\Omega  \\
&\propto v^3
\frac{\sinh(\mu v/\sigma^2)}{\mu v/\sigma^2}
\exp\left[-\frac{v^2+\mu^2}{2\sigma^2}\right]\ .
\end{aligned}
\end{equation}
We show the derivation of the distributions \eqref{eq:4.1 v maxwell} and \eqref{eq:4.1 flux distribution} in Appendix \ref{sec:A. estimation of AQN flux}. As a consistency check of the isotropized distribution in equation~\eqref{eq:4.1 flux distribution} we also carry out calculations in which the wind direction is fixed in space (Appendix \ref{sec:fixed wind}) and obtain consistent results.

To generate AQN initial conditions, we uniform-randomly choose a point on the surface of Earth and then generate an AQN with velocity sampled from distribution \eqref{eq:4.1 flux distribution}. We remove AQNs with initial velocity pointing away from the Earth as these correspond to AQNs that would have already experienced an Earth interaction. Following this sampling scheme, we generate $10^{6}$ AQNs and compute $q(r,v_{\rm AQN})$, the heat emission profile of axions at a given location within the Earth, $r$, and originating from an AQN with local velocity $v_{\rm AQN}$ as defined by Eq. \eqref{eq:3.3 q(r,v)}. As explained in Sec. \ref{subsec:3.3 axion emission}, this function is a key element to the calculation of the axion density distribution $\rho_\mathrm{a}(r)$. Furthermore in the next subsection, we will see that it is $q(r,v_{\rm AQN})$ that solely determines the number of axions we simulate at a given point in the phase space $(r,v_{\rm AQN})$.

For individual AQNs, Eqs. \eqref{eq:3.1 AQN DEs} are solved numerically, and a 2D histogram of the $(r,v_{\rm AQN})$ phase space is computed, where each point in a given trajectory is weighted by $\dot{m}(t)$ to give the total amount of annihilation energy going into axions. We refer to Appendix \ref{sec: heat emission} for further details .


\subsection{Computations of the density distribution of trapped axions}
\label{subsec:4.3 Find rho_a(r)}

The density distribution of trapped axions $\rho_\mathrm{a}(r)$ is directly related to $p_\mathrm{a}(r)$, the probability of finding a trapped axion at a given distance $r$, given by Eq. \eqref{eq:3.3 rho_a(r)}. Similar to the calculation of the AQN heat profile $q(r,v_{\rm AQN})$, we obtain $p_\mathrm{a}(r)$ by simulation, but the numerical procedure is more sophisticated. First, it is clear that the number of axions to simulate at a given point in the phase space $(r,v_{\rm AQN})$ is proportional to two weighting factors -- the rate of axion production [i.e. the heat profile function $q(r,v_{\rm AQN})$], and the rate of producing trapped axions [i.e. the spectrum $\rho_{\rm obs}(r)$ and the local escape speed $v_{\rm esc}$]. Therefore, we sample axion initial conditions from the distribution within the integral
\begin{equation}
\label{eq:4.3 p_a(r)}
\begin{aligned}
p_\mathrm{a}(r)
&\sim\langle N_\mathrm{a}^{\rm trap}\rangle  \\
&\propto\int_{0}^{\infty}\rmd v_{\rm AQN}\int_0^{R_{\oplus}}\rmd r~
q(r,v_{\rm AQN})  \\
&\quad\times\int_0^{v_{\rm esc}(r)}\rmd v_\mathrm{a}
\frac{v_{\rm AQN}}{v_\mathrm{a}}
\rho_{\rm rest}(v_\mathrm{a}) \ .
\end{aligned}
\end{equation}
The interpretation of Eq. \eqref{eq:4.3 p_a(r)} is as follows: we first choose the initial conditions $(r,v_{\rm AQN})$ with probability weighted by $q(r,v_{\rm AQN})v_{\rm AQN}/\langle v_{\rm AQN}\rangle$, and then we shoot an axion at the given position $r$ at a random angle,with a speed in the range $[0,v_{\rm esc}(r)]$ weighted by $\langle v_{\rm AQN}\rangle/v_\mathrm{a}\rho_{\rm rest}(v_\mathrm{a})$.

Monte Carlo sampling can be performed in 2D space due to the azimuthal symmetry. At $\mathbf{r}(t=0)=(0,r)$, the velocity
\begin{equation}
\label{eq:4.3 axion ICs}
\mathbf{v}_\mathrm{a}(t=0)=v_\mathrm{a}(\sin\theta,\cos\theta)
\end{equation} is randomly generated using the following distributions:
\begin{equation}
\label{eq:4.3 axion ICs ass}
v_\mathrm{a}\sim\frac{\langle v_{\rm AQN}\rangle}{v_\mathrm{a}}\rho_{\rm rest}(v_\mathrm{a})\ ;\quad
\cos\theta\sim{\rm uniform}(-1,1)\ .
\end{equation}
Then, we obtain the initial velocity of the sampled axion:
\begin{equation}
\label{eq:4.3 axion ICs ass 2}
v_r=v_\mathrm{a}\cos\theta,\qquad v_\theta=v_\mathrm{a}\sqrt{1-\cos^2\theta}\ .
\end{equation}
Note that the sign of $v_\theta$ is irrelevant, as $p_\mathrm{a}(r)$ preserves symmetry of time reversal.

Unlike for the AQN, the equations of motion for the axion are simple and obey an energy and angular momentum conservation laws, see Eq. \eqref{eq:3.3 axion DEs}. We use these conservation laws, together with the fact that the probability, $p_\mathrm{a}^{\rm (one)}(r)$, to find a single trapped axion at a distance, $r$ from Earth must be inversely proportional to the radial velocity $v_r$ of that axion at that distance:
\begin{equation}
\label{eq:4.3 p_a(r)_2}
p_\mathrm{a}^{\rm (one)}(r)=N\frac{1}{v_r}\ ,
\end{equation}
where $N$ is a normalization, and $v_r$ can be explicitly written in terms of $r$ using the conservation laws for angular momentum and energy:
\begin{subequations}
	\label{eq:4.3 p_a(r)_2 ass dot(r)}
	\begin{equation}
	\label{eq:4.3 p_a(r)_2 ass dot(r)_eqns}
	H(r,v_r,v_\theta)
	=\frac{1}{2}m_\mathrm{a}v_r^2+\frac{1}{2}m_\mathrm{a}\frac{l^2}{r^2}+U(r)
	=E\ ;
	\end{equation}
	\begin{equation}
	\label{eq:4.3 p_a(r)_2 ass dot(r)_Elv}
	l=r(t)v_\theta(t)|_{t=0}\ ;
	\quad 
	E=H|_{t=0}\ .
	\end{equation}
\end{subequations}
We then conclude that
\begin{equation}
\label{eq:4.3 p_a(r)_3}
p_\mathrm{a}^{\rm (one)}(r)
=\frac{N}{\sqrt{2[E-U(r)]/m_\mathrm{a}-l^2/r^2}}\ .
\end{equation}
The full $p_\mathrm{a}(r)$ is obtained by averaging over sufficient numbers of $p_\mathrm{a}^{\rm (one)}(r) $ using the Monte Carlo technique.

\subsection{Computations of  the axion spectrum on the Earth surface}
\label{subsec:4.4 Find F(va)}

Finally, we describe the method to calculate $F(v_\mathrm{a})$, the axion speed spectrum on the Earth surface. The spectrum is important for detection instrument as the de Broglie wavelength of the trapped axion $\lambda_\mathrm{a}=h/m_\mathrm{a}v_\mathrm{a}$ is sensitive to the local speed in the non-relativistic limit.

We follow a similar approach as in the previous subsection. Once the initial conditions are known, it is easy to find the speed of an axion at the surface of the Earth from the equation of motion \eqref{eq:4.3 p_a(r)_2 ass dot(r)}:
\begin{equation}
\label{eq:4.4 v_a(R_Earth)}
v_\mathrm{a}(R_\oplus)\equiv|\mathbf{v}_\mathrm{a}(R_\oplus)|
=\sqrt{\frac{2}{m_\mathrm{a}}\left[E-U(R_\oplus)\right]}\ .
\end{equation}
We can then obtain the axion speed profile, $F(v_\mathrm{a})$, by computing a histogram of the axions with weighting inversely proportional to the orbital period of each axion (axions with shorter periods pass through the surface more often). The period of a given trajectory is found from \eqref{eq:4.3 p_a(r)_2 ass dot(r)}:
\begin{equation}
\label{eq:4.4 dot(r)}
\frac{\rmd r}{\rmd t}
=\sqrt{\frac{2}{m_\mathrm{a}}\left[E-U(r)\right]-\frac{l^2}{r^2}}\ .
\end{equation}
Therefore, the (half-)period is obtained by numerical integration
\begin{equation}
\label{eq:2.4 dot(r) - period}
\frac{1}{2}T
=\int_{R_{\rm min}}^{R_{\rm max}}\frac{dr}{\sqrt{2[E-U(r)]/m_\mathrm{a}-l^2/r^2}}\ .
\end{equation}

\section{Axion profiles: density distribution and velocity spectrum}
\label{sec:5. axion profiles}

\subsection{Heat emission profile}
\label{subsec:5.1 AQN profiles}
 
\begin{figure}[h]
\centering
\captionsetup{justification=raggedright}
\includegraphics[width=1\linewidth]{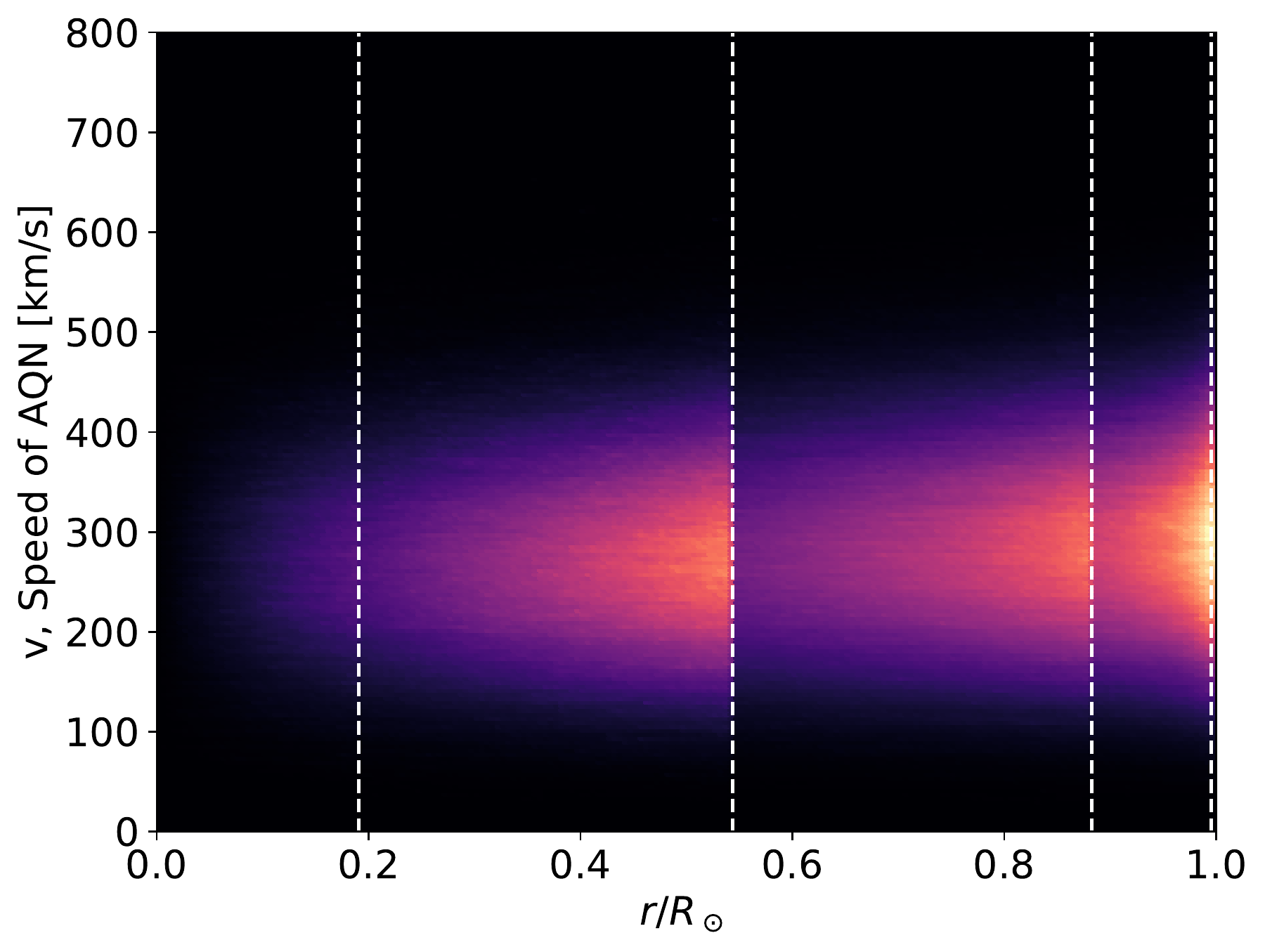}
\caption{Probability density for the heat-emission profile of axions within the Earth: $q(r,v_{\rm AQN})$ for $\alpha=(1.2, 2.5)$, $B_{\min}=3\times10^{24}$. We show the radius, $r$, from the center at which axions are emitted and the velocity of the AQN, $v_{\rm AQN}$, responsible for generating those axions. The color intensity corresponds to probability density. The dashed lines denote radii of the different layers of the Earth, where the abrupt changes in density translate to abrupt changes in heat of axion emission. This figure demonstrates that most axion heat is emitted close to the surface of the Earth, but that there is a significant tail to low depths. There is a particular spike in axion emission at the boundary between the mantle and core, at $r\simeq 0.55 R_\oplus$, where the density jumps by more than a factor of 2. $10^6$ AQN samples were used to generate this figure.}
\label{fig:5.1 q_rv alphaPW}
\end{figure}

We first discuss the results of the simulation ($10^6$ samples) for the heat-emission profile of axions: $q(r,v_{\rm AQN})$. In our simulations  we use a number of parameters such as the AQN baryon-charge distribution (parameters $\alpha$ and $B_{\rm min}$), $\epsilon$ (describing the strength of the interaction of AQNs with earth's material) and the flux distribution of incoming AQNs (characterized by the galactic wind and DM velocity dispersion). Surprisingly, we find that our results are quite insensitive to these parameters. Therefore, in the main body of the paper we only present the case with $\alpha=(1.2,2.5)$ and $B_{\rm min}=3\times10^{24}$ shown in Fig. \ref{fig:5.1 q_rv alphaPW}, while  leaving other cases with different parameters to Appendix \ref{sec:other plots}. 


  
From Fig. \ref{fig:5.1 q_rv alphaPW}, we first observe a Maxwellian-like distribution with mean speed of approximately 300 $\kmps$ and dispersion 200 $\kmps$ in all cases. In addition, we specifically note that $q(r,v_{\rm AQN})$ looks like a stack of multiple Maxwellian-like distributions, such that an abrupt change appears at some specific distance (e.g. $r\sim0.9$ and 0.5$R_\oplus$). One can see, from Table \ref{table:3.1 Labels for each layers}, that these jumps correspond to the successive layers of the Earth interior. Consequently, whenever an AQN moves into a new layer with an abrupt change of local density, the mass loss drastically changes. Another observations is that very few AQNs reach the core of the Earth, as the dominant portion of the AQNs   propagate at the distances $r\geq 0.5 R_{\oplus}$, i.e. far  away from the center of the Earth. This is a geometrical effect that stems from the fact that there is more volume in the outer part of the Earth compared to the inner part.

As shown in Appendix \ref{sec:other plots}, we reach very similar conclusion for almost all effects we considered. The heat emission profile is insensitive to the baryon-charge distribution of AQN models and to the velocity distribution of incoming DM flux. The effective cross section, parametrized by $\epsilon$, obviously increases the number of annihilation events expressed as $\Delta B/B$ in Table  \ref{table:5.2 summary of some results}. However, the corresponding increase of $\epsilon$ does not lead to significant modification of the axion density on the surface $\rho_\mathrm{a}(R_{\oplus})$, which is the main observable of the system. We have verified that this conclusion of insensitivity to variety of physical parameters is a generic feature of the system and this feature will be further elaborated in next subsections.

\subsection{Axion density distribution}
\label{subsec:5.2 Axion density}

The key results of our simulations are summarized in Table \ref{table:5.2 summary of some results}. From the mass loss, we obtain the expected number of trapped axions $\langle N_\mathrm{a}^{\rm trap}\rangle$ per AQN by Eq. \eqref{eq:3.3 N_a_trap} and substitute this value into Eq. \eqref{eq:3.3 rho_a(r)}, the energy density distribution of axions is therefore
\begin{equation}
\label{eq:5.2 rho_a(r)}
\rho_\mathrm{a}(r)
\simeq \frac{\rho_\mathrm{a}(R_\oplus)}{(r/R_\oplus)^2}\frac{p_\mathrm{a}(r)}{p_\mathrm{a}(R_\oplus)}
\left(\frac{v_{\rm AQN}}{220 \kmps}\right)
\left(\frac{\rho_{\rm DM}}{0.3\rm GeV/cm^3}\right)\ ,
\end{equation}
where $p_\mathrm{a}(r)$ is computed using the Monte Carlo simulations discussed in the previous section.

We can parametrize our results for $\rho_\mathrm{a}(r)$ using a simple analytical fit as follows 
\begin{equation}
\label{eq:5.2 fit pa(r)}
\left(\frac{r}{R_\oplus}\right)^2\frac{\rho_\mathrm{a}(r)}{\rho_\mathrm{a}(R_\oplus)}
\simeq\left\{
\begin{aligned}
&a_1 x^{n_1} e^{-kx^{m}}\ ,  &x<1  \\
&\frac{1}{2}\left[a_1 x^{n_1} e^{-kx^{m}}+a_2 x^{-n_2}\right] ,  &x=1\\
&a_2 x^{-n_2}\ ,  &x>1
\end{aligned}
\right.
\end{equation}
where $x\equiv r/R_\oplus$, with parameters chosen to be $(a_1,n_1,k_1,m_1)=(3.11919, 1.58484, 1.11709, 2.06927)$ and $(a_2,n_2)=(1.000051, 1.56429)$ for the best fit. Note that there exists a discontinuous jump at $r=R_\oplus$, which is due to the steep change of density near the Earth surface, because axion emission always happens inside the earth. The corresponding plot is presented in Fig. \ref{fig:5.2 rho_a alphaPW}. Similar plots obtained with different choice of parameters are shown in Appendix \ref{sec:other plots}. It is found that the approximation \eqref{eq:5.2 fit pa(r)} fits the density profile of all models with $\epsilon\simeq1$ over a range of radii $r\in[0.5, 5]R_\oplus$ with a maximum deviation of 10\%.

We would like to make following comments:

$\bullet $ We note that the \emph{shape} of $\rho_\mathrm{a}(r)$ is not sensitive to the parameters ($\alpha$, $B_{\rm min}$, and flux distribution) of the AQN mass-distribution function; only its magnitude shows a modest variation. This feature was anticipated before the computations: Indeed, the axions inside the Earth are non-interacting particles (except for the gravity). However, the gravitation becomes weak near the inner core in comparison with the surface of the Earth. Thus, one can treat axions as a free ideal gas with a probability density scaling as $\sim r^2$ near $r\lesssim R_\oplus$, and as $\sim r^{-2}$ outside the radiation source. We can do this because the axion-density profile is a statistical count of occupation states when the entropy is maximized, and therefore it should not depend on the location of the heat source inside the system (Earth). This argument is generic, model independent and is supported by explicit computations that we have parameterized by simple analytical formula \eqref{eq:5.2 fit pa(r)}.

  \exclude{
  given that the cross section scales like $\epsilon^2$ and naively one might then expect the profile to scale in the same way. The insensitivity can be understood from the fact that the cross section also scales with AQN radius, as $R^2$, and $R$ decreases as the annihilation proceeds. 
}

$\bullet $ The amplitude of the axion-density profile increases with $\epsilon$. However, the  observable density   $\rho_\mathrm{a}(R_\oplus)$ on the earth's surface shows very modest changes  
  when   $\epsilon$ varies as discussed in Appendix 
 \ref{sec:other plots}. 
 \exclude{The shape of the axion-density profile $\rho_\mathrm{a}(r)$ is not very sensitive to $\epsilon$, similar to the behaviour of $q(r,v_{\rm AQN})$ discussed in the previous subsection, although $\Delta B/B$ increases when $\epsilon$ increases, as expected.}
 Qualitative explanation of this effect is as follows.   In general, a larger $\epsilon$ results in a peak in the density profile closer to the surface of the Earth because more AQNs annihilate closer to the surface of the Earth due to a larger effective cross section. Nevertheless, the observable $\rho_\mathrm{a}(R_\oplus)$ shows only modest variations with respect to these changes as mentioned above. This is not a surprise because $\rho_\mathrm{a}(r)$ is predominantly determined by the local annihilation rate of AQNs, and therefore the local density of the Earth. Consequentially, even for the large $\epsilon$ cases, $\rho_\mathrm{a}$ becomes more uniform in the inner layers, but drops quickly near the surface due to the reduced rate of local AQN annihilation.

\exclude{\Mead{Why is this? The AQN loose much more mass for higher $\epsilon$.}\Liang{I modify the last two sentences, hope it explains clearly this time.}}

$\bullet $ It is instructive to compare this result with the crude estimate \eqref{eq:1. rho_a} presented in Ref. \cite{Liang:2018ecs}: there is about 3 orders of magnitude deviation from that naive estimate. This can easily be understood from the work described here: First, \cite{Liang:2018ecs} assumed that $\Delta B/B\sim 1$ (i.e., most incident AQN are mostly annihilated), however, our simulations show that $\Delta B/B\sim 0.1$ for a typical model with $\epsilon\simeq 1$ (see Table \ref{table:5.2 summary of some results}). Second, the emitted axions form an extended cloud in space around the Earth, such that the axion density remains high even far away from Earth, at the distances $r\sim 5 R_{\oplus}$, for instance see Fig. \ref{fig:5.2 rho_a alphaPW} and corresponding figures in Appendix \ref{sec:other plots}. It is the physical size of the extended cloud that is the main source for the discrepancy with the naive estimate in \cite{Liang:2018ecs}, where it was assumed the axion density dropped sharply beyond the surface of the Earth. Finally, there is a suppression factor related to averaging over inclination angles between the AQN velocity and the surface of the Earth that was missing from \cite{Liang:2018ecs}. The order of magnitude estimate from \cite{Liang:2018ecs} assumes $\dot{N}\simeq nA v_{\rm AQN}$. The proper treatment of the inclination angle is given in Appendix \ref{sec:A. estimation of AQN flux} and produces an additional suppression factor $\sim\frac{1}{4}$ such that corresponding expression assumes the form  $\dot{N}\simeq \frac{1}{4} nA v_{\rm AQN}$.



The results of our Monte Carlo simulations are presented in Table \ref{table:5.2 summary of some results}. The main result from the point-of-view of axion-search experiments is the axion density on the surface of the Earth $\rho_\mathrm{a}(R_\oplus)$\footnote{We do not expect any large enhancement factors due to the gravitational focusing of DM, which may briefly increase the incident AQN flux by factor $10^6$, as advocated in \cite{Patla:2013vza,Bertolucci:2017vgz}. This is because \emph{trapped} axions are accumulated during 4.5 billion of years and enhancement effects due to gravitational focusing are of short temporal duration and we do not think they will significantly change the mean AQN impact rate over the long accumulation time. 
\exclude{\textbf{KYLE: Sorry, maybe I missed the resolution of this, how can we say that the duration of focusing events will be short without knowing how the DM distribution differs from isotropic?}\Liang{Yes DM distribution is largely unknown, but we are pretty sure the crude picture (mean speed $\mu\sim220\kmps$) is qualitatively correct. As long as such assumption stands, we know the estimation in [72] for the duration of focusing should be reliable. The duration of focusing in [72] is estimated to be within 30 minutes for moderate lensing $\sim10^4$ and a few seconds for maximum lesing ($\sim10^6$). Apparently the duration is extremely short comparing to 4.5Gyr.}\textbf{KYLE: As far as I can tell [72] assumes a single point source for an additional dark matter stream. In [73] the assumption is that the DM has a preferred arrival direction and the time scale is much longer. }\Liang{Say in [73] the time scale is much longer (a few days) under some very specific alignment, it would be still negligibly short comparing to accumulation in 4.5 Gyrs. }\textbf{KYLE: But it would be a few days per earth orbit. So any boost larger than $10^2$ would make it the dominant contribution for as long as the DM flow remains coherent.} \Liang{First, the actual amplification can be far weaker than the advertisement in [73]. There may be amplification up to $\sim10^6$ at some specific spot on Earth. But in our case, we are using the entire Earth as the detector, so the `proper' amplification is the integral over the whole Earth surface. If the amplification quickly drops off beyond that specific spot (which is very likely the case), the enhancement is meaningless. Beyond this, I remain highly skeptical with the duration/amplitude of enhancement in [73]. I cannot see where the duration of focusing comes from (no derivation nor citation). Also, [72,73] obtain such amplification by assuming a transparent sun, such that the deflection angle can be more than a few order of magnitude enhanced because they can set the impact parameter whatever small as they want,i.e. they assume the DM streams pass through the inner core of sun without any interaction! While this may be correct for WIMPs or axions, this definitely does not apply to our case. Lastly, gravitation focusing requires perfect alignment, unfortunately the plane of our solar system has a large fixed angle $63\%$ against the direction of galactic wind, so the focusing point may be always outside solar system. Therefore, the gravitation focusing is very unlikely. At least comparing to the accumulation over 4.5 Gyrs. } \textbf{KYLE: I agree that [73] takes a set up that maximizes the impact of focusing. My only point was that if one were to choose a scenario that maximizes the effect one could probably come up with a model that changes our predictions. Of course such a model would have to have a very finely tuned velocity distribution. I also agree that at this time it makes no sense to consider anything beyond the usual isotropic velocity distribution so comments on potential focusing are largely irrelevant. I might be tempted to stress that focusing is largely irrelevant in the the conventional model with no preferred direction for the DM flow, but am essentially okay with the text as is.}}}


\be
\label{final_estimate}
\rho_\mathrm{a}(R_\oplus)\sim 10^{-4}{\rm  \frac{GeV}{cm^3}}, ~~~ \la v_\mathrm{a}(R_\oplus)\ra \simeq 8\kmps.
\ee
\exclude{
Accounting for  the DM wind  enhances the numerical value for the density (\ref{final_estimate}) approximately by factor of  two, i.e.
\be
\label{final_estimate_wind}
\rho_\mathrm{a}(R_\oplus, {\bm \mu})\sim 2\cdot 10^{-4}{\rm  \frac{GeV}{cm^3}}, ~~~    \mu   \simeq ~{220\kmps}.
\ee
}

There is no remaining freedom in the model to significantly modify our final results in eq. (\ref{final_estimate}). In this scenario, the AQN framework is {\it rigid  and predictive}. This is because dark- and visible-matter densities in the AQN framework must always satisfy the relation (\ref{Omega}), irrespective of the axion mass $m_\mathrm{a}$. This is in contrast with conventional estimates for the galactic-axion density, which strongly depends on the axion mass and scale as $m_\mathrm{a}^{-7/6}$. 
 
The axion velocity distribution has a maximum close to $v_\mathrm{a}(R_\oplus)\simeq 8\kmps$ (see next subsection \ref{subsec:5.3 Axion speed spectrum}). The corresponding wavelength $\lambda_\mathrm{a}\sim {\hbar}/({m_\mathrm{a} v_\mathrm{a}})$ is approximately $30$ times greater than for galactic axions, which have a typical velocity of about $\sim 220\kmps$. Therefore, coherent effects can be maintained for a longer time period compared to those for conventional galactic axion searches. One may hope that the feature of having a large coherence length, $\lambda_\mathrm{a}\sim v_\mathrm{a}^{-1}$, could play a key role in the design of instruments, capable of discovering such gravitationally trapped axions (see concluding Section \ref{sec:6.conclusion} for more comments).

Indeed, the local axion density (\ref{final_estimate}) for the trapped axions is approximately three orders of magnitude smaller than conventional galactic axion density (assuming that the galactic axions with a given mass $m_a$ saturate the DM density). If the experimental sensitivity on the axion-photon coupling is
inversely proportional to the square root of the local axion density\footnote{this case is realized when the design of the axion search experiment is  based on coherent accumulation of the signal.}, one should expect a suppression of order $\sqrt{10^{-3}}\sim 1/30 $ or so in the amplitude. However, the enhancement factor due to the longer wave length  
$(220\kmps) /(8\kmps) \sim 30$ may cancel the  suppression factor  in the amplitude mentioned above.
It remains to be seen if this potentially strong enhancement factor can be practically realized in real experiments.
\begin{table*}
	\caption{Summary of some results. Our baseline model is $B_{\rm min}=3\times10^{24}$, $\epsilon=1$ unless otherwise specified. For each model we show: $\langle \Delta m_{\rm AQN}\rangle$, the change in mass of a typical antimatter AQN as a result of interaction with the Earth; ${\langle\Delta B\rangle}/{\langle B\rangle}$, the typical loss of baryon charge of a typical AQN (also the fractional mass loss); $\rho_\mathrm{a}(R_\oplus)$, the magnitude of the axion-halo density profile at the surface of the Earth after 4.5 Gyrs of accumulation. \exclude{\Mead{For the $\epsilon=3$ model the ${\langle\Delta B\rangle}/{\langle B\rangle}$ goes up to $84\%$ compared to $34\%$ for the standard model. Why does the axion density profile not increase by $84/34$? I'm sure you explained this to me before, but I can't remember what you said and probably it should be spelled out in the text.}\Liang{It is due to the insufficiency of local AQN annihilation. Also see the comments and modification of the text in your preceding question.   }} }
	\begin{tabular}{cccrcc} 
	\hline\hline 
	$\langle B\rangle$ &$\alpha$ & Other parameters & $\langle \Delta m_{\rm AQN}\rangle$ [kg] & ${\langle\Delta B\rangle}/{\langle B\rangle}$ & $\rho_\mathrm{a}(R_\oplus)~\rm[GeV\,{cm^{-3}}]$ \\\hline
	$8.84\times10^{24}$ &2.5 & --       & $ {5.01\times10^{-3}}$                 &  ${33.9\%} $                    &  $
	{2.36\times10^{-4}}$                                        \\
	$8.84\times10^{24}$ &2.5 & $\boldsymbol{\mu}=-\mu\hat{\mathbf{z}}$       & ${4.96\times10^{-3}}$                 &  ${33.5\%} $                    &  ${2.29
	\times10^{-4}}$                                        \\
	$8.84\times10^{24}$ &2.5 & Solar gravitation\footnote{We consider an additional $42.1\kmps$ AQN impact velocity from AQN falling to Earth from infinity in the gravitational well of the Sun, this gives an additional velocity to the AQN that is always in the direction of travel. See Appendix \ref{sec:other plots} for more details}      & ${5.01\times10^{-3}}$                 &  ${33.9\%} $                    &  $
	{2.74\times10^{-4}}$                                        \\
	$8.84\times10^{24}$ &2.5 & $\epsilon=3$        & ${1.24\times 10^{-2}}$                 & ${84.0\%}$                                         &   ${3.73\times10^{-4}}$                                        \\
	$2.43\times10^{25}$ &2.0 & --       & ${8.34\times10^{-3}}$                     & ${20.5\%}$                                         & ${1.52\times10^{-4}}$                \\
	$2.43\times10^{25}$ &2.0 & $\boldsymbol{\mu}=-\mu\hat{\mathbf{z}}$       & ${8.15\times10^{-3}}$                     & ${20.1\%}$                                         & ${1.43\times10^{-4}}$                \\
	${4.25\times10^{25}}$ &(1.2, 2.5) & $B_{\rm min}=10^{23}$       & ${1.08\times10^{-2}}$                &  ${15.1\%} $                                    & ${1.17\times10^{-4}}$                                        \\
	${1.05\times10^{26}}$ &(1.2, 2.5) & --      & ${2.57\times10^{-2}}$                     & ${14.6\%}$                                         & ${1.14\times10^{-4}}$                                         \\
	${1.05\times10^{26}}$ &(1.2, 2.5) & $\boldsymbol{ \mu}=-\mu\hat{\mathbf{z}}$       & ${2.48\times10^{-2}}$                     & ${14.1\%}$                                         & ${1.07\times10^{-4}}$                                         \\
	\hline
\end{tabular}
	\label{table:5.2 summary of some results}
\end{table*}

\begin{figure}[h]
\centering
\captionsetup{justification=raggedright}
\includegraphics[width=1\linewidth]{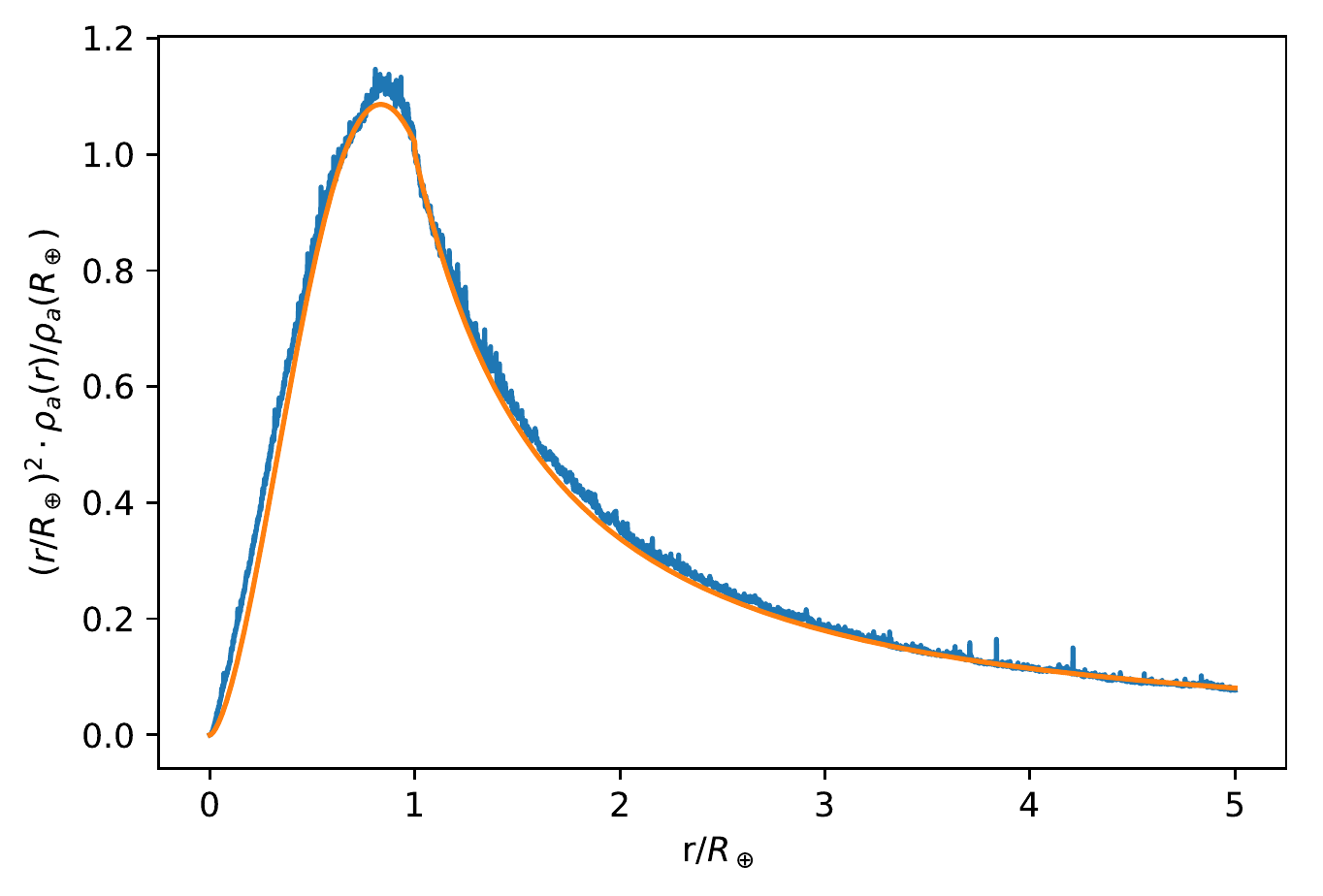}
\caption{The axion-density profile around the Earth, $\rho_\mathrm{a}(r)$, corresponding to an incident AQN distribution with $\alpha=(1.2, 2.5)$, $B_{\min}=3\times10^{24}$. The simulation (blue) is compared with the analytical fit (orange). $2\times10^5$ axion samples were in the simulation.}
\label{fig:5.2 rho_a alphaPW}
\end{figure}

\subsection{Axion-velocity spectrum features on the surface of the Earth}
\label{subsec:5.3 Axion speed spectrum}

In  this subsection we derive the axion-velocity spectrum $F(v_\mathrm{a})$ at the surface of the Earth. Before discussing the simulation results, we note that the axion spectrum $F(v_\mathrm{a})$ can be estimated based on a simple toy model. Surprisingly, the estimated profile gives a consistent and precise description which is in good agreement with Monte Carlo numerical simulations. The arguments presented below strongly suggest that $F(v_\mathrm{a})$ is not sensitive to any parameters of the incoming AQN flux, and $F(v_\mathrm{a})$ is expected to be similar regardless of the  distribution functions computed in previous sections. The main reason for this insensitivity to parameters is that the basic physics of low-energy axions is governed by gravity rather than by any specific features related to the AQNs emitting these axions. 


We start our argument by rewriting eq. (\ref{eq:4.4 v_a(R_Earth)}) as 
\begin{equation}
\label{eq:5.3 va}
\begin{aligned}
v_\mathrm{a}(r)=\sqrt{\frac{2}{m_\mathrm{a}}[E-U(r)]}\ ,
\end{aligned}
\end{equation}
and introduce following notation 
\begin{equation}
\label{eq:5.3 notations}
v_\mathrm{a}\equiv v_\mathrm{a}(R_\oplus),~r_0\equiv r(0),~u_0\equiv v_\mathrm{a}(0)\ .
\end{equation}
The escape velocity at the surface of the Earth is given by
\begin{equation}
\label{eq:5.3 notations 2}
v_\oplus\equiv v_{\rm esc}(R_\oplus)
=\sqrt{\frac{2GM_\oplus}{R_\oplus}}
\simeq 11 ~\rm \kmps\ .
\end{equation}
The energy in eq. \eqref{eq:5.3 va} can now be written as
\begin{equation}
\label{eq:5.3 denote E}
E
=\frac{1}{2}u_0^2+U(r_0)
=\frac{1}{2}m_\mathrm{a}u_0^2-\frac{1}{2}m_\mathrm{a}v_{\rm esc}^2(r_0)\ .
\end{equation}

We now construct a toy model using a number of approximations. First, we assume that the Earth is point mass, which simplifies the gravitation potential:
\begin{equation}
\label{eq:5.3 uniform U}
U(r)
\simeq-\frac{GM_\oplus m_\mathrm{a}}{r}
=-\frac{1}{2}m_\mathrm{a}v_\oplus^2\frac{R_\oplus}{r}\ .
\end{equation}
This is correct for all distances above the surface of the Earth. As the next step, we use  Eqs. \eqref{eq:5.3 denote E} and \eqref{eq:5.3 uniform U} to simplify Eq. \eqref{eq:5.3 va} and derive
\begin{equation}
\label{eq:5.3 va 2}
v_\mathrm{a}(r)\simeq\sqrt{u_0^2-v_{\rm esc}^2(r_0)+v_\oplus^2\frac{R_\oplus}{r}}\ ,
\end{equation}
which should be valid for $r\geq R_\oplus$. From this simplified case, one can extract many useful relations that must be valid for all radii above the surface of the Earth. For example, by setting $v_\mathrm{a}(r)=0$ in Eq. \eqref{eq:5.3 va 2}, one can estimate the maximum radial distance $r_{\rm max}$ for a trapped axion as follows
\begin{equation}
\label{eq:5.3 r_max}
r_{\rm max}(u_0)
\simeq R_\oplus \frac{v_\oplus^2}{v_{\rm esc}^2(r_0)-u_0^2}\ .
\end{equation}
This relation allows to estimate the orbital period for these axions:
\begin{equation}
\label{eq:5.3 T}
T(u_0)
\simeq\frac{2r_{\rm max}}{v_\mathrm{a}}
\simeq\frac{2R_\oplus}{v_\mathrm{a}}
\frac{v_\oplus^2}{v_{\rm esc}^2(r_0)-u_0^2}\ ,
\end{equation}
where $v_\mathrm{a}\equiv v_\mathrm{a}(R_\oplus)$ is the speed of a given axion at the surface of the Earth. Also, at $r=R_\oplus$, the parameter $u_0$, the axion velocity at the center of the Earth, can be expressed as a function of $v_\mathrm{a}$ using Eq. \eqref{eq:5.3 va 2} as follows:
\begin{equation}
\label{eq:5.3 u0}
u_0(v_\mathrm{a})
\simeq\sqrt{v_\mathrm{a}^2+v_{\rm esc}^2(r_0)-v_\oplus^2}\ .
\end{equation}
We can now estimate the axion spectrum $F^{\rm (est)}(v_\mathrm{a})$ at the surface of the Earth:
\begin{equation}
\label{eq:5.3 F(va)_est 1}
\begin{aligned}
F^{\rm (est)}(v_\mathrm{a})
&\propto\rho_{\rm obs}(u_0)\cdot\frac{1}{T(u_0)}\cdot
\left(\frac{du_0}{dv_\mathrm{a}}\right)\ ,
\end{aligned}
\end{equation}
where $u_0$ is given by Eq. \eqref{eq:5.3 u0}. Expression (\ref{eq:5.3 F(va)_est 1}) comes from the fact that $F(v_\mathrm{a})$ must be proportional to the axion emission spectrum $\rho_{\rm obs}$ and the frequency of the orbital period $1/T$. The last term $(\rmd u_0/\rmd v_\mathrm{a})$ is the Jacobian of the transformation from the basis of $u_0$ to $v_a$.


Substituting Eq. \eqref{eq:5.3 u0} into Eq. \eqref{eq:5.3 F(va)_est 1}, the spectral axion distribution (not normalized) at the surface of the Earth can be written as:
\begin{equation}
\label{eq:5.3 F(va)_est 2}
F^{(\rm est)}(v_\mathrm{a})
\sim v_\mathrm{a}^2\left(1-\frac{v_\mathrm{a}^2}{v_\oplus^2}\right)
\sqrt{\frac{v_\mathrm{a}^2+v_{\rm esc}^2(r_0)}{v_\oplus^2}-1}\ ,
\end{equation}
where $v_{\rm esc}(r_0)$ is the only unknown parameter affecting the form of $F^{(\rm est)}(v_\mathrm{a})$. However, the local escape velocity $v_{\rm esc}(r_0)$ varies only in a narrow range, from $11\kmps$ (from the surface of the Earth) to $15\kmps$ (from the center of the Earth). Therefore, the spectrum  $F^{(\rm est)}(v_\mathrm{a})$ given by (\ref{eq:5.3 F(va)_est 2}) depends only weakly on $v_{\rm esc}(r_0)$.  

\begin{figure}[h]
\centering
\captionsetup{justification=raggedright}
\includegraphics[width=1\linewidth]{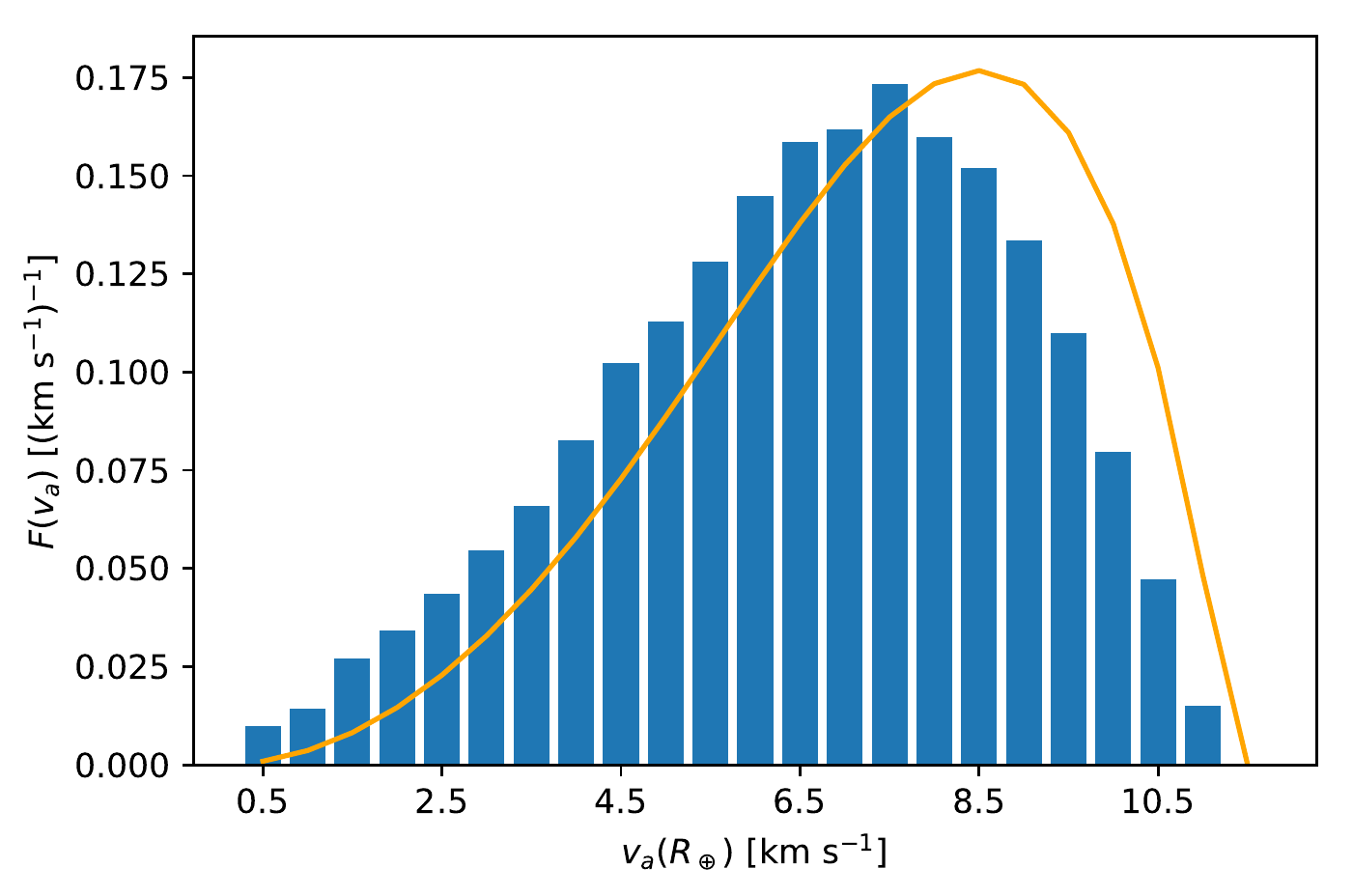}
\caption{The axion velocity-distribution profile on the surface of the Earth $F(v_\mathrm{a})$ vs. $v_\mathrm{a}(R_\oplus)$: $\alpha=(1.2, 2.5)$, $B_{\min}=3\times10^{24}$. The simulation (blue) is compared with the analytical estimation from our simple toy model $F^{(\rm est)}(v_\mathrm{a})$ (orange) with $v_{\rm esc}(r_0)=13.5 \kmps$. The distribution peaks $\simeq7\kmps$ and there is (obviously) a sharp cut off at the escape speed $\simeq 11\kmps$. $2\times10^5$ axion samples were used for the simulation.}
\label{fig:5.3 F(va) alphaPW}
\end{figure}

Fig. \ref{fig:5.3 F(va) alphaPW} shows that our simple estimate, Eq. (\ref{eq:5.3 F(va)_est 2}), and the simulations are in very good agreement. In Appendix \ref{sec:other plots} we show that the agreement holds for different choices of the AQN model parameters. The reason for this insensitivity to large variations in AQN parameters (such as $\alpha$, $B_{\rm min}$, $\epsilon$, incident-flux distribution) is because the velocity spectrum of the gravitationally-bound axions is mostly determined by the gravity of the Earth, rather than by any features of the AQNs. Therefore, the analytical expression (\ref{eq:5.3 F(va)_est 2}) remains a valid description of the spectrum, where all specific features related to  the AQNs are hidden in $v_{\rm esc}(r_0)$ which can be assumed to be a constant computed as an average over the entire ensemble of axions emitted from any point in the interior of the Earth by any AQN of any size at any moment. One can therefore use $\la v_{\rm esc}(r_0)\ra \simeq 13.5 \kmps$ in Eq. (\ref{eq:5.3 F(va)_est 2}), which is in good agreement with numerical simulations.

\section{Conclusions and future directions}
\label{sec:6.conclusion}

The main results of our work can be summarized as:

{\bf a}. We computed the energy-density profile of trapped axions as a function of the distance from the center of the Earth using numerical Monte Carlo simulations. These axions are produced when antimatter AQNs incident on Earth annihilate with the rocky material in the interior. The trapped axions represent the tiny, low-velocity tail of the axions emitted during this annihilation process, a velocity distribution that otherwise peaks at around $0.5c$. However, the number of trapped axions can be appreciable since they build up over the lifetime of the Earth. Results for the density profile are shown in Fig. \ref{fig:5.2 rho_a alphaPW}, and in Figs. \ref{fig:E plots - q(rv)_alpha2.0_dist2}, \ref{fig:E plots - q(rv)_alpha2.5_Epsilon3} in Appendix \ref{sec:other plots} for different parameters of the incident AQN population ($\alpha$ and $B_{\rm min}$, governing the AQN-mass distribution; $\epsilon$ governing the AQN-Earth material cross section; incident flux velocity distribution). The final value we calculate for the axion density at the surface of the Earth after 4.5 Gys, $\rho (R_\oplus)$, is not very sensitive to these AQN parameters, which suggests that one can use the estimate (\ref{final_estimate}) as a baseline for the trapped-axion density that is quite independent of the details of the AQN model. We refer the reader to Table \ref{table:5.2 summary of some results} for an overview of the sensitivity of the results to the AQN parameters.

{\bf b}. We computed the trapped-axion velocity-distribution $F(v_\mathrm{a})$, shown on Fig. \ref{fig:5.3 F(va) alphaPW}, and Figs. \ref{fig:E plots - F(va)_alpha2.0_dist2}, \ref{fig:E plots - F(va)_alpha2.5_Epsilon3} in Appendix \ref{sec:other plots} for different parameters of the incident AQN population. Once again, we find that $F(v_\mathrm{a})$ is not sensitive to the incident AQN parameters, implying that the formula (\ref{eq:5.3 F(va)_est 2}) can be used as a good approximation for the velocity distribution of the gravitationally bound axions that is quite independent of the details of the AQN model. One can express the important features of the trapped-axion system in terms of the de Broglie wavelength $\lambda_\mathrm{a}$:
\be
\label{lambda}
\frac{\lambda_\mathrm{a}}{2\pi}\equiv \frac{\hbar}{m_\mathrm{a}v_\mathrm{a}}\simeq 50\,{\rm m}\left(\frac{m_\mathrm{a}}{10^{-4}\rm eV}\right)^{-1}\left(\frac{v_\mathrm{a}}{11\kmps}\right)^{-1}\ ,
\ee
which is much greater than the typical wavelength of the conventional galactic axions because $v_\mathrm{a}\leq v_{\oplus}\approx 11\kmps$ in formula (\ref{lambda}), whereas typical galactic axions have $v_\mathrm{a}\simeq 200\kmps$. This key feature of gravitationally-trapped axions may be a decisive element that may drastically increase the discovery potential for these axions. This is because axion-search instruments that are based on idea of {\it coherence} could be made more sensitive, and could collect the signal for much longer period of time without losing the {\it coherent} features. These experiments may be able to overcome the relatively small axion density we calculate in eq. (\ref{final_estimate}) in comparison with conventional estimates for the galactic axions, due to the possibility that the signal may be collected over a longer period of time.


For example, the conventional cavity quality factor $Q\approx 10^{5}-10^{6}$ is mostly constrained  by the finite bandwidth ${\cal{O}}(\beta^2)\sim 10^{-6}$ for galactic axions with $\beta\equiv v/c\sim 10^{-3}$. Galactic axions can be thought as a highly monochromatic radiation with a coherence time corresponding to $10^6$ oscillation periods and coherence length of roughly $10^3$ times the axion Compton length $\lambda_C= {\hbar}/({m_\mathrm{a}c})$, see e.g. the analysis \cite{Chaudhuri:2018rqn} of the DM radio experiment and recent review \cite{Battesti:2018bgc}. This should be compared with the gravitationally trapped axions when $\beta_{\oplus}\equiv v_{\oplus}/c\simeq 3.6\times 10^{-5}$, and therefore factor $Q\sim  10^{10}$ could be potentially four orders of magnitude greater than for the galactic axions. This huge enhancement factor is due to the fact that gravitationally trapped axions are much more monochromatic than galactic axions as the coherence length is roughly $\lambda_\mathrm{a}/\lambda_C\sim 10^{5}$ of the axion Compton length,
to be contrasted with factor $10^3$ for the galactic axions. 

We mentioned in Sec.\ref{sec:1. introuction} a number of instruments existing, upgraded (or planning to be upgraded), designed (but not build yet), or under consideration that in principle are capable of achieving the precision required to detect gravitationally-trapped axions. Along with the list of instruments presented in Introduction 
we also want to mention another possible design, advocated in \cite{Cao:2017ocv}, which is based on quantum coherence of the  propagating axions, and which may also benefit from the large coherence length (\ref{lambda}) mentioned above. Finally, we want to mention
the idea advocated in \cite{Goryachev:2017wpw,Jeong:2017hqs,Melcon:2018dba}  that the combination of large number of cavities can drastically enhance the axion signal. The  design advocated in \cite{Goryachev:2017wpw,Jeong:2017hqs,Melcon:2018dba} may  benefit from   large wave length (\ref{lambda}) of the gravitationally bound axions.

It is the central result of this work that all the questions formulated in the Introduction as items 1-6 have been successfully answered: we have derived the axion density on the surface of the Earth, where detection can be made (as stated in item {\bf ``a"}   above) and the velocity-spectral features of these gravitationally trapped axions (as stated in item {\bf ``b"} above). We have also demonstrated that these results are not very sensitive to the incident AQN distribution model, nor to the interaction pattern between the AQNs in the interior of the Earth. Therefore, we think our results are solid and robust predictions within the AQN framework. We emphasize once again that the corresponding properties (\ref{final_estimate}), (\ref{eq:5.3 F(va)_est 2}) and (\ref{lambda}) are very distinct from conventional galactic axions. In particular, the spectral features (\ref{eq:5.3 F(va)_est 2}) and (\ref{lambda}) describe the gravitationally trapped axions that are produced in the interior of the Earth; and no other model we know of would generate such spectral properties. The {\it rigidity  and predictiveness} of the model leads to very limited freedom and flexibility for any modification of (\ref{final_estimate}), (\ref{eq:5.3 F(va)_est 2}) and (\ref{lambda}). 

Why should we take this AQN model seriously? First of all, this model is consistent with all available cosmological, astrophysical, satellite and ground based constraints, where AQNs could leave a detectable electromagnetic signature. While the model was initially invented to explain the observed relation $\Omega_{\rm DM}\sim \Omega_{\rm visible}$, it may also explain a number of other (naively unrelated) phenomena, such as the  excess of galactic emission in different frequency bands as discussed in section \ref{subsec:2.1 AQN structure}. The AQN model may also offer resolutions to some other astrophysical mysteries: the so-called ``Primordial Lithium Puzzle" \cite{Flambaum:2018ohm},  the so-called ``The Solar Corona Mystery"  \cite{Zhitnitsky:2017rop,Raza:2018gpb}, the recent EDGES observations of a stronger than anticipated 21 cm absorption features \cite{Lawson:2018qkc}, to name just a few.\footnote{In fact, there is  some conceptual similarity between the ``Primordial Lithium Puzzle" and the axions-emission problem which is the topic of  the present work. In both cases, a fraction of the constituents is hidden inside the AQNs. The Li nuclei are captured by AQNs after BBN at $T\simeq 20$ keV as discussed in \cite{Flambaum:2018ohm}, while off-shell axions are hidden in the form of axion domain walls during the formation time at $T\simeq 100$ MeV as discussed in Section \ref{sec:1. introuction}. In first case it leads to the depletion of the visible Li- nuclei  as argued in \cite{Flambaum:2018ohm}. In the second case the originally hidden axions might become observable propagating on-shell particles as a result of AQN annihilation events similar to those discussed in the present work.} These cosmological puzzles could  be resolved within AQN framework with the {\it same set of physical parameters} advocated in the present work. The only parameter that changes is $\epsilon$ which scales the effective cross section as in expression (18).


In this respect, the proposal advocated in this work: to search for gravitationally trapped axions that are inevitably produced as a result of the annihilation of incident antimatter AQN with the matter of the Earth is a {\it direct} manifestation of the AQN model. In fact, the observation of these axions, with their very distinct density and velocity-spectral properties would be the smoking gun supporting the entire AQN framework. 


\exclude{
\section*{Contributions}
AM helped with the numerical work and with writing the paper.
}

\section*{Acknowledgments}
AZ is thankful to Konstantin Zioutas for questions which motivated these studies. AZ also thanks Kent Irwin and Yannis Semertzidis for discussions on possible enhancement factors  due to the large wave length (\ref{lambda}) which drastically increases the coherence time scale. We are also  thankful to  David Marsh, Alex Millar, Jan Schuette-Engel, Pierre Sikivie   
and Mike Tobar for comments, correspondence  and discussions. This work was supported in part by the National Science and Engineering Research Council of Canada. AM acknowledges support from the Horizon 2020 research and innovation programme of the European Union under the Marie Sk\l{}odowska-Curie grant agreement No. 702971.

\label{lastpage}

\appendix
\section{Constraint from IceCube}
\label{sec:Constraint from IceCube}
Here we estimate the constraint  of the AQN model from IceCube's observation. In short, depending on the sensitivity of the detector, we find a mean baryon charge    $\langle B\rangle\lesssim 1.6\times10^{24}$  is excluded up to $3.5\sigma$ level of significance for $\eta=1$.
For $\eta=3$ similar constraint  reads  $\langle B\rangle\lesssim 4.7\times10^{24}$, see definition for the parameter $\eta$ below.   We present the details of such estimation as follows.

\begin{figure}[h]
\centering
\includegraphics[width=0.6\linewidth]{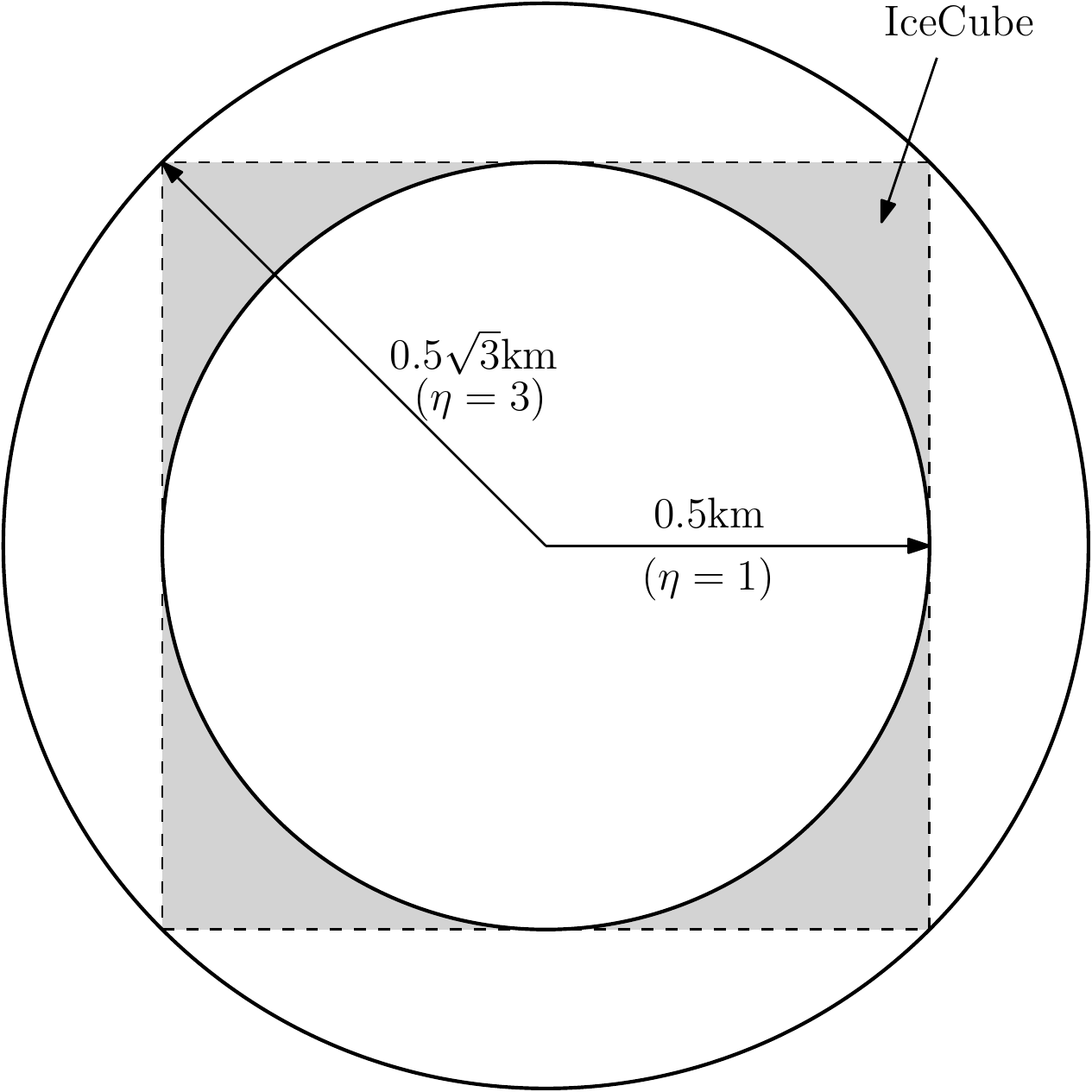}
\caption{Definition of the $\eta$ parameter.}
\label{fig:IceCube eta}
\end{figure}

To estimate the hit rates of AQNs to the IceCube, we assume majority of AQNs pass through the Earth in the end without annihilation, which is in agreement with the simulation results in the present work.\footnote{To be more specific, simulation indicates the trajectories of AQNs are mostly as like free motions and annihilations only slow down the AQNs by a small fraction $\sim25$-50 $\kmps$. Since slower velocity results in smaller flux, the simplification of zero velocity loss (i.e. free motion) is in fact putting us to a stricter constraint from IceCube.} Then, we conclude the flux of AQNs hitting the Icecube must be approximately isotropic in all directions. To simplify the calculation without loss of generality, we further approximate IceCube as a spherical detector with diameter $\sqrt{\eta}\cdot(1\rm km)$. As shown in Fig. \ref{fig:IceCube eta}, we introduce a parameter $\eta$ to account for the effective cross section of the detector. For the modest estimation, we may expect $\eta=1$, while $\eta=3$ is clearly the maximal cross section ever possible. In all cases, the reasonable range of $\eta$ is from 1 to 3, where $\eta=3$ is a strict upper limit of IceCube's sensitivity. Using average flux density of AQNs computed in Appendix \ref{sec:A. estimation of AQN flux}, we obtain the total hits expected to IceCube in 10 years from Eq. \eqref{eq:D Nflux 3}:
\begin{equation}
\label{eq:4.1 Nflux IceCube}
\begin{aligned}
\langle N_{\rm IceCube}^{\rm 10yr}\rangle
&=4\pi(0.5{\rm km})^2\eta\cdot 10{\rm yr}\cdot
\frac{\langle \dot{N}\rangle}{4\pi R_\oplus}  \\
&\simeq 1.3037\eta
\left(\frac{\rho_{\rm DM}}{0.3{\rm GeV\cdot cm^{-3}}}\right)
\left(\frac{10^{25}}{\langle B\rangle}\right)\ .
\end{aligned}
\end{equation}
Assuming a Poisson distribution, the probability of observing $k$ events within 10 years is 
\begin{equation}
\label{eq:4.2 Poison distribution}
{\rm Prob}(k)
=\frac{\lambda^k}{k!}e^{-\lambda},\quad
\lambda=\langle N_{\rm IceCube}^{\rm 10yr}\rangle.
\end{equation}
The probability of seeing zero event in IceCube over 10 years is summarized in Table \ref{table: IceCube_eta}. As an rough estimate, we choose the average value between $\eta=1$ and 3
\begin{equation}
\label{eq:4.2 B bound_final}
\langle B\rangle\gtrsim3\times10^{24}
\end{equation}
as the constraint from IceCube. This  is precisely the value  presented  in the main body of the paper in eq. (\ref{direct}).
\begin{table} [h]
\caption{Probability of seeing zero event in IceCube over 10 years} 
\centering 
\begin{tabular}{c|cccc} 
\hline\hline 
$\eta$ &$\langle B\rangle$ & ~Hit number~ & Prob($k$=0) [\%]~~ & Significance \\\hline
 & $1.6\times10^{24}$ & 8.148                           & 0.02893                        & $3.44\sigma$          \\
1 &$5.0\times10^{24}$   & 2.607                           & 7.37260                        & $1.45\sigma$          \\
 &$1.0\times10^{25}$   & 1.304                           & 27.15253~~                       & $0.61\sigma$    \\\hline
 &$4.7\times10^{24}$ & 8.321                           & 0.02432                       & $3.49\sigma$          \\
3 &$5.0\times10^{24}$   & 7.822                           & 0.04007                       & $3.35\sigma$          \\
&$1.0\times10^{25}$   & 3.911                           & 2.00185                      & $2.05\sigma$    \\\hline
\end{tabular}
\label{table: IceCube_eta} 
\end{table}

\section{Spectral properties in the rest frame}\label{rest-frame}
In this appendix, we fulfill the technical details in Sec. \ref{subsec:3.2 axion emission spectrum in laboratory frame}. In Eq. \eqref{eq:3.2 rho}, the $H_l(\tilde{p},\delta)$ is the partial wave expansion known as follows \cite{Liang:2018ecs}: 
\begin{widetext}
\begin{subequations}
\label{eq:3.2 a_solution ass}
	\begin{equation}
	\label{eq:3.2 a_solution ass_H}
	\begin{aligned}
	H_l(\tilde{p},\delta)
	&\equiv\sum_{n=0}^\infty\sum_{k=0}^n\frac{e^{-k\delta}}{2^n}
	\frac{(n+1)!}{k!(n-k)!}\frac{(-1)^k}{(k+1)^{l+3}}
	\Gamma(l+\frac{3}{2})f\left(\frac{1}{2}(l+3),\frac{1}{2}(l+4),
	l+\frac{3}{2};\frac{-(\tilde{p}/m_\mathrm{a})^2}{(k+1)^2}\right),
	\end{aligned}
	\end{equation}
	\begin{equation}
	\label{eq:3.2 a_solution ass_f}
	\begin{aligned}
	f\left(\frac{1}{2}(l+3),\frac{1}{2}(l+4),
	l+\frac{3}{2};\frac{-(\tilde{p}/m_\mathrm{a})^2}{(k+1)^2}\right)
	&\simeq\frac{1}{\Gamma(l+\frac{3}{2})}\left[
	1-\frac{(l+3)(l+4)}{4(k+1)^2}
	\frac{\Gamma(l+\frac{3}{2})}{\Gamma(l+\frac{5}{2})}(\tilde{p}/m_\mathrm{a})^2
	+{\cal O}(\tilde{p}/m_\mathrm{a})^4\right],  \\
	\end{aligned}
	\end{equation}
\end{subequations}
\end{widetext}
where $f(a,b,c;z)$ is defined to be the regularized Gauss hypergeometric function ${}_2F_1(a,b,c;z)$, i.e. $f(a,b,c;z)\equiv\frac{1}{\Gamma(c)}{}_2F_1(a,b,c;z)$, see Refs. \cite{Abramowitz15:1972,Abramowitz6:1972} and recent article \cite{DiLella:2002ea}. We also note a useful fact in Eq. \eqref{eq:3.2 a_solution ass_f} that $f(a,b,c;z)$ has a simple quadratic behaviour in the non-relativistic limit $z\rightarrow0$.

Next, we turn to the normalization factor $N(\delta)$. According to Eq. \eqref{eq:3.2 a_solution ass}, the normalization factor obviously depends on $\delta$. We refer the detailed calculation to the original work \cite{Liang:2018ecs}, while quoting results as follows
\begin{equation}
\label{eq:3.2 N}
N(\delta)=\left\{
\begin{aligned}
&0.434, &\delta=0.0  \\
&0.616, &\delta=0.5  \\
&0.798, &\delta=1.0
\end{aligned}
\right.
\end{equation}
In this work, we choose the intermediate value $\delta=0.5$ in the simulation.

\section{Estimation of AQN flux}
\label{sec:A. estimation of AQN flux}
We compute the average flux of AQN (in number density). We start with the standard definition of flux
\begin{equation}
\label{eq:D Nflux 1}
\dot{N}=n\mathbf{A}\cdot\mathbf{v},
\end{equation}
where $n$ is the number density of AQNs, $A=4\pi b^2$ is the cross section of the impact, and $v$ is the velocity of AQN flux. Now we take the average:
\begin{equation}
\label{eq:D Nflux 2}
\langle\dot{N}\rangle
=\langle n\rangle\langle\mathbf{A}\cdot\mathbf{v}\rangle,
\end{equation}
where we take the average number density of an antimatter AQN
\begin{equation}
\label{eq:D Nflux 2 ass_n}
\langle n\rangle
=\frac{3}{5}\frac{\rho_{\rm DM}}{m_\mathrm{p}\langle B\rangle}
\end{equation}
and the incident flux vector is also averaged out as
\begin{equation}
\label{eq:D Nflux 2 ass_Av}
\begin{aligned}
\frac{\langle \mathbf{A}\cdot\mathbf{v}\rangle}{4\pi R_\oplus^2}
&=2\pi\int_0^\infty dv~v^2\int_{0}^{1}d(\cos\theta)
\left(1+\frac{2GM_\oplus}{R_\oplus v^2}\right)
\times  \\
&\quad\times v\cos\theta\cdot
\langle f_{\mathbf{v}}(\mathbf{v})\rangle_{\Omega}  \\
&\simeq 68.6 \kmps\left(\frac{\mu}{220\kmps}\right)
\end{aligned}
\end{equation}
where in the last step, we take $\sigma=110\kmps$, $M_\oplus=5.972\times 10^{24}\rm kg$, and $R_\oplus=6371\rm km$, and the impact parameter $b$ for the Earth is taken into account:
\begin{equation}
\label{eq:1.3 b}
b
=R_\oplus\sqrt{1+\frac{2GM_\oplus}{R_\oplus v^2}}
=1.0013R_\oplus.
\end{equation}
As an additional note, in Eq. \eqref{eq:D Nflux 2 ass_Av} we have taken the angular average over all direction of the galactic wind as
\begin{equation}
\label{eq:D isotropic f}
\begin{aligned}
\langle f_{\mathbf{v}}(\mathbf{v})\rangle_\Omega
&=\frac{1}{4\pi}\int 
d\Omega
\frac{1}{(2\pi \sigma^2)^{3/2}}\exp\left[
-\frac{(\mathbf{v}-{\bm \mu})^2}{2\sigma^2}\right]  \\
&=\frac{1}{(2\pi\sigma^2)^{3/2}}
\frac{\sinh(\mu v/\sigma^2)}{\mu v/\sigma^2}
\exp\left[-\frac{v^2+\mu^2}{2\sigma^2}\right],  \\
\end{aligned}
\end{equation}
which gives the distribution \eqref{eq:4.1 v maxwell} in Sec. \ref{subsec:4.1 AQNs initial}. Also we can read off the flux spectrum from Eqs. \eqref{eq:D Nflux 2} and \eqref{eq:D Nflux 2 ass_Av} as
\begin{equation}
\label{eq:D isotropic flux spectrum}
\frac{d}{dv}\langle\dot{N}\rangle_\Omega
=\frac{1}{4}n\cdot 4\pi R_\oplus^2\cdot 4\pi v^3\
\langle f_{\rm v}({\rm v})\rangle_\Omega
\end{equation}
which gives the flux distribution \eqref{eq:4.1 flux distribution}.

We therefore obtain the hit rate per unit area on Earth surface
\begin{equation}
\label{eq:D Nflux 3}
\begin{aligned}
\frac{\langle\dot{N}\rangle}{4\pi R_\oplus^2}
&=\frac{4.15\times10^{23}}{\langle B\rangle}
{\rm km^{-2}yr^{-1}}
\left(\frac{\rho_{\rm DM}}{0.3{\rm \frac{GeV}{cm^3}}}\right)
\left(\frac{\mu}{220\kmps}\right)
\end{aligned}
\end{equation}
where we take $m_\mathrm{p}=938.27$ MeV and the total hit rate is therefore
\begin{equation}
\label{eq:D Nflux 3 tot}
\begin{aligned}
\langle\dot{N}\rangle
\simeq2.12\times10^{7}{\rm yr^{-1}}
\left(\frac{\rho_{\rm DM}}{0.3{\rm \frac{GeV}{cm^3}}}\right)
\left(\frac{\mu}{220\kmps}\right)
\left(\frac{10^{25}}{\langle B\rangle}\right).
\end{aligned}
\end{equation}
Therefore, we can find the expected number of AQNs that have impac the Earth within 4.5 Gyr is 
\begin{equation}
\label{eq:D N_AQN_4.5Gyr}
\begin{aligned}
\langle N_{\rm AQN}^{\rm 4.5Gyr}\rangle
&=\langle\dot{N}\rangle\cdot4.5{\rm Gyr}  \\
&\simeq9.52\times 10^{16}
\left(\frac{\rho_{\rm DM}}{0.3{\rm \frac{GeV}{cm^3}}}\right)
\left(\frac{\mu}{220\kmps}\right)
\left(\frac{10^{25}}{\langle B\rangle}\right).
\end{aligned}
\end{equation}

\section{Estimation of AQN flux (fixed wind)}
\label{sec:fixed wind}
We also present calculation of AQN flux for fixed wind direction. 

\begin{figure}[h]
	\centering
	\captionsetup{justification=raggedright}
	\includegraphics[width=0.7\linewidth]{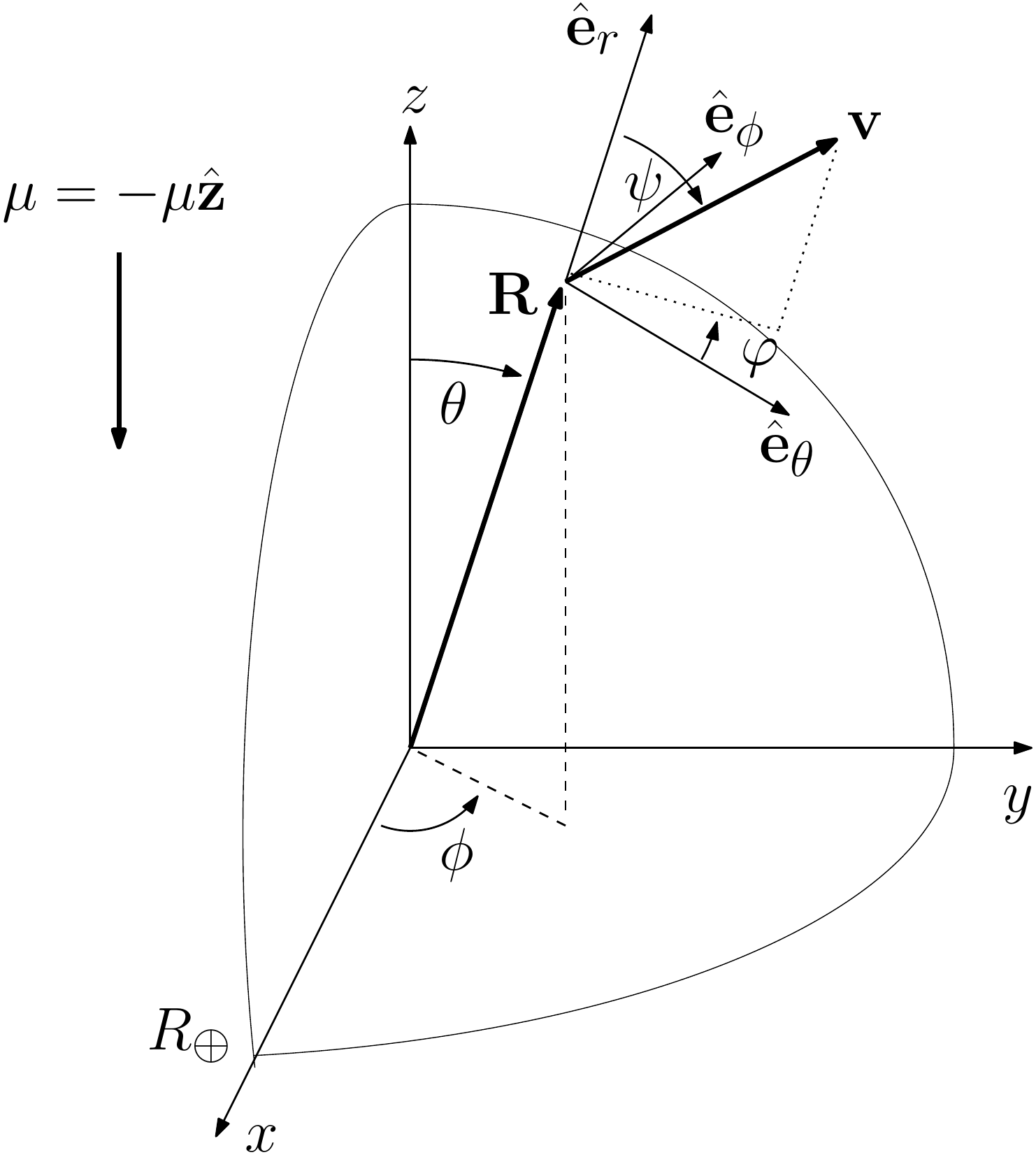}
	\caption{Coordinate system used in the calculation of AQN flux (fixed wind)}
	\label{fig:fixed wind}
\end{figure}

As shown in Fig. \ref{fig:fixed wind}, The coordinate system is set up as shown in Fig. \ref{fig:fixed wind}, which implies the following relation
\begin{equation}
\label{eq:coords}
\begin{bmatrix}
\hat{\mathbf{e}}_r  \\
\hat{\mathbf{e}}_\theta \\
\hat{\mathbf{e}}_\phi  \\
\end{bmatrix}
=
\begin{bmatrix}
&\sin\theta\cos\phi	&\sin\theta\sin\phi	&\cos\theta  \\
&\cos\theta\cos\phi	&\cos\theta\sin\phi	&-\sin\theta  \\
&-\sin\phi	&\cos\phi	&0  \\
\end{bmatrix}
\begin{bmatrix}
\hat{\mathbf{x}}  \\
\hat{\mathbf{y}}  \\
\hat{\mathbf{z}}  \\
\end{bmatrix}.
\end{equation}
Starting with the general velocity distribution
\begin{equation}
\label{eq:f}
f_{\mathbf{v}}(\mathbf{v})
=\frac{1}{(2\pi\sigma^2)^{3/2}}
\exp\left[
-\frac{1}{2\sigma^2}(\mathbf{v}-{\bm \mu})^2
\right]
\end{equation}
and noting that 
\begin{equation}
\label{eq:f ass_vV}
\begin{aligned}
\hat{\mathbf{v}}\cdot\hat{\bm \mu}
&=-\cos\psi\cos\theta+\sin\psi\cos\varphi\sin\theta
\end{aligned}
\end{equation}
from the relation \eqref{eq:coords}, we conclude
\begin{equation}
\label{eq:f_2}
\begin{aligned}
f(\mathbf{v})
&=\frac{1}{(2\pi\sigma^2)^{3/2}}
\exp\left[-{\frac{v^2+\mu^2}{2\sigma^2}}\right]\times\\
&\quad\times\exp\left[
-\frac{v\mu}{\sigma^2}
(\cos\psi\cos\theta-\sin\psi\cos\varphi\sin\theta)
\right]
\end{aligned}
\end{equation}
and the flux of AQN is
\begin{equation}
\label{eq:flux2}
\begin{aligned}
\dot{N}
&=nR^2\int_0^\pi d\theta\sin\theta\int_0^{2\pi}d\phi
\int_{{\hat{\mathbf{e}}_r}\cdot\hat{\mathbf{v}}\leq0} d^3\mathbf{v}~
\hat{\mathbf{e}}_r\cdot \mathbf{v}f(\mathbf{v})  \\
&\propto
\int_0^\pi d\theta
\int_0^\infty dv\int_{\frac{\pi}{2}}^\pi d\psi\int_0^{2\pi}d\varphi~
v^3e^{-\frac{v^2}{2\sigma^2}}\times  \\
&\quad\times\sin\theta\sin\psi\cos\psi\times  \\
&\quad\times\exp\left[{-\frac{v\mu}{\sigma^2}
(\cos\psi\cos\theta-\sin\psi\cos\varphi\sin\theta)}
\right]
\end{aligned}
\end{equation}
where in the last step we drop out all numerical factors as normalization will be applied in the simulation, and the $\phi$ component is simply integrated out due to the azimuthal symmetry of setup. 

\begin{figure}[h]
\centering
\captionsetup{justification=raggedright}
\includegraphics[width=0.9\linewidth]{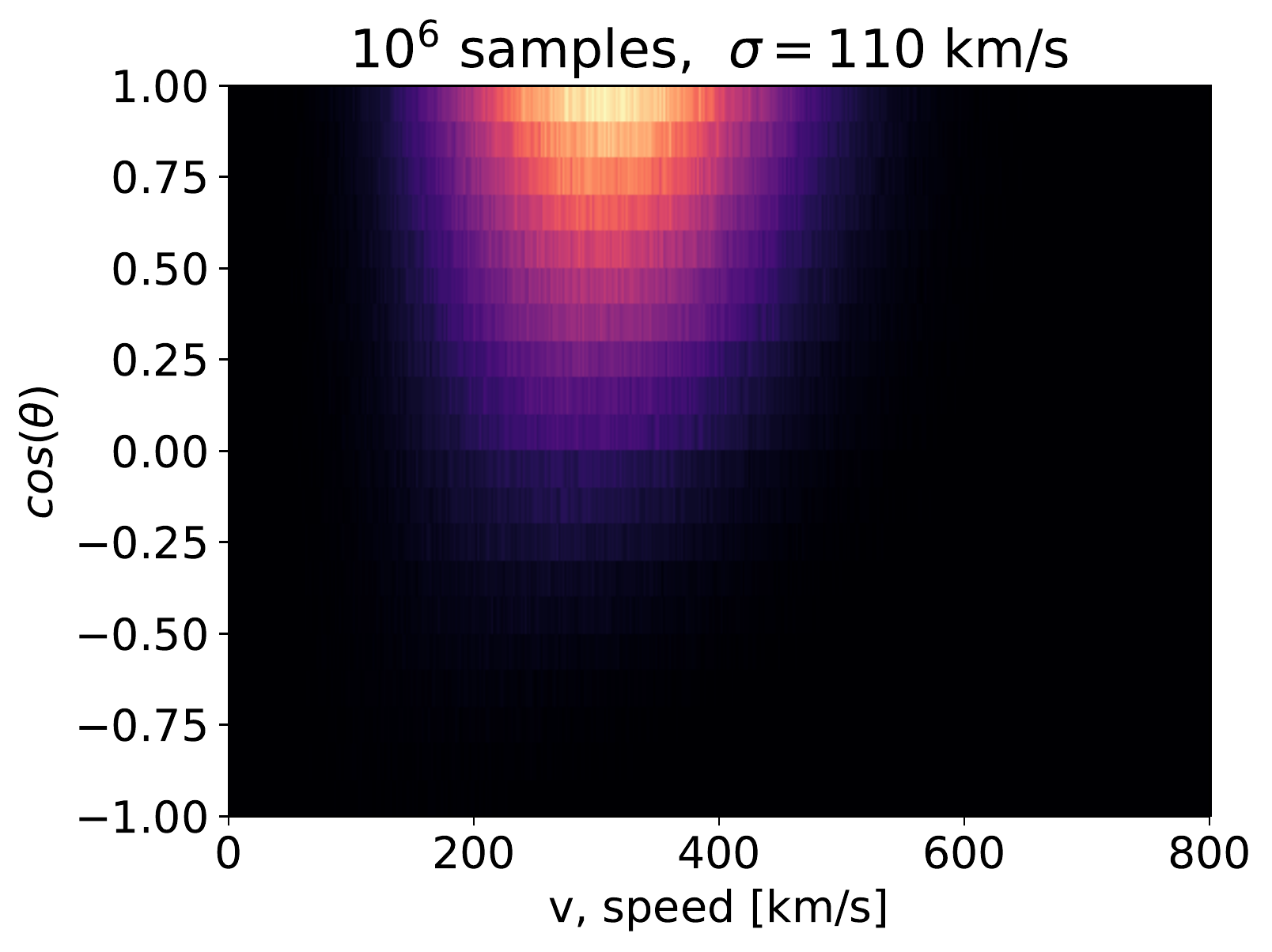}
\caption{The flux distribution (fixed wind) profile obtained by Monte Carlo simulation ($10^6$ samples).Here $\mu=220 \kmps$, $\sigma=110\kmps$ .}
\label{fig:fixed wind Monte Carlo}
\end{figure}

The integral over $\psi$ and $\varphi$ can be performed numerically by Monte Carlo method, as presented in Fig. \ref{fig:fixed wind Monte Carlo}.

\section{Heat emission profile simulation}
\label{sec: heat emission}
This appendix describes the computation details solving for the heat emission profile $q(r,v_{\rm AQN})$. 
First, the simulation implements numerical solving of differential equations \eqref{eq:3.1 AQN DEs} using Python for individual AQNs, and for each of these AQN trajectories through the Earth the simulation returns the solution in the form of the following array
\[
\begin{bmatrix}
\mathbf{r}(t_i) & v(t_i) & \dot{m}(t_i) & m(t_i) \\
\mathbf{r}(t_i+\Delta t) & v(t_i+\Delta t) & \dot{m}(t_i+\Delta t) & m(t_i+\Delta t) \\
\mathbf{r}(t_i+2\Delta t) & v(t_i+2\Delta t) & \dot{m}(t_i+2\Delta t) 
& m(t_i+2\Delta t) \\
\vdots &\vdots	&\vdots &\vdots \\
\mathbf{r}(t_f) & v(t_f) & \dot{m}(t_f) & m(t_f)
\end{bmatrix}
\]
where each row corresponds to a time $t$ for the trajectory at which sampling has been done, and denotes the radial position $\mathbf{r}(t)$, the speed  $v(t)$, the rate of mass loss $\dot{m}(t)$ and the instantaneous mass $m(t)$, where $t_i$ and $t_f$ give the AQN's time of entry into the Earth and the time of exit respectively. 

By doing the histogram statistics of the first two columns with weighting factor $\dot{m}$ in the third column, we obtain the unnormalised $q(r,v_{\rm AQN})$. To do so, we used NumPy's $hist2d$ function to bin the first two columns into a 200$\times$200 pixel grid using weights $\dot{m}(t)$. The simulation for an AQN trajectory was set to have a default sampling rate of $2$ sample points per second. For a given AQN trajectory in the phase space $(r,v_{\rm AQN})$, at sufficiently high velocities this would lead to gaps (empty pixels) being seen in the trajectory on the weighted 2D histogram plot. To tackle this, we use Python's SciPy package to perform spline interpolation on our points for the radial distances $\mathbf{r}(t)$, instantaneous velocities $v(t)$ and instantaneous masses $m(t)$ on the interval [$t_i$,$t_f$] and demand a sufficient number of sample points such that for the highest $v(t)$ for the trajectory we would obtain at least one point per pixel. We use these and Eq. \eqref{eq:3.1 mass loss} to compute the corresponding $\dot{m}(t)$ (directly interpolating $\dot{m}(t)$ is unreliable due to it not being a smooth function). This procedure changes the number of sampling points per second and makes it different for every AQN trajectory, and hence when plotting $q(r,v_{\rm AQN})$ we multiply the weighting factor $\dot{m}$ by 1/(sample points per second) to obtain the new weighting factor $\dot{m}(t)$/(sample points per second) for points corresponding to each AQN. By plotting sufficiently many such weighted trajectories on the 2D histogram, we obtain $q(r,v_{\rm AQN})$. For our simulations, we have chosen to do this for $10^6$ AQNs. 

To normalize $q(r,v_{\rm AQN})$, we should use the information in the fourth column. We first build up a new array for the total $N$ samples of AQNs
\[
\begin{bmatrix}
& m_1(t_i)  & m_1(t_f) \\
& \vdots & \vdots \\
& m_N(t_i)  & m_N(t_f)
\end{bmatrix}
\]
with each row recorded the mass loss of one sample AQN. Then we can estimate $\langle \Delta m_{\rm AQN}\rangle$,  the total mass loss per AQN on average, using the general formula
\begin{equation}
\label{eq:4.2 Delta m_AQN}
\langle\Delta m_{\rm AQN}\rangle
=\frac{\sum_{k=1}^{N}[m_k(t_i)-m_k(t_f)]}{\sum_{k=1}^{N}m_k(t_i)}
\cdot m_\mathrm{p}\langle B\rangle.
\end{equation}
We therefore obtain the normalized $q(r,v_{\rm AQN})$ from condition \eqref{eq:3.3 q(r,v)}.

\section{On sensitivity of the main results  to the AQN's parameters}
\label{sec:other plots}
The main goal of this appendix is to argue that the main results of this work are not very sensitive to the parameters of the model, such as size distribution of AQN (represented by parameters $\alpha$ and $B_{\rm min}$),   $\epsilon$  (describing the strength of the interaction of AQNs with material), flux distribution of AQNs (depending on the direction of galactic wind ${\bm \mu}$ and gravitation attraction from the Solar system $v_{\rm esc}\simeq 42.1\kmps$). 

Before presenting the simulation results, we describe the parameters considered in the work as summarized in Table \ref{table:5.2 summary of some results}. First, we choose the set of $(\alpha,B_{\rm min})$ in Table \ref{table:2.2 mean B} such that $\langle B\rangle\gtrsim10^{25}$, 
subjecting to the constraint \eqref{direct} from IceCube. Next, we choose different values of $\epsilon=1$ and 3 to study the dependence of the effective cross section of AQN, as discussed in Sec. \ref{subsec:3.1 annihilation of AQNs impacting the Earth}. Additionally, we consider three different flux distributions of AQNs: isotropic averaging, fixed-wind direction, and acceleration by solar gravitation. Since the first two are fully described in the main text as well as Appendices \ref{sec:A. estimation of AQN flux} and \ref{sec:fixed wind}, we only comment on the third distribution here. The gravitation of the solar system implies an additional escape velocity $\sim42.1\kmps$ at Earth distance. To take into account of such effect, we add an additional magnitude to the initial speed for each AQN by $\sim42.1\kmps$ in the simulation based on the isotropic distribuiton \eqref{eq:4.1 flux distribution}. 

\begin{figure*}[!htp]
	\centering
	\begin{subfigure}[b]{0.465\linewidth}
		\includegraphics[width=\linewidth]{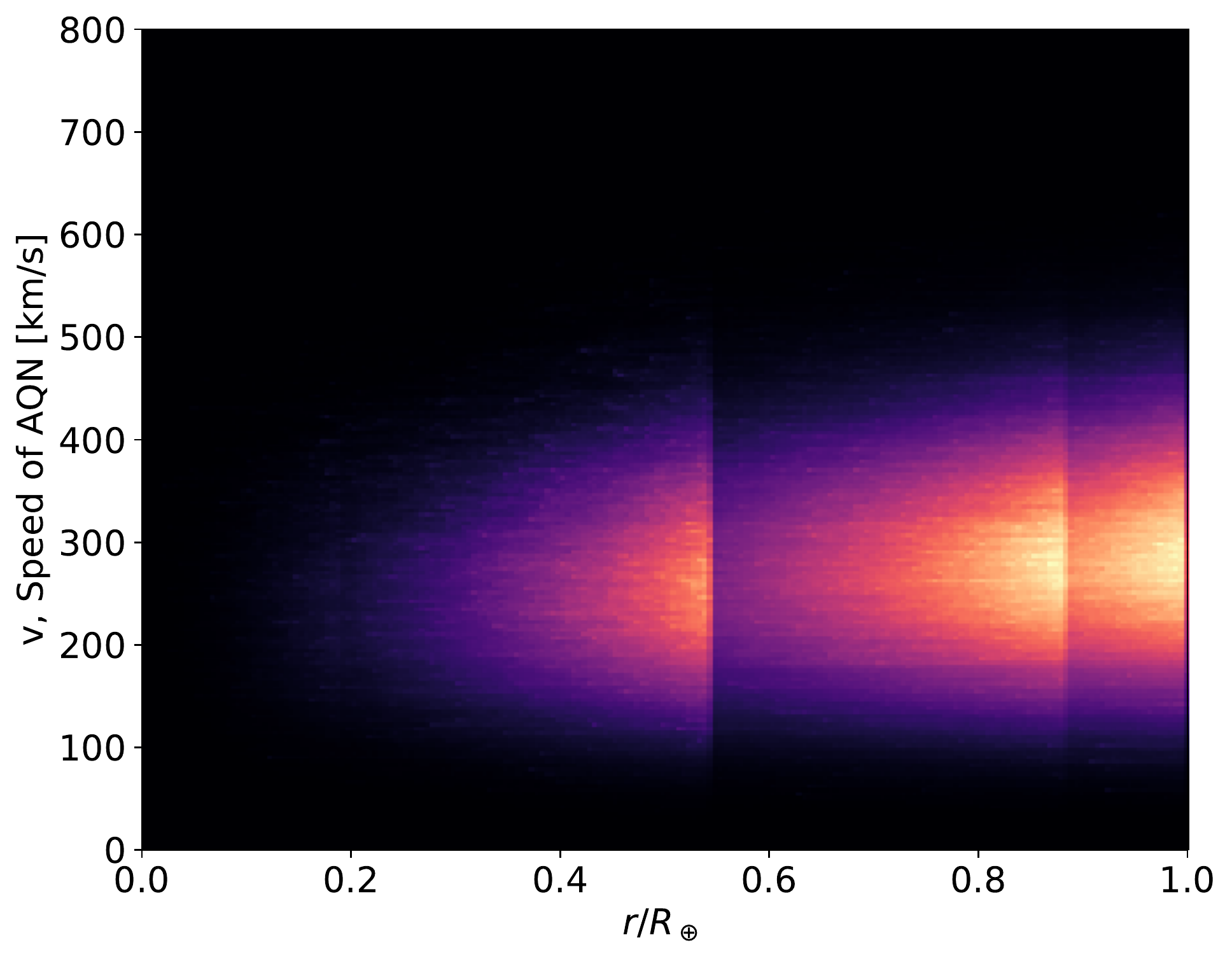}
		\caption{ $\alpha=2.0$, $\epsilon=1$, ${\bm \mu}=-\mu\hat{\mathbf{z}}$ }
	\label{fig:E plots - q(rv)_alpha2.0_dist2}
	\end{subfigure}
	\begin{subfigure}[b]{0.47\linewidth}
	\includegraphics[width=\linewidth]{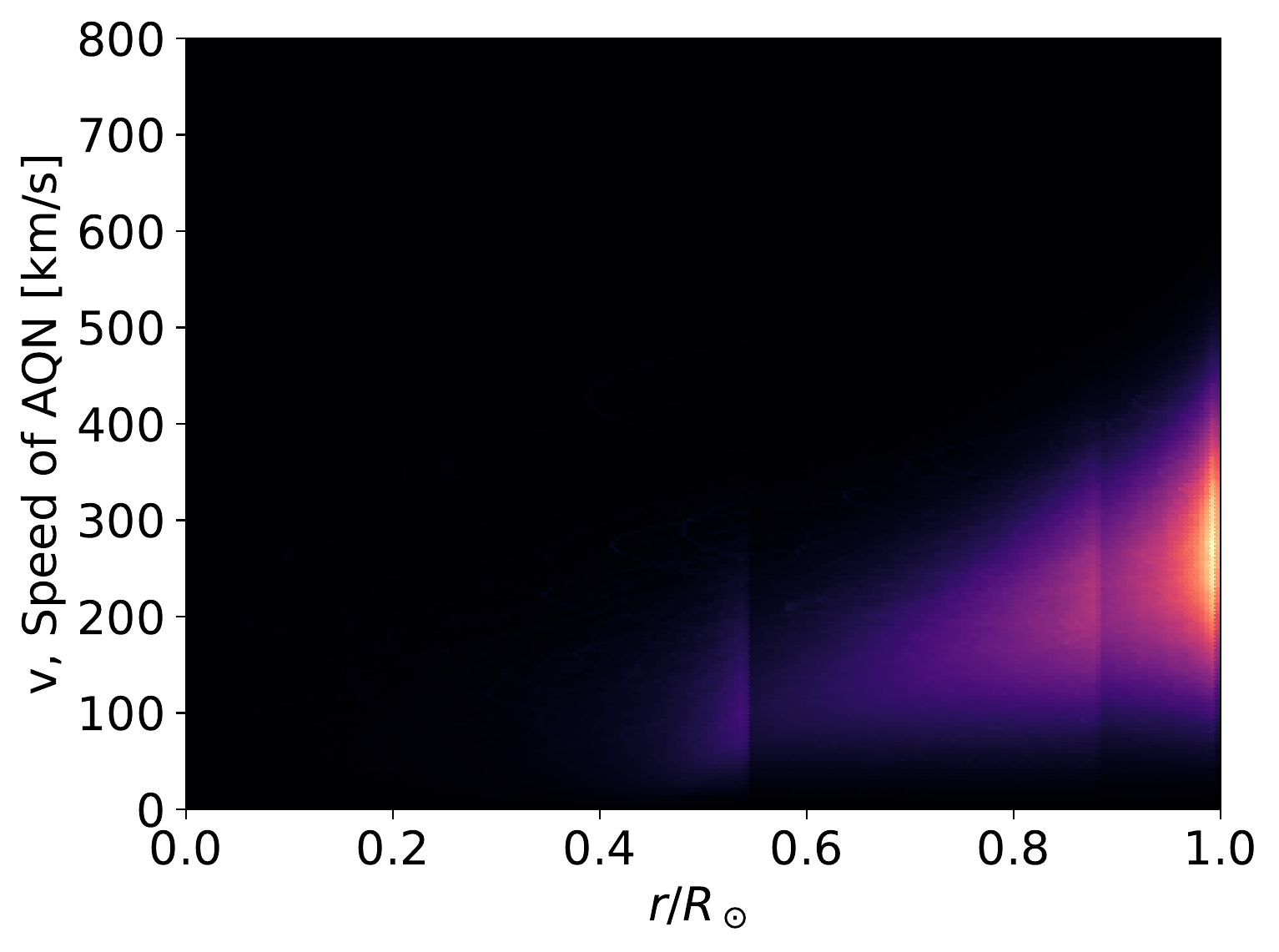}
	\caption{ $\alpha=2.5$, $\epsilon=3$}
	\label{fig:E plots - q(rv)_alpha2.5_Epsilon3}
	\end{subfigure}
	\begin{subfigure}[b]{0.47\linewidth}
	\includegraphics[width=\linewidth]{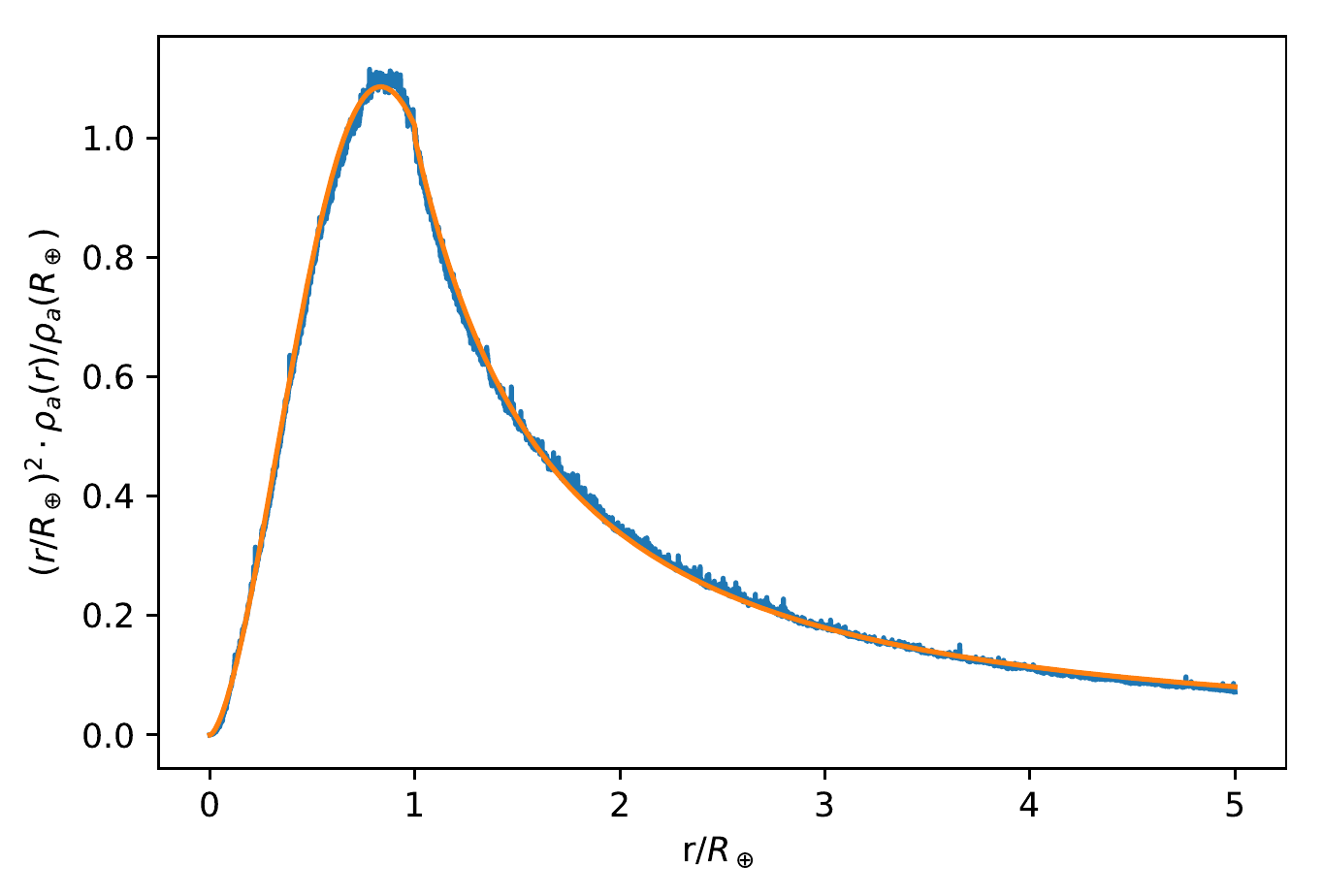}
	\caption{ $\alpha=2.0$, $\epsilon=1$, ${\bm \mu}=-\mu\hat{\mathbf{z}}$}
	\label{fig:E plots - rhoa(r)_alpha2.0_dist2}
	\end{subfigure}
	\begin{subfigure}[b]{0.47\linewidth}
	\includegraphics[width=\linewidth]{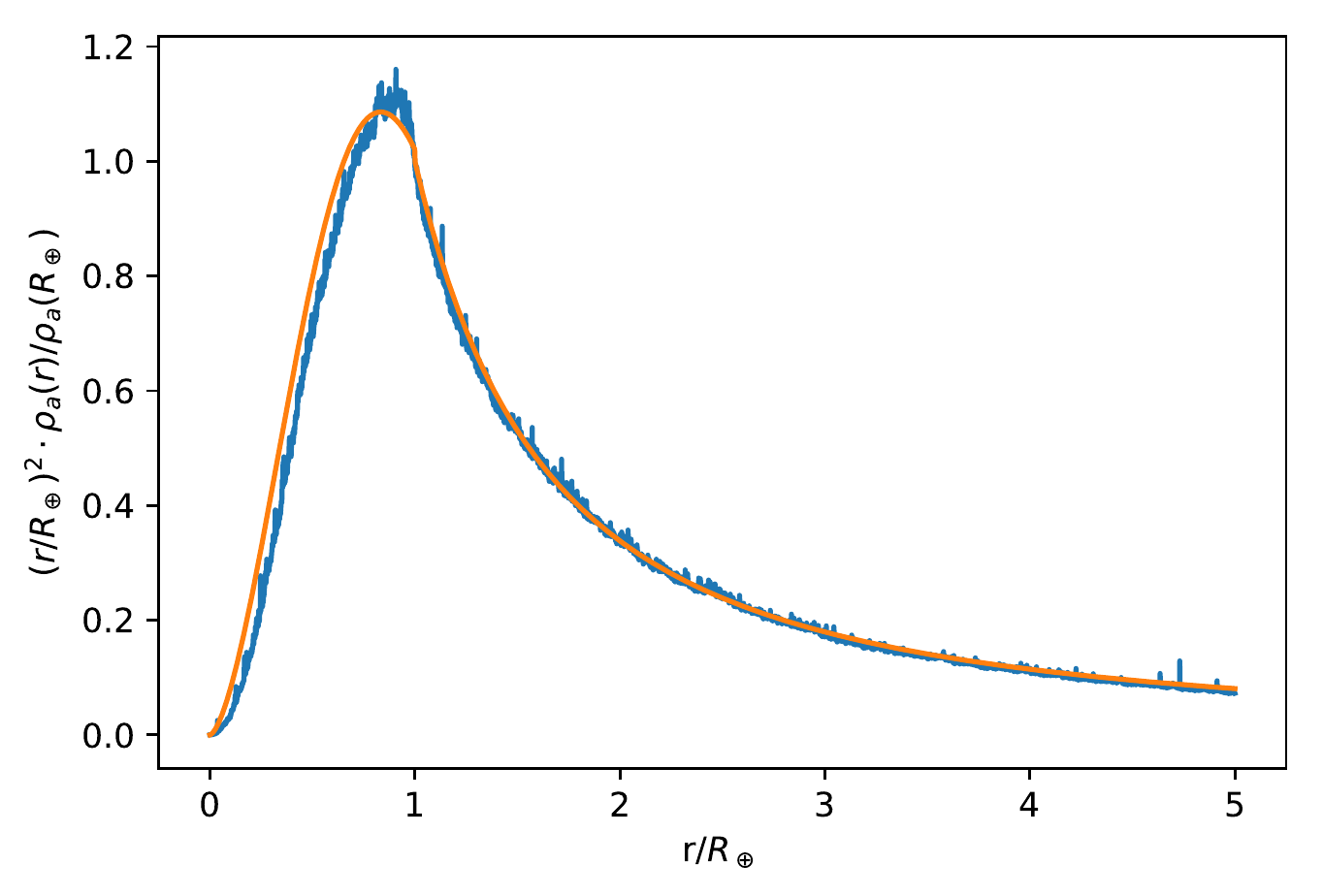}
	\caption{ $\alpha=2.5$, $\epsilon=3$}
	\label{fig:E plots - rhoa(r)_alpha2.5_Epsilon3}
	\end{subfigure}
	\begin{subfigure}[b]{0.47\linewidth}
	\includegraphics[width=\linewidth]{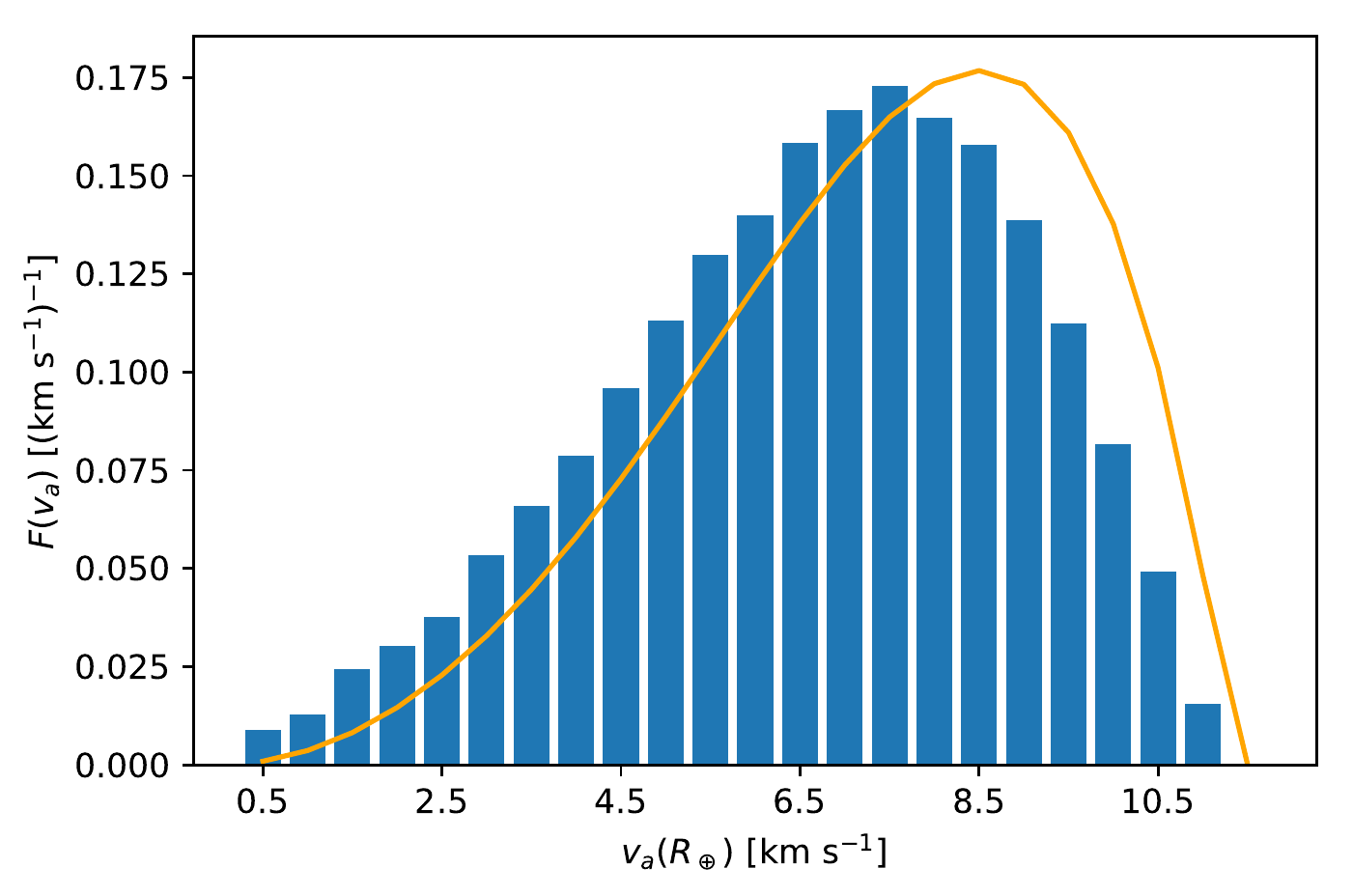}
	\caption{ $\alpha=2.0$, $\epsilon=1$, ${\bm \mu}=-\mu\hat{\mathbf{z}}$}
	\label{fig:E plots - F(va)_alpha2.0_dist2}
	\end{subfigure}
	\begin{subfigure}[b]{0.47\linewidth}
	\includegraphics[width=\linewidth]{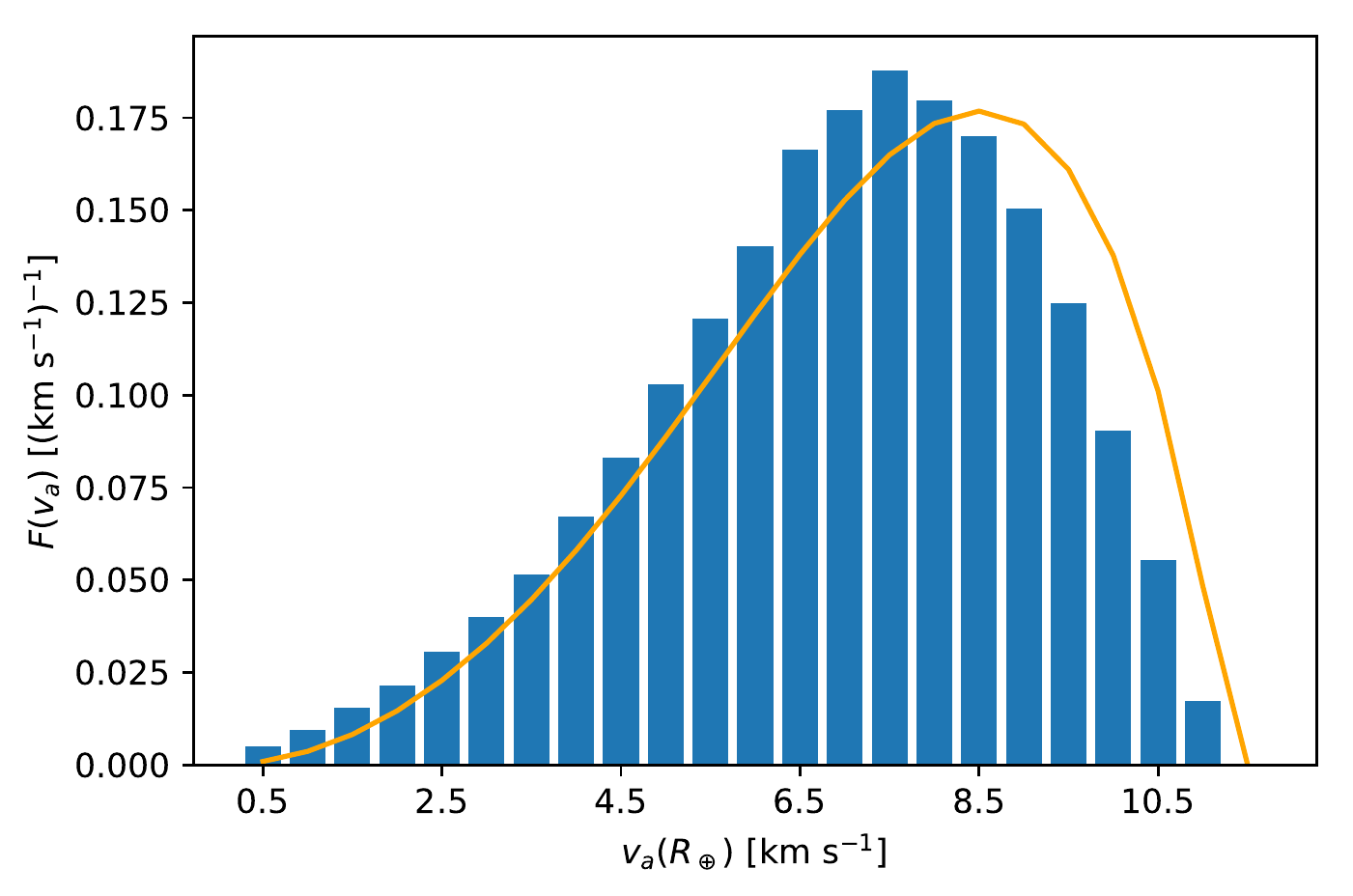}
	\caption{ $\alpha=2.5$, $\epsilon=3$}
	\label{fig:E plots - F(va)_alpha2.5_Epsilon3}
	\end{subfigure}
	\caption{Summary of simulations ($B_{\rm min}=3\times10^{24}$). The AQN heat emission profiles $q(r,v_{\rm AQN})$ are presented in the first row ($10^6$ samples), following by the axion density distribution $\rho_a(r)$	and the velocity spectrum $F(v_\mathrm{a})$ on Earth surface [with $v_{\rm esc}(r_0)=13.5\kmps$] respectively (blue, $2\times10^5$ samples). The analytical fits (orange) are also presented as in the main text. In general, all choices of parameters in Table. \ref{table:5.2 summary of some results} share similar features on the left column. The moderate exception is the case $\epsilon=3$ as presented on the right column here.	}
	\label{fig:E plots}
\end{figure*}

As mentioned earlier, we find the simulation results are largely similar in all cases, with one moderate exception $\epsilon=3$. For the sake of brevity,  we only present one common case [$(\alpha,B_{\rm min})=(2.0,3\times10^{24})$, $\epsilon=1$ and fixed-wind distribution] and the case with $\epsilon=3$ here. As shown in Fig. \ref{fig:E plots}, the heat emission profiles $q(r,v)$ is presented in the first row, following by by the axion density distribution $\rho_a(r)$ and the velocity spectrum $F(v_\mathrm{a})$ on Earth surface. 

We only comment on the sensitivity to the effective cross section $\epsilon$ (column on the right), as the common case (column on the left) presents the same features already discussed in the main text. It is found for a larger $\epsilon$, the heat distribution become  distorted in favor of a slower speed and more annihilation events  closer to the Earth surface.  This behaviour is well anticipated  as annihilation becomes more efficient. However, these modifications of the heat emission profile $q(r,v)$ have almost no effect on the axion density distribution $\rho_\mathrm{a}(R_{\oplus})$ and the axion spectral velocity distribution $F(v_\mathrm{a})$ on the surface of the Earth as we discuss below. 

We now move to the axion density distributions (Figs. \ref{fig:E plots - rhoa(r)_alpha2.0_dist2} and \ref{fig:E plots - rhoa(r)_alpha2.5_Epsilon3}) and velocity spectrums (Figs. \ref{fig:E plots - F(va)_alpha2.0_dist2} and \ref{fig:E plots - F(va)_alpha2.5_Epsilon3}), which are the main observable. We reach  a conclusion  that the simulation results are in general insensitive to almost all effects we considered. In fact, it is specifically insensitive to the baryon charge distribution of AQN models or the velocity distribution of DM flux. The effective cross section $\sim \epsilon$  modifies  the heat distribution as mentioned  above. However, this modification almost has no effect on the axion density $\rho_\mathrm{a}(R_{\oplus})$ and spectral properties $F(v_\mathrm{a})$ on the surface of the Earth, which are the relevant  parameters for the observations.

\exclude{
\begin{figure*}[!htp]
	\centering
	\begin{subfigure}[b]{0.47\linewidth}
		\includegraphics[width=\linewidth]{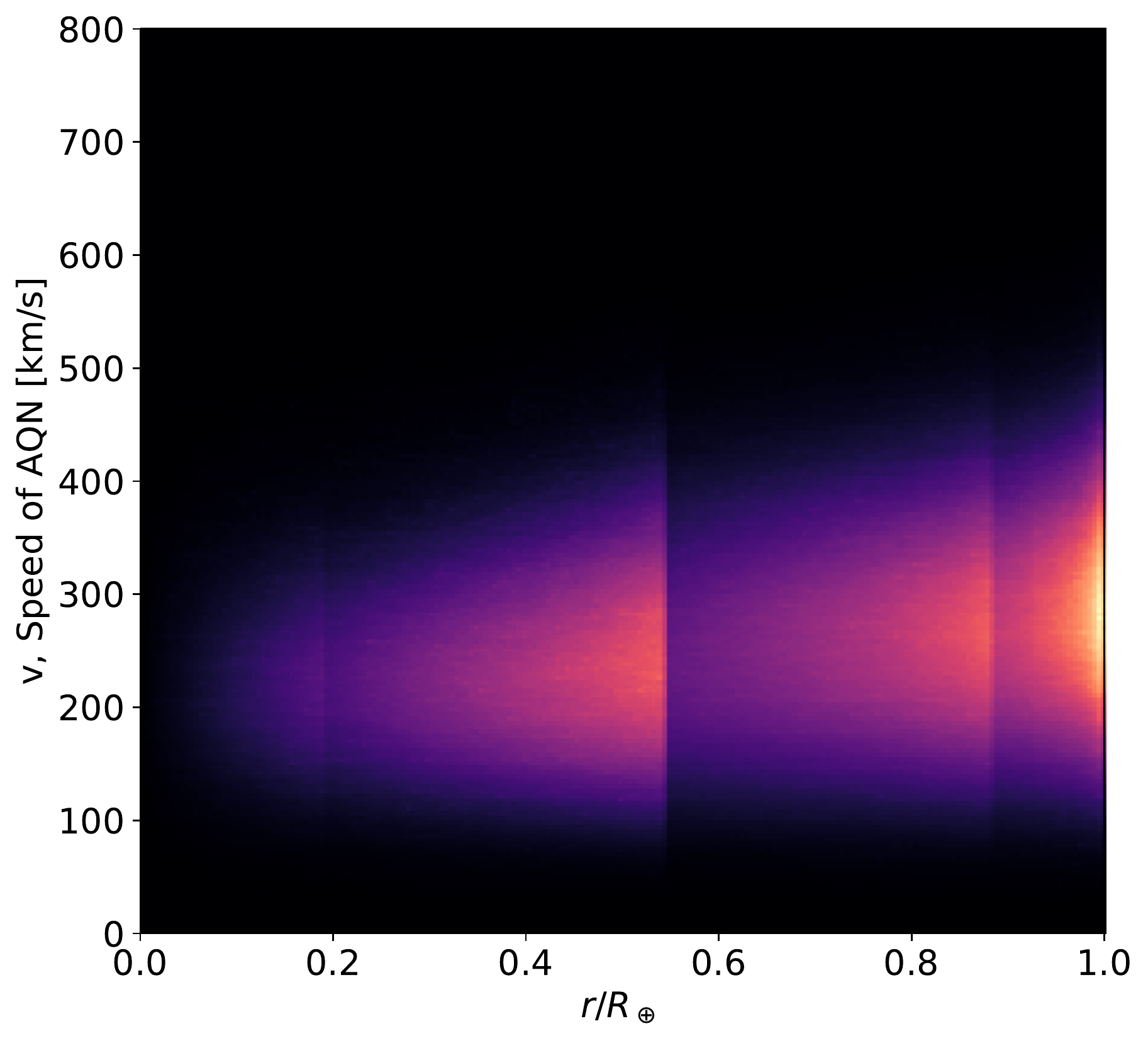}
		\caption{$\alpha=2.5$}
	\end{subfigure}
	\begin{subfigure}[b]{0.47\linewidth}
		\includegraphics[width=\linewidth]{AlexDist2_BMin3e24_Alpha25_Epsilon1_MillionAQN_200by200_spline_q_rv_.pdf}
		\caption{$\alpha=2.5$, ${\bm \mu}=-\mu\hat{\mathbf{z}}$}
	\end{subfigure}
	\begin{subfigure}[b]{0.47\linewidth}
	\includegraphics[width=\linewidth]{AlexDist1_BMin3e24_Alpha25_Epsilon1_MillionAQN_unstacked_vescp_spline.pdf}
	\caption{$\alpha=2.5$, $v_{\rm esc}=42.1 \kmps$}
	\end{subfigure}
	\begin{subfigure}[b]{0.47\linewidth}
	\includegraphics[width=\linewidth]{AlexDist1_BMin3e24_Alpha25_Epsilon3_MillionAQN_unstacked_spline.pdf}
	\caption{$\alpha=2.5$, $\epsilon=3$}
	\end{subfigure}
	\caption{2D histograms of $q(r,v_{\rm AQN})$, $B_{\rm min}=3\times10^{24}$ and $\epsilon=1$ unless specified. $10^6$ samples in the simulation.}
	\label{fig:E q_rv alpha2.5}
\end{figure*}

\begin{figure*}[h]
	\centering
	\begin{subfigure}[b]{0.47\linewidth}
	\includegraphics[width=\linewidth]{AlexDist1_BMin3e24_Alpha20_Epsilon1_MillionAQN_unstacked_spline.pdf}
	\caption{$\alpha=2.0$}
	\end{subfigure}
	\begin{subfigure}[b]{0.47\linewidth}
		\includegraphics[width=\linewidth]{AlexDist2_BMin3e24_Alpha2_Epsilon1_MillionAQN_200by200_spline_q_rv_.pdf}
		\caption{$\alpha=2.0$, ${\bm \mu}=-\mu\hat{\mathbf{z}}$}
	\end{subfigure}
	\begin{subfigure}[b]{0.47\linewidth}
		\includegraphics[width=\linewidth]{AlexDist1_BMin1e23_AlphaPieceWise_Epsilon1_MillionAQN_spline.pdf}
		\caption{$\alpha=(1.2, 2.5)$, $B_{\rm min}=10^{23}$}
	\end{subfigure}
	\begin{subfigure}[b]{0.47\linewidth}
		\includegraphics[width=\linewidth]{AlexDist2_BMin3e24_AlphaPieceWise_Epsilon1_MillionAQN_unstacked_spline.pdf}
		\caption{$\alpha=(1.2, 2.5)$, ${\bm \mu}=-\mu\hat{\mathbf{z}}$}
	\end{subfigure}
	\caption{2D histograms of $q(r,v_{\rm AQN})$, $B_{\rm min}=3\times10^{24}$ and $\epsilon=1$ unless specified. $10^6$ samples in the simulation.}
	\label{fig:E q_rv Others}
\end{figure*}

\begin{figure*}[h]
	\centering
	\captionsetup{justification=raggedright}
	\begin{subfigure}[b]{0.47\linewidth}
		\includegraphics[width=\linewidth]{pa_r_2e5axions_BMin3e24_Alpha25_dist1-fit.pdf}
		\caption{$\alpha=2.5$}
	\end{subfigure}
	\begin{subfigure}[b]{0.47\linewidth}
		\includegraphics[width=\linewidth]{pa_r_2e5axions_BMin3e24_Alpha25_dist2-fit.pdf}
		\caption{$\alpha=2.5$, ${\bm \mu}=-\mu\hat{\mathbf{z}}$}
	\end{subfigure}
	\begin{subfigure}[b]{0.47\linewidth}
	\includegraphics[width=\linewidth]{pa_r_2e5axions_BMin3e24_Alpha25_dist1_Vesc-fit.pdf}
	\caption{$\alpha=2.5$, $v_{\rm esc}=42.1 \kmps$}
	\end{subfigure}
	\begin{subfigure}[b]{0.47\linewidth}
	\includegraphics[width=\linewidth]{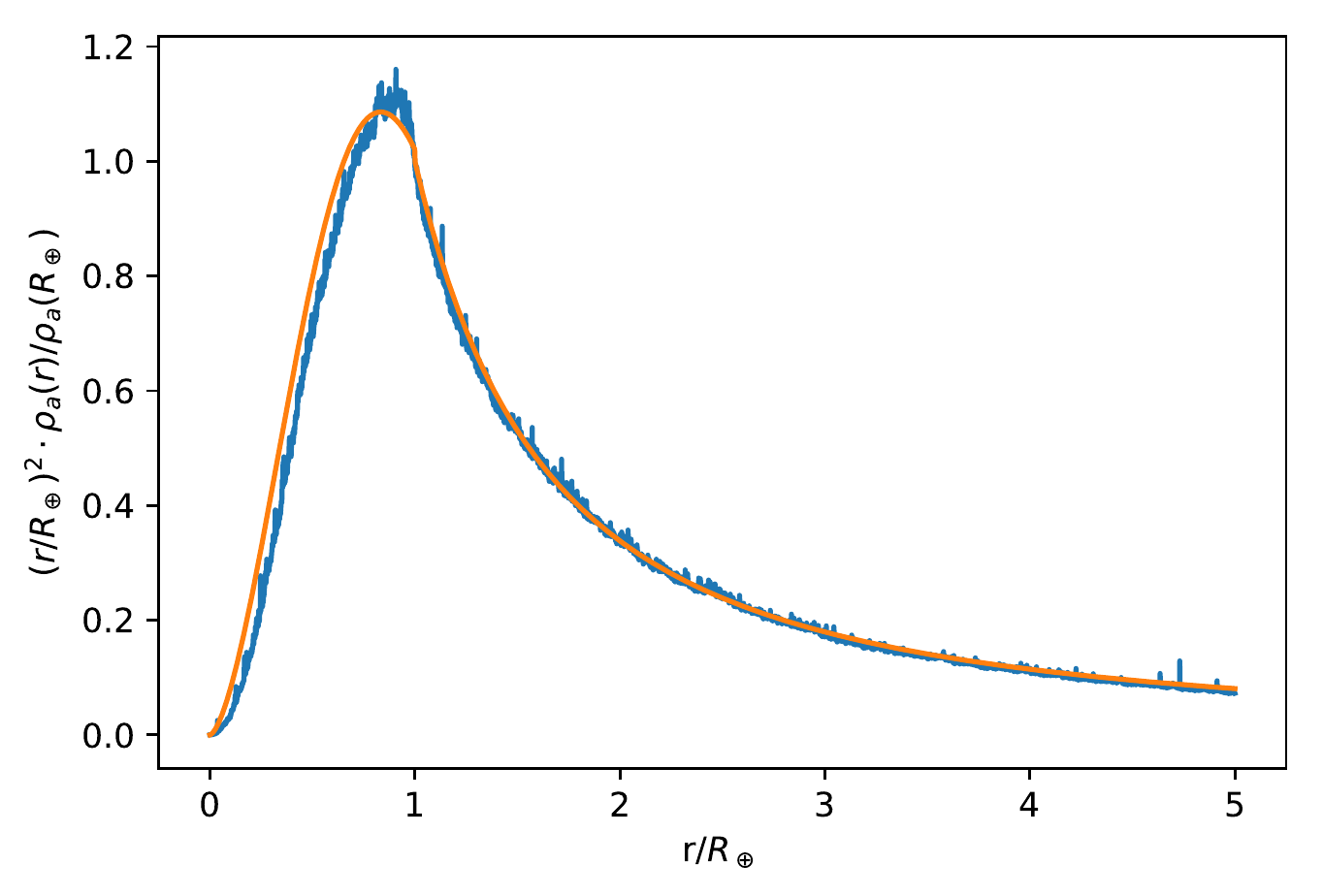}
	\caption{$\alpha=2.5$, $\epsilon=3$}
	\end{subfigure}
	\caption{$\rho_a(r)$ vs. $r$, $B_{\rm min}=3\times10^{24}$ and $\epsilon=1$ unless specified. The simulation (blue) is compared with the analytical fit (orange). $2\times10^5$ samples in the simulation.}
	\label{fig:E rho_a alpha2.5}
\end{figure*}

\begin{figure*}[h]
	\centering
	\captionsetup{justification=raggedright}
	\begin{subfigure}[b]{0.47\linewidth}
	\includegraphics[width=\linewidth]{pa_r_2e5axions_BMin3e24_Alpha20_dist1-fit.pdf}
	\caption{$\alpha=2.0$}
	\end{subfigure}
	\begin{subfigure}[b]{0.47\linewidth}
		\includegraphics[width=\linewidth]{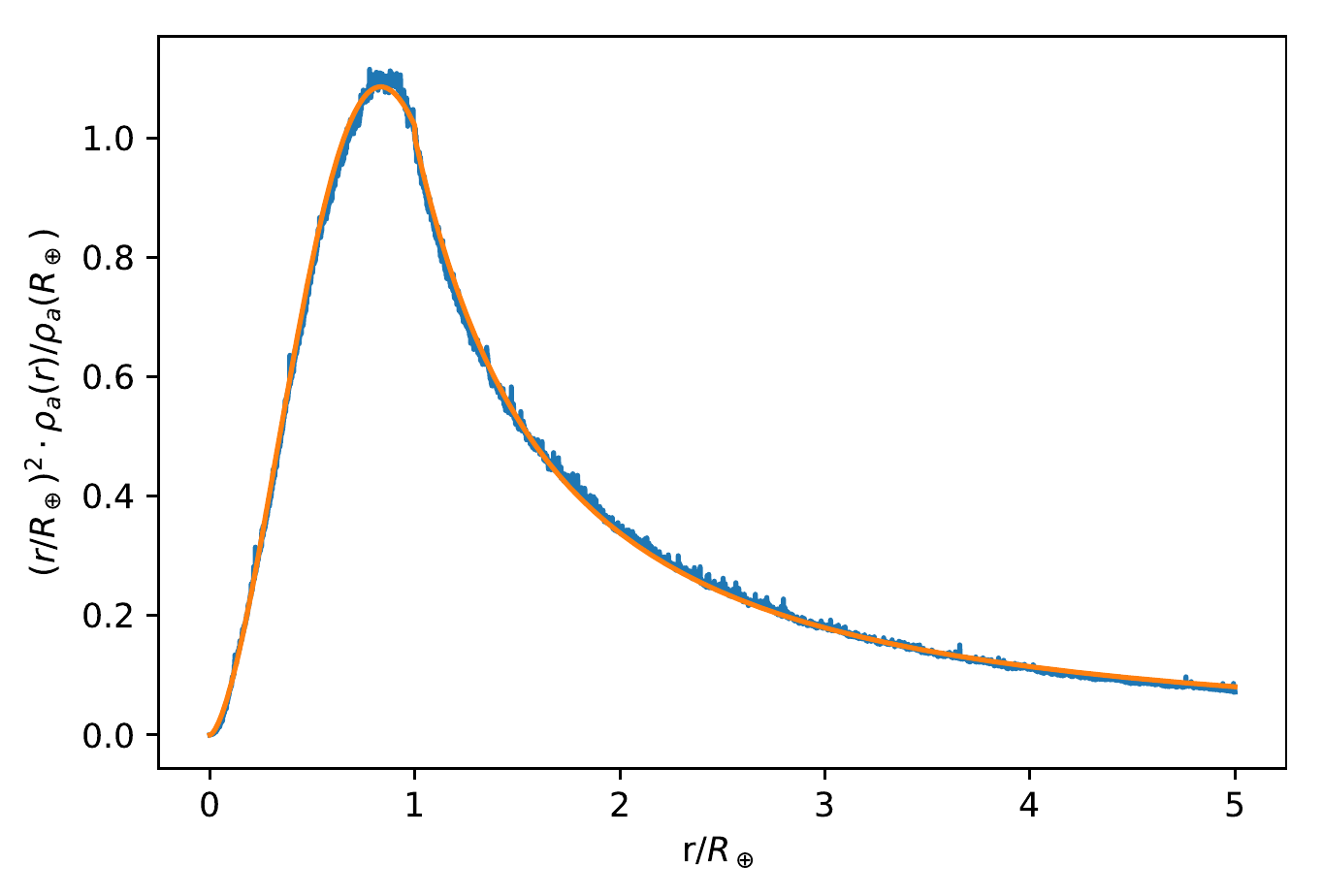}
		\caption{$\alpha=2.0$, ${\bm \mu}=-\mu\hat{\mathbf{z}}$}
	\end{subfigure}
	\begin{subfigure}[b]{0.47\linewidth}
		\includegraphics[width=\linewidth]{pa_r_2e5axions_BMin1e23_AlphaPW_dist1-fit.pdf}
		\caption{$\alpha=(1.2, 2.5)$, $B_{\rm min}=10^{23}$}
	\end{subfigure}
	\begin{subfigure}[b]{0.47\linewidth}
		\includegraphics[width=\linewidth]{pa_r_2e5axions_BMin3e24_AlphaPW_dist2-fit.pdf}
		\caption{$\alpha=(1.2, 2.5)$, ${\bm \mu}=-\mu\hat{\mathbf{z}}$}
	\end{subfigure}
	\caption{$\rho_\mathrm{a}(r)$ vs. $r$, $B_{\rm min}=3\times10^{24}$ and $\epsilon=1$ unless specified. The simulation (blue) is compared with the analytical fit (orange). $2\times10^5$ samples in the simulation.}
	\label{fig:E rho_a Others}
\end{figure*}

\begin{figure*}[h]
	\centering
	\captionsetup{justification=raggedright}
	\begin{subfigure}[b]{0.47\linewidth}
		\includegraphics[width=\linewidth]{F_va_2e5axions_BMin3e24_Alpha25_dist1.pdf}
		\caption{$\alpha=2.5$}
	\end{subfigure}
	\begin{subfigure}[b]{0.47\linewidth}
		\includegraphics[width=\linewidth]{F_va_2e5axions_BMin3e24_Alpha25_dist2.pdf}
		\caption{$\alpha=2.5$, ${\bm \mu}=-\mu\hat{\mathbf{z}}$}
	\end{subfigure}
	\begin{subfigure}[b]{0.47\linewidth}
	\includegraphics[width=\linewidth]{F_va_2e5axions_BMin3e24_Alpha25_dist1_Vesc.pdf}
	\caption{$\alpha=2.5$, $v_{\rm esc}=42.1 \kmps$}
	\end{subfigure}
	\begin{subfigure}[b]{0.47\linewidth}
	\includegraphics[width=\linewidth]{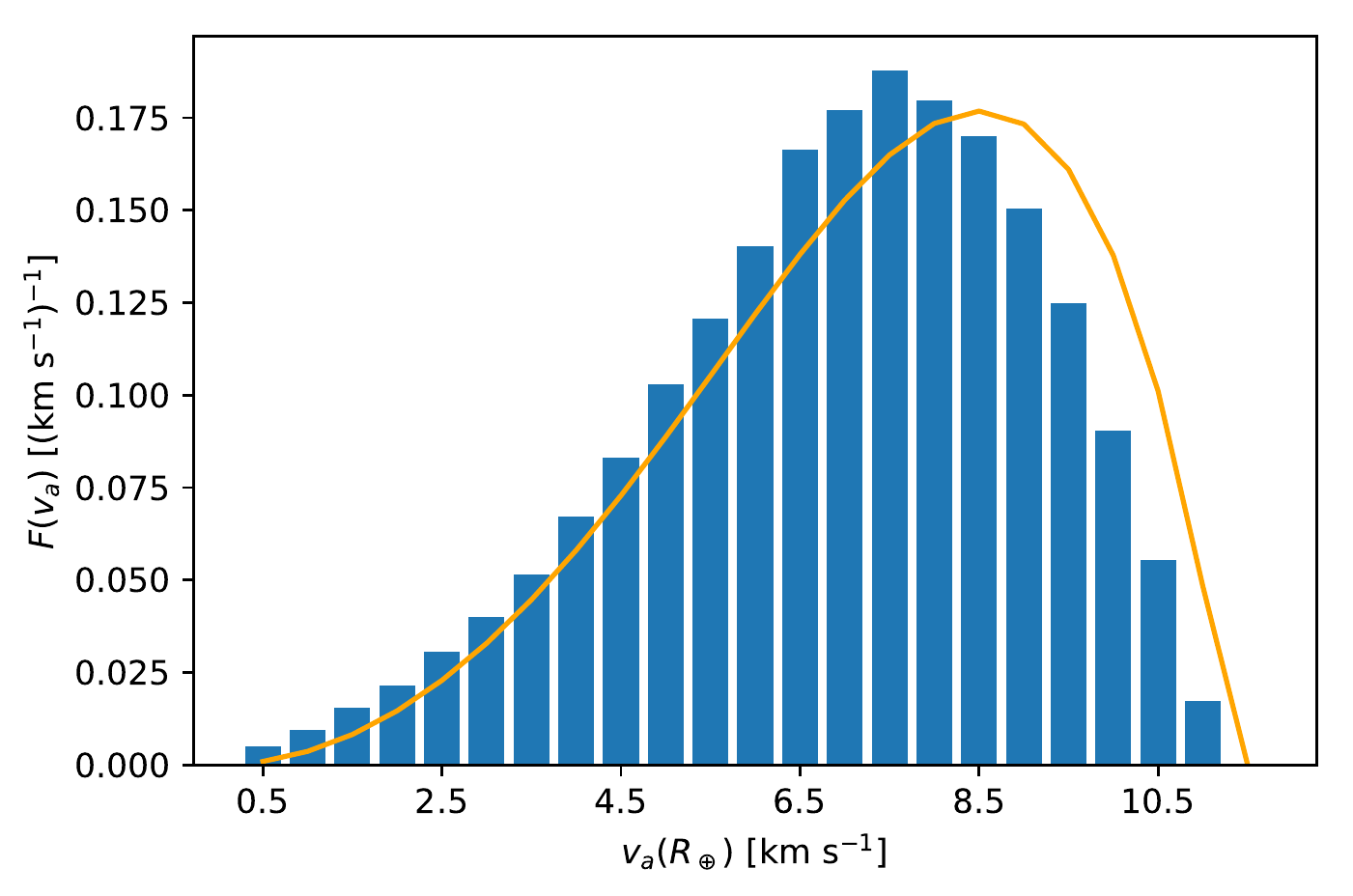}
	\caption{$\alpha=2.5$, $\epsilon=3$}
	\end{subfigure}
	\caption{Velocity profile on Earth surface $F(v_\mathrm{a})$ vs. $v_\mathrm{a}(R_\oplus)$, $B_{\rm min}=3\times10^{24}$ and $\epsilon=1$ unless specified. The simulation (blue) is compared with the analytical estimation $F^{(\rm est)}(v_\mathrm{a})$ (orange) with $v_{\rm esc}(r_0)=13.5\kmps$. $2\times10^5$ samples in the simulation.}
	\label{fig:E F(va) alpha2.5}
\end{figure*}

\begin{figure*}[h]
	\centering
	\captionsetup{justification=raggedright}
	\begin{subfigure}[b]{0.47\linewidth}
	\includegraphics[width=\linewidth]{F_va_2e5axions_BMin3e24_Alpha20_dist1.pdf}
	\caption{$\alpha=2.0$}
	\end{subfigure}
	\begin{subfigure}[b]{0.47\linewidth}
		\includegraphics[width=\linewidth]{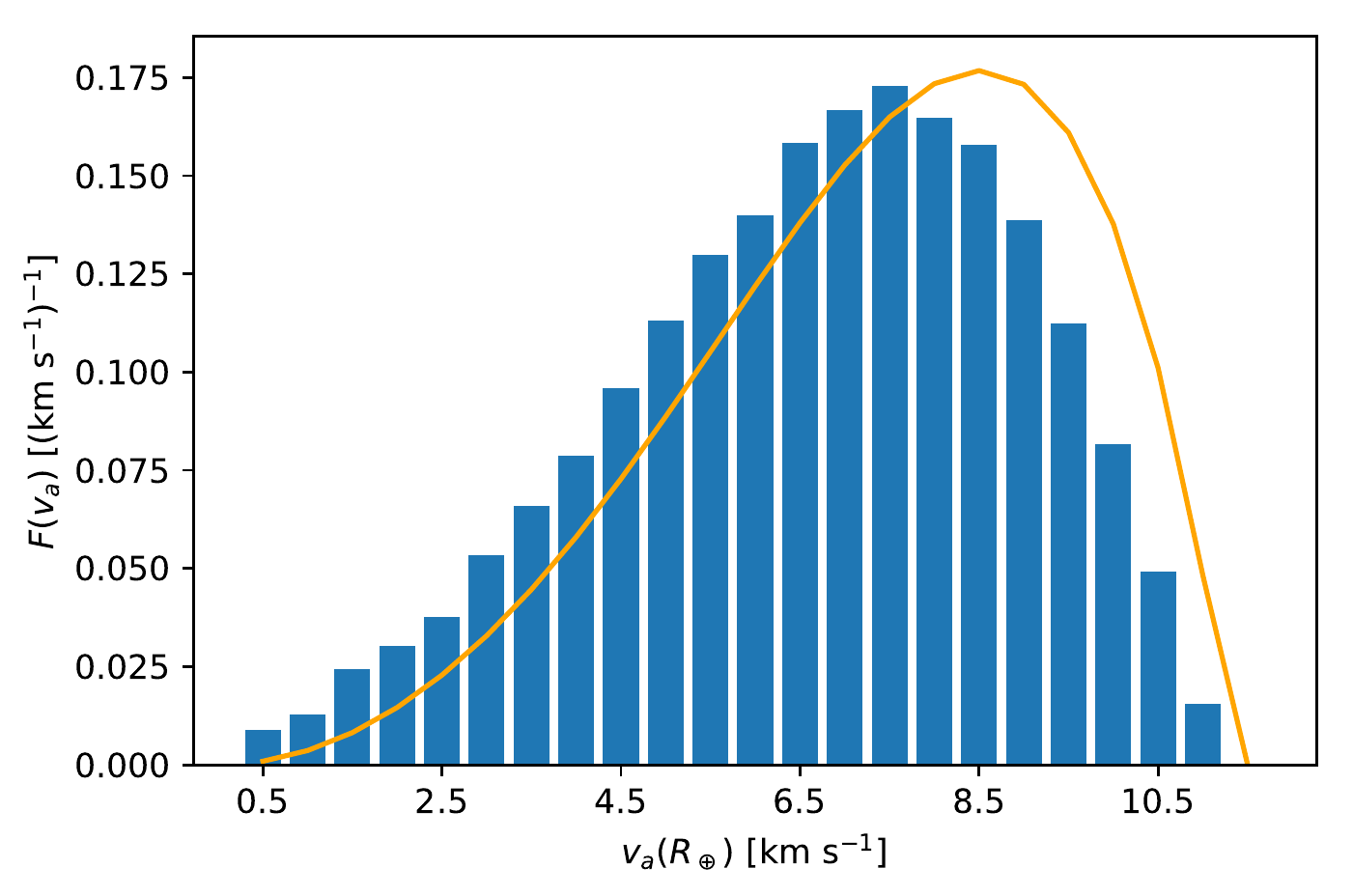}
		\caption{$\alpha=2.0$, ${\bm \mu}=-\mu\hat{\mathbf{z}}$}
	\end{subfigure}
	\begin{subfigure}[b]{0.47\linewidth}
	\includegraphics[width=\linewidth]{F_va_2e5axions_BMin1e23_AlphaPW_dist1.pdf}
		\caption{$\alpha=(1.2, 2.5)$, $B_{\rm min}=10^{23}$}
	\end{subfigure}
	\begin{subfigure}[b]{0.47\linewidth}
		\includegraphics[width=\linewidth]{F_va_2e5axions_BMin3e24_AlphaPW_dist2.pdf}
		\caption{$\alpha=(1.2, 2.5)$, ${\bm \mu}=-\mu\hat{\mathbf{z}}$}
	\end{subfigure}
	\caption{Velocity profile on Earth surface $F(v_\mathrm{a})$ vs. $v_\mathrm{a}(R_\oplus)$, $B_{\rm min}=3\times10^{24}$ and $\epsilon=1$ unless specified. The simulation (blue) is compared with the analytical estimation $F^{(\rm est)}(v_\mathrm{a})$ (orange) with $v_{\rm esc}(r_0)=13.5\kmps$. $2\times10^5$ samples in the simulation.}
	\label{fig:E F(va) Others}
\end{figure*}

}

\FloatBarrier
 
\bibliography{axion_halo}

\end{document}